\documentclass[twocolumn,twocolappendix]{aastex7}

\usepackage{amsmath}
\usepackage{amssymb}
\usepackage{CJK}

\begin{document}

\title{The Balmer Break and Optical Continuum of Little Red Dots from Super-Eddington Accretion}

\begin{CJK*}{UTF8}{gbsn}

\author[0000-0003-2488-4667]{Hanpu Liu (刘翰溥)}
\affiliation{Department of Astrophysical Sciences, Princeton University, 4 Ivy Lane, Princeton, NJ 08544, USA}
\email[show]{hanpu.liu@princeton.edu}

\author[0000-0002-2624-3399]{Yan-Fei Jiang (姜燕飞)}
\affiliation{Center for Computational Astrophysics, Flatiron Institute, New York, NY 10010, USA}
\email{yjiang@flatironinstitute.org}

\author[0000-0001-9185-5044]{Eliot Quataert}
\affiliation{Department of Astrophysical Sciences, Princeton University, 4 Ivy Lane, Princeton, NJ 08544, USA}
\email{quataert@princeton.edu}

\author[0000-0002-5612-3427]{Jenny E. Greene}
\affiliation{Department of Astrophysical Sciences, Princeton University, 4 Ivy Lane, Princeton, NJ 08544, USA}
\email{jgreene@astro.princeton.edu}

\author[0000-0002-0463-9528]{Yilun Ma (马逸伦)}
\affiliation{Department of Astrophysical Sciences, Princeton University, 4 Ivy Lane, Princeton, NJ 08544, USA}
\email{yilun@princeton.edu}

\begin{abstract}

The physical origin of Little Red Dots (LRDs)—compact extragalactic sources with red rest-optical continua and broad Balmer lines—remains elusive. The redness of LRDs is likely intrinsic, suggesting optically thick gas emitting at a characteristic effective temperature of $\sim5000{\rm~K}$. Meanwhile, many LRD spectra exhibit a Balmer break, often attributed to absorption by a dense gas shell surrounding an AGN. Using semi-analytical atmosphere models and radiation transport calculations, we show that a super-Eddington accretion system can give rise to a Balmer break and a red optical color simultaneously, without invoking external gas absorption for the break or dust reddening. The break originates from a discontinuity in opacity across the Balmer limit, similar to that of early-type stars, but the lower photosphere density of super-Eddington systems, $\rho<10^{-9}{\rm~g~cm^{-3}}$, implies a significant opacity contrast even at a cool photosphere temperature of $\sim5000{\rm~K}$. Furthermore, while accretion in the form of a standard thin disk requires fine tuning to match the optical color of LRDs, an alternative scenario of a geometrically thick, roughly spherical accretion flow implies an effective temperature $4000{\rm~K}\lesssim T_{\rm eff}\lesssim6000{\rm~K}$ that is very insensitive to the accretion rate (analogous to the Hayashi line in stellar models). The continuum spectra from the latter scenario align with the Balmer break and optical color of currently known LRDs. We discuss predictions of our model and the prospects for more realistic spectra based on super-Eddington accretion simulations.   

\end{abstract}


\section{Introduction} 
\label{sec:intro}
Since its launch, the James Webb Space Telescope (JWST) has revealed a previously unrecognized population: compact sources that exhibit broad emission lines and a distinctive V-shaped continuum, characterized by blue rest-frame ultraviolet (UV) and red rest-frame optical colors \citep[e.g.,][]{Labbe2023,Labbe2025,Kokorev2024,Greene2024,Kocevski2024,Matthee2024}. These objects, now colloquially referred to as ``Little Red Dots" (LRDs), have prompted intense investigation as the community seeks to understand their physical nature.

The origin of the redness of LRDs remains an open question. Spectroscopically selected LRDs have rest-optical colors comparable to blackbodies of $T\sim4000$--$5000{\rm~K}$ \citep[e.g., see Figure~1 in][where a 5000 or 4000~K blackbody would have a rest-frame color of $\rm B-R=1.20$ or 1.86 in their definition]{Setton2024}. At first, dust attenuation of an AGN, a stellar population, or mixture of both seemed a straightforward candidate explanation, and spectral continuum fits typically infer $A_V>1$ \citep[e.g.,][]{Kocevski2023,Greene2024}. However, the presence of dust is stringently constrained by the nondetection of its thermal emission in the rest-mid- and far-infrared \citep{Williams2024,Akins2024,Labbe2025,Wang2025,Setton2025,Xiao2025,Casey2025}, indicating that the rest-optical continuum emission is likely intrinsically red rather than dust-reddened. In addition, the strikingly high abundance of LRDs at high redshifts \citep[a few percent of the number density of UV-selected galaxies at $z\sim5$, e.g.,][]{Kokorev2024,Matthee2024,Greene2024,Zhuang2025} implies that the optical redness must be produced without fine-tuning. 

Another puzzling property of LRDs is the common presence of a Balmer break in their spectra. Even for objects that do not exhibit a strong spectral discontinuity, the turnover points of the V-shaped continuum appear to be consistently located at the Balmer limit \citep{Setton2024}. The break feature suggests that an evolved stellar population may dominate the observed flux at $\lambda_{\rm rest}\sim4000{\rm~\AA}$.   However, this picture is not easily reconciled with the luminous Balmer emission lines \citep{Wang2024,Ma2025}, and the central stellar density implied by an evolved stellar population model is rarely seen, if at all, at lower redshifts \citep[e.g.,][]{Baggen2024,Labbe2024,Ma2025}. Moreover, the Balmer breaks of some recently reported objects are too strong to be explained by standard stellar population models \citep{Naidu2025,deGraaff2025} and demand a non-stellar origin. A high covering fraction of dense gas, where neutral hydrogen is collisionally excited to the state of principal quantum number $n=2$, is proposed to produce the break feature without stars \citep{Inayoshi2025}, and has since been applied to interpret objects showing strong breaks \citep[e.g.,][]{Ji2025,Naidu2025,deGraaff2025,Taylor2025}. This scenario provides a phenomenological solution to the Balmer break but lacks a dynamical picture of the inferred dense gas shell. Moreover, these models have limited predictive power for the optical color, which remains sensitive to the assumed incident spectrum and gas conditions.

Independently, increasing attention has been given to the hypothesis that LRDs are powered by super-Eddington accretion onto a central supermassive black hole \citep{Pacucci2024,Lambrides2024,Inayoshi2024,Madau2024,Trinca2024}. Super-Eddington accretion was proposed in part to reconcile the broad permitted emission lines seen in LRDs with features that differ from typical Type-I AGNs, including the weakness in X-ray emission \citep{Yue2024,Ananna2024,Maiolino2025} and the scarcity of UV/optical variability \citep{Kokubo2024,ZhangZ2025}. However, whether super-Eddington accretion can in fact produce the optical redness and the Balmer breaks of LRDs is underexplored. This motivates us to explore simple super-Eddington models in the context of the continuum signatures in JWST observations.  

\begin{figure}
    \centering
    \includegraphics[width=0.46\textwidth]{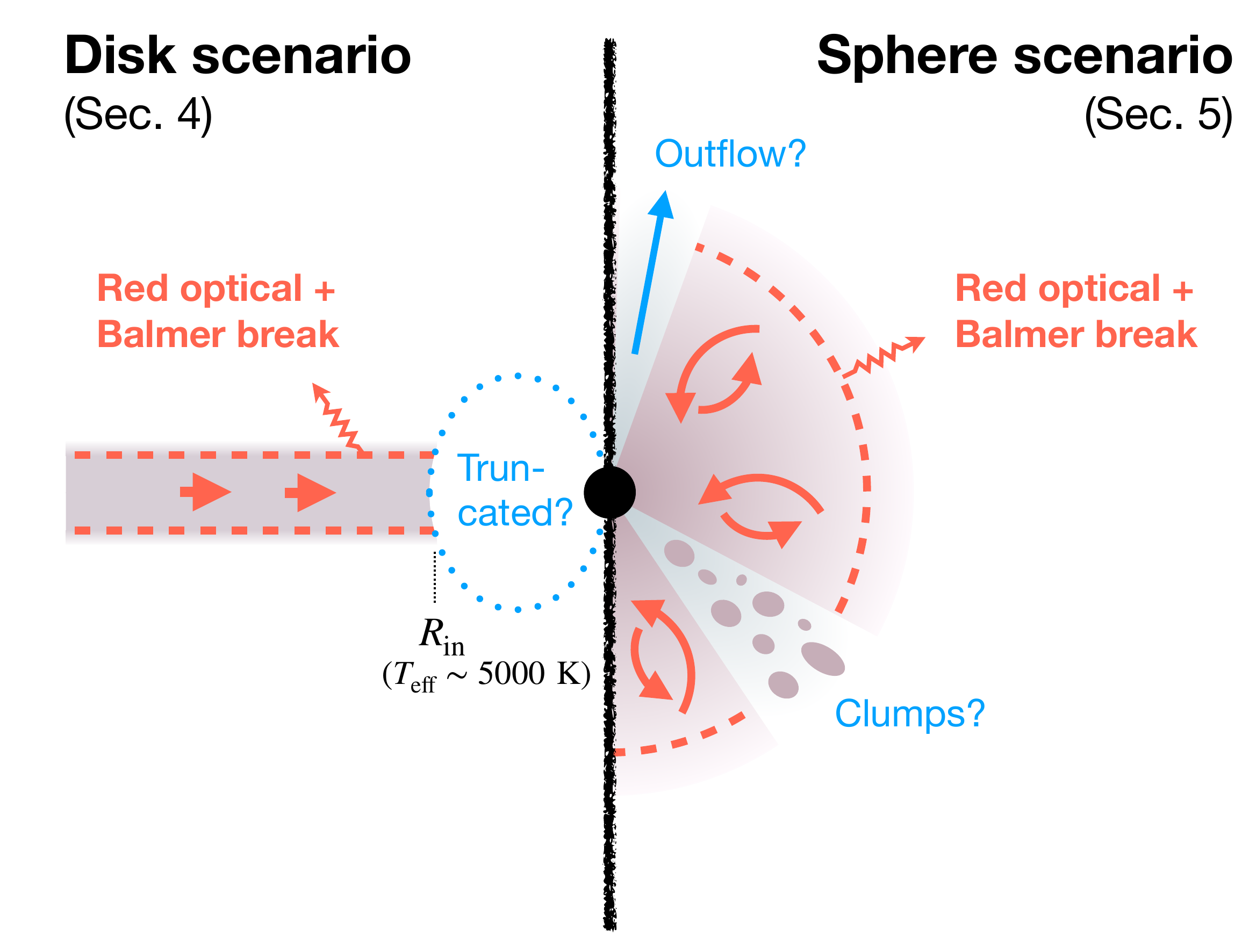}
    \caption{Cartoon showing the two scenarios considered in this work. Straight or curved arrows indicate direction of gas flow. Dashed lines represent the photosphere. Wiggle arrows indicate photons. \textit{Left}: A standard thin disk in super-Eddington accretion, which we investigate in Section~\ref{sec:disk}. It reproduces the red optical color and Balmer break of LRDs if the disk has an inner truncation radius where the effective temperature is $\sim5000{\rm~K}$, which is fine-tuned. \textit{Right}: A spherical gas profile undergoing turbulent accretion, which we investigate in Section~\ref{sec:sphere}. This model gives rise to a red optical continuum and the Balmer break over a large range of gas densities (a proxy for a wide range of accretion rates). Components with question marks, i.e., truncation in the disk scenario and outflow and clumps in the sphere scenario, are discussed in Section~\ref{sec:conclusions}.}
    \label{fig:cartoon}
\end{figure}

In this work, we construct idealized semi-analytical models for the gas around a supermassive black hole undergoing super-Eddington accretion and investigate the resulting UV/optical continuum emission. Figure~\ref{fig:cartoon} sketches the models considered. First, we study a standard geometrically thin accretion disk, as shown in the left half of the cartoon. Such a thin disk may be present at super-Eddington accretion rates at large radii. This disk must, however, be truncated at an unfavorably fine-tuned inner radius to reproduce the redness of LRDs. An alternative picture, more consistent with super-Eddington simulations \citep[e.g.,][]{Ohsuga2005,Jiang2014,McKinney2014,Sadowski2014,Jiang2019,Hu2022}, is that a combination of inflow and outflow is distributed over a wide range of polar angles. In this initial exploratory work, we approximate this with a spherical model, as indicated on the right in Figure~\ref{fig:cartoon}.

We investigate whether super-Eddington accretion flows with these simple geometries can reproduce the observed properties of LRDs. Using radiation transport simulations, we compute the gas temperature and emergent continuum flux ab initio, capturing the optical color and the Balmer break self-consistently. Our results revolve around two themes. First, if super-Eddington accretion produces a red optical color from a cool photosphere, then a Balmer break is to be expected from this very photosphere. This is true for both the disk and sphere scenarios and stems from the low density of the gas at the photosphere, $<10^{-9}{\rm~g~cm^{-3}}$, compared to that of most main-sequence stars. Second, the cool photosphere itself is a natural outcome of the sphere scenario, in which the effective temperature is very insensitive to the accretion rate in the range $4000{\rm~K}\lesssim T_{\rm eff}\lesssim6000{\rm~K}$ due to the temperature dependence of the opacity law \citep[see a similar scenario in][which we became aware of near the completion of this manuscript]{Kido2025}.

In the rest of this paper, Section~\ref{sec:Balmer_break} provides the physical grounds for our first theme, where we use a toy model to show that the Balmer break can emerge from a cool, low-density photosphere. We describe in Section~\ref{sec:numerical_methods} our numerical methods common for the disk and sphere models. In Sections~\ref{sec:disk} and \ref{sec:sphere}, we construct analytical density structures for the disk and sphere scenarios, respectively, and report our simulation results of the emergent continuum spectra. In particular, readers mainly interested in the spectral results may move directly to Section~\ref{subsec:disk_spec} and \ref{subsec:sphere_multi}. We summarize our results, compare this study to previous works, and outline future directions in Section~\ref{sec:conclusions}.  We have made our model continuum spectra of the sphere scenario publicly available on Zenodo\footnote{\url{https://doi.org/10.5281/zenodo.17204644}}. Throughout this work, we refer to radiation frequencies and wavelengths in the rest frame unless otherwise noted. We adopt the \citet{PlanckColl2016} cosmological parameters, i.e., $\Omega_m =0.307$, $\Omega_\Lambda =0.693$, $\Omega_b =0.0486$, and $H_0=67.7{\rm~km~s^{-1}~Mpc^{-1}}$. 

\section{Balmer break from cool, low-density photospheres}
\label{sec:Balmer_break}
The Balmer break is well studied in the stellar context: it is associated with a discontinuity in opacity in wavelength space. The same radiation microphysics should apply to supermassive black hole accretion flows \citep[e.g.,][who showed that spectral discontinuities could be present in theoretical AGN disk spectra]{Hubeny2000}. However, among main-sequence stars, only those with $T_{\rm eff}\sim10^4{\rm~K}$ show a strong Balmer break, raising the question of how this feature can be reconciled with the cooler optical color temperatures ($\sim5000{\rm~K}$) observed in LRDs. In this section, we build a toy model to show that the same opacity law responsible for the Balmer break in early-type stellar atmospheres can also produce this feature in photospheres at $T\sim5000{\rm~K}$. This occurs if the photospheric density is sufficiently low, which can result from super-Eddington accretion flows.

In general, non-blackbody spectral features may arise from an atmosphere with a temperature gradient along the line of sight. Define a frequency-dependent radiation temperature, $T_{{\rm rad},\nu}$. It is determined by the gas temperature at the frequency-dependent photosphere, i.e., the location where the monochromatic optical depth, $\tau_\nu$, becomes of order unity. If the optical depth strongly varies with frequency, the photosphere locations of two nearby frequencies may differ substantially, which then implies a varying radiation temperature with frequency. In the special case of the Balmer break, the optical depth variation originates from the photoionization opacity as a neutral hydrogen atom with principal quantum number $n=2$ absorbs a photon of wavelength $\lambda<\lambda_{\infty}=3645{\rm~\AA}$ and becomes ionized. 

We illustrate the Balmer break using the following toy model. In a 1D spatial coordinate in $z$, we assume a plane-parallel gas slab of constant density $\rho$ for $z>0$ and vacuum for $z<0$. The gas temperature is given by a simple relation, $T^4(z)\propto z$, which would be exact in the diffusive energy transport regime satisfying local thermal equilibrium (LTE) with a constant Rosseland mean opacity. The radiation temperature at a given frequency is then equal to the gas temperature at the photosphere location, $z_{\rm ph,\nu}$, given by $\rho\kappa_{{\rm eff},\nu}z_{\rm ph,\nu}=2/3$, where $\kappa_{{\rm eff},\nu}\equiv\sqrt{3\kappa_{a,\nu}(\kappa_{a,\nu}+\kappa_{s,\nu})}$ is the effective opacity accounting for the absorption and scattering opacities $\kappa_{a,\nu}$ and $\kappa_{s,\nu}$ \citep{Rybicki1979}. The photosphere (or, more precisely, thermalization surface) defined with the effective opacity characterizes where the radiation becomes thermalized with the gas, and we keep using the term ``photosphere'' in this sense throughout this paper. For simplicity, we assume that $\kappa_{{\rm eff},\nu}$ is constant along the slab even though it would change with temperature in reality. 

The above assumptions relate the radiation temperature to the effective opacity: $T_{{\rm rad},\nu}\propto z_{\rm ph,\nu}^{1/4}\propto \kappa_{{\rm eff},\nu}^{-1/4}$. To characterize the Balmer break, we choose two representative wavelengths at $3600{\rm~\AA}$ and $4000{\rm~\AA}$ and calculate the effective continuum opacity ratio, $\kappa_{{\rm eff}}(4000{\rm~\AA})/\kappa_{{\rm eff}}(3600{\rm~\AA})$, to indicate the strength of the Balmer break. A value of this ratio being significantly less than unity will imply a large excess of $T_{\rm rad}$ at $4000{\rm~\AA}$ relative to $3600{\rm~\AA}$ and hence a strong break. The details of our opacity calculations are described in Section~\ref{subsec:opacity}. 

\begin{figure}
    \centering
    \includegraphics[width=0.49\textwidth]{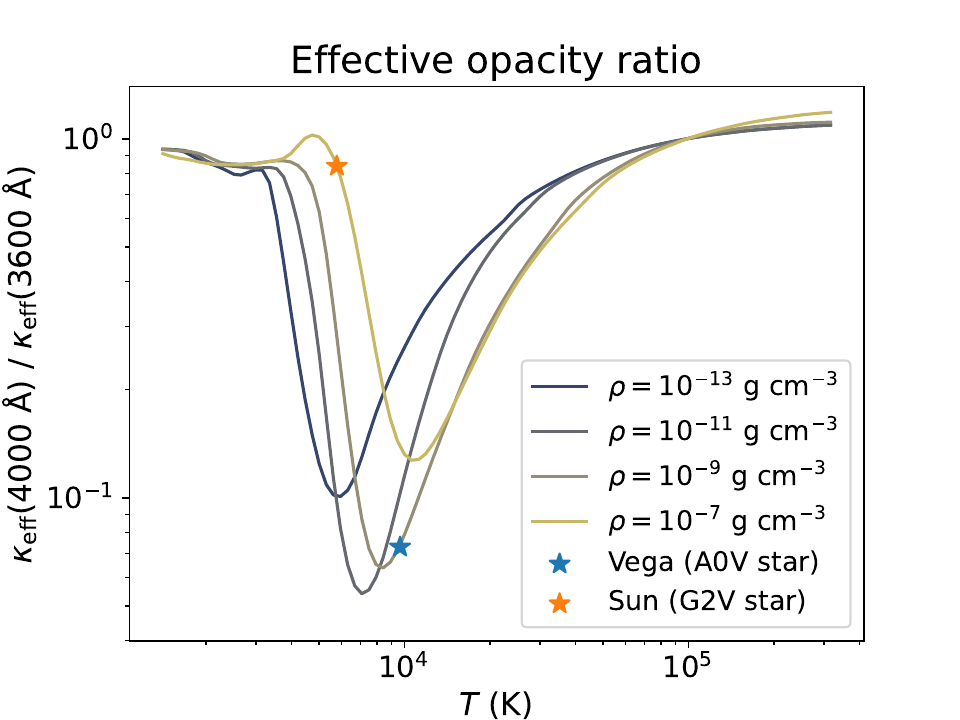}
    \caption{The ratio of two continuum effective opacities (accounting for absorption and scattering), $\kappa_{{\rm eff}}(4000{\rm~\AA})/\kappa_{{\rm eff}}(3600{\rm~\AA})$, for different temperature and density. A ratio significantly below one tends to produce a Balmer break. Star symbols mark the parameters corresponding to the star Vega and the Sun. The optimal temperature to produce a Balmer break increases with density.}
    \label{fig:kappa_ratio}
\end{figure}

Figure~\ref{fig:kappa_ratio} shows the effective opacity ratio as a function of temperature and density, assuming an ideal gas of metallicity $Z=0.1~Z_\odot$. Each curve has a significant trough, indicating the photosphere temperature favorable to producing the Balmer break at a given atmosphere density. From now on, for a given density, we term the temperature corresponding to the minimum of the trough the ``optimal temperature'' for the Balmer break. A typical early A-type main-sequence star, e.g., Vega, has a photosphere gas density of the order $10^{-9}{\rm~g~cm^{-3}}$ \citep{Dreiling1980}, where Figure~\ref{fig:kappa_ratio} indicates an optimal temperature of $\sim9000{\rm~K}$. The star has an effective temperature of the same value. This agrees with the observed strong breaks of such stars. In contrast, the Sun does not show a strong Balmer break with a photosphere temperature of $\sim6000{\rm~K}$ and a density of the order $10^{-7}{\rm~g~cm^{-3}}$ \citep{Holweger1974}. The crucial message from the figure is that the trough shifts to lower temperature as the characteristic atmosphere density decreases; as the density is reduced to $<10^{-10}{\rm~g~cm^{-3}}$, even an atmosphere temperature of $\sim5000{\rm~K}$ will give rise to a low opacity ratio and hence to a Balmer break. 

The positive correlation between the density and the optimal temperature for the Balmer break is known for early-type stars \citep[e.g., see Figure~10 in][]{Mihalas1965} and can be understood from thermal equilibrium. The difference in the absorption opacity on either side of the Balmer break is equal to the photoionization opacity of the $n=2$ hydrogen at $\sim3600{\rm~\AA}$, $\kappa_{2\to\infty}$:
\begin{equation}
    \kappa_{2\to\infty} = \frac{n_2\sigma_{2\to\infty}}{\rho} = \frac{X\sigma_{2\to\infty}}{m_p}\frac{n_2}{n_{\rm H}}\,, \label{eq:kappa_Balmer}
\end{equation}
where $n_2$ is the number density of neutral hydrogen at $n=2$, $n_{\rm H}$ is the total hydrogen number density, $X$ is the mass fraction of hydrogen ($X=0.74$ for solar composition, \citealt{Asplund2009}), $m_p$ is the proton mass, and $\sigma_{2\to\infty}$ is the ionization cross-section of the $n=2$ hydrogen, which depends only on atomic physics. Equation~(\ref{eq:kappa_Balmer}) implies that $\kappa_{2\to\infty}$ is proportional to the fraction $n_2/n_{\rm H}$ for a given chemical composition. On the one hand, a high temperature tends to populate the $n=2$ level according to the Boltzmann equation. On the other hand, most of the hydrogen will be ionized if the temperature is too high, as indicated by the Saha equation (assuming a system consisting only of electrons, protons, and neutral hydrogen),
\begin{equation}
    \frac{x^2}{1-x} = \left(2\pi m_e k_BT\over h^2\right)^{3/2}{e^{-I_0/k_BT}\over n_{\rm H}}\,, \label{eq:saha}
\end{equation}
where $x$ is the degree of ionization, $I_0=13.60{\rm~eV}$ is the ionization potential of ground-state hydrogen, and $m_e, k_B$, and $h$ are the electron mass, the Boltzmann constant, and the Planck constant. The fraction $n_2/n_{\rm H}$ will decrease as $x$ approaches unity. Therefore, the optimal temperature is close to the ionization temperature, $T_{\rm ion}$, at which the left-hand side of Equation~(\ref{eq:saha}) becomes of order unity. Equation~(\ref{eq:saha}) then suggests that $T_{\rm ion}$ positively correlates with $n_{\rm H}$. This qualitatively explains the trend in Figure~\ref{fig:kappa_ratio}.

Another notable feature in Figure~\ref{fig:kappa_ratio} is that the minimum effective opacity ratio of each curve is consistently of the order $0.1$ even as the optimal temperature changes from $\sim10000{\rm~K}$ to $\sim5000{\rm~K}$ across different densities. This result appears counterintuitive, since the fraction $n_2/n_{\rm H}$ and hence $\kappa_{2\to\infty}$ steeply increase with temperature as $\exp(-3I_0/4k_BT)$ before ionization occurs. However, near the optimal temperature, we observe that the continuum absorption opacity at $\sim4000{\rm~\AA}$ is dominated by the photoionization opacity of hydrogen with $n=3$. The latter similarly increases with temperature as $\exp(-8I_0/9k_BT)$. Therefore, both the numerator and the denominator of the opacity ratio are diminished at lower density and temperature, and their ratio, which is proportional to $\exp(-5I_0/36k_BT)$, is relatively insensitive to temperature. This maintains a strong opacity contrast and makes a Balmer break theoretically plausible at an optimal temperature of $\sim5000{\rm~K}$. 

The atmosphere model in this section is simplistic, but the opacity ratio in Figure~\ref{fig:kappa_ratio} only depends on atomic physics and the equation of state. The trend applies to more realistic atmospheres as long as LTE holds. The Balmer break is compatible with a cool, $\sim5000{\rm~K}$ photosphere if the gas density at the photosphere is sufficiently low. We will show in Sections~\ref{sec:disk} and \ref{sec:sphere} that such a low density is a natural outcome of super-Eddington accretion, which explains the optical redness and the Balmer break of the LRDs simultaneously. 

\section{Numerical Methods}
\label{sec:numerical_methods}

\begin{figure*}
    \centering
    \includegraphics[width=0.8\textwidth]{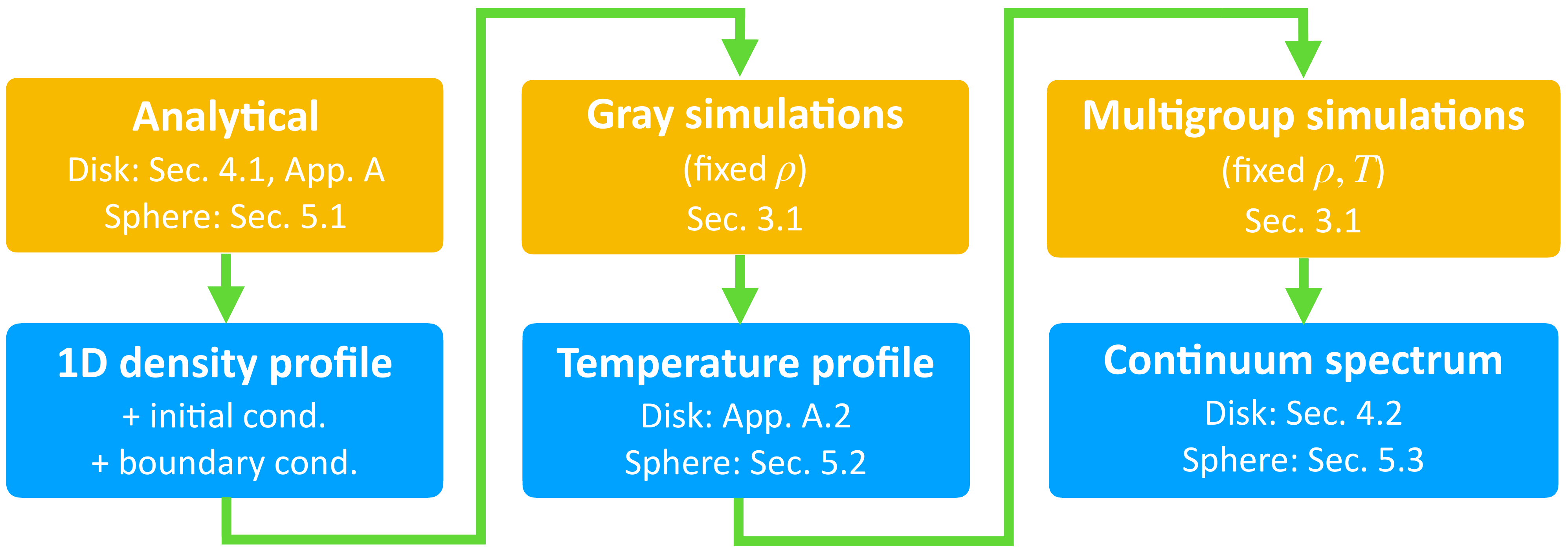}
    \caption{Workflow in this paper. Methods are outlined in yellow blocks, and results are outlined in blue blocks. ``Sec.'', ``app.'', and ``cond.'' stand for ``section'', ``appendix'', and ``condition''.}
    \label{fig:workflow}
\end{figure*}
Our continuum modeling is divided into three steps and schematically shown in Figure~\ref{fig:workflow}. The first step is the analytical construction of a hydrodynamic atmosphere profile. This is derived from either a standard accretion disk or a spherical flow; see Sections~\ref{sec:disk} and \ref{sec:sphere} respectively. After this step, we will have obtained the density, temperature, and radiation flux as a function of location. However, our analytical temperature profile is based on the diffusion approximation, which breaks down in the optically thin region. This motivates the second step, the numerical evaluation of the temperature by evolving the radiation and gas thermodynamics simultaneously toward a steady state while keeping the gas density fixed. The simulations in this step are performed under the gray approximation; i.e., the radiation intensities and opacities are independent of frequency. This sets the stage for the final step, in which we fix all hydrodynamic quantities and conduct multigroup radiation transport simulations to obtain the continuum spectra. 

Below, in Section~\ref{subsec:method_Athena}, we describe the simulation setups in the second and third steps, which we apply to both disk and spherical models. Our radiation calculations require opacity as input, and we outline our opacity calculations in Section~\ref{subsec:opacity}.

\subsection{From atmosphere models to the continuum spectra}
\label{subsec:method_Athena}
To obtain the steady-state temperature profile of an atmosphere model, we conduct 1D simulations using the non-relativistic gray radiation transport module in \textsc{Athena++} \citep{Stone2020,Jiang2021}. In the disk case, each simulation represents a narrow annulus at different radii of the disk. We use a Cartesian coordinate that spans the vertical direction and 32 fixed discrete angles for the radiation. In the sphere case, each simulation represents a radial profile, and we use a spherical polar coordinate and 32 discrete radiation angles that rotate with respect to the local coordinate direction.  In both the sphere and disk cases, we fix the density profile from analytical models and only evolve the temperature and the radiation field. The velocity is set to zero, although a small radial velocity  is expected in the spherical flow model (see Equation~(\ref{eq:sphere_velocity})): we test that including the radial velocity does not significantly change the temperature structure or the spectrum but requires longer time to converge. The temperature profile is initialized with the analytical result. The radiation intensities are initialized using the two-stream approximation such that the local radiation energy density and flux are equal to $aT^4$ ($a$ being the radiation density constant) and the analytical flux profile, respectively. In the disk case, we set a reflective inner boundary condition on radiation at the disk midplane and an outgoing outer boundary condition at a height where the density has reached a floor value of $10^{-16}{\rm~g~cm^{-3}}$. In the sphere case, we fix the radiation field at the initial value at the inner boundary, a small radius where the continuum is very optically thick, and use an outgoing boundary condition at a large, optically thin radius. We use root grids of 2048--6144 cells depending on specific runs, with adaptive mesh refinement to resolve locations with large temperature gradients. We terminate the simulation when the relative change of the location where $T=5000{\rm~K}$, an approximate proxy of the photosphere location, varies by less than one percent per logarithmic time. 

The simulation requires opacities as input to solve the radiation transport equation. Specifically, the scattering opacity $\kappa_s$, Rosseland mean opacity $\kappa_r$ (equivalent to $\kappa_s+\kappa_a$ in \citealt{Jiang2021}), and Planck mean opacity $\kappa_p$ are needed. Under the assumption of LTE, these opacities depend only on the local gas density and temperature. We prepare a table of these three opacities as a function of density and temperature beforehand (see Section~\ref{subsec:opacity}), covering $10^{-16.5}{\rm~g~cm^{-3}}<\rho<10^{-7.5}{\rm~g~cm^{-3}}$ and $1.5\times10^3{\rm~K}<T<{3\times10^5}{\rm~K}$. During the simulation, the opacities are updated with the temperature evolution by linear interpolation from the table in logarithmic space. In the sphere case, we may encounter regions with density or temperature below the table range, and we simply use the nearest value in the table. Opacities at low density may deviate from LTE, and low-temperature opacities are likely strongly modified by dust. However, such regions only exist far beyond the optical continuum photosphere in the sphere models.

After obtaining the numerical temperature profiles, we conduct multigroup radiation transport simulations using \textsc{Athena++} \citep{Jiang2022} to calculate the continuum spectrum. This calculation solves the radiation transport equation for multiple wavelength groups and thus predicts a spectrum from the model atmosphere. All hydrodynamic quantities, including the temperature, are fixed. We initialize the intensities to follow the Planck distribution defined by the local temperature with a frequency-integrated flux given by the analytical model. For the wavelength groups, we use a wavelength grid spanning $1250{\rm~\AA}<\lambda<13000{\rm~\AA}$ and a resolution of $50{\rm~\AA}$ in UV and 100--$200{\rm~\AA}$ in optical to near-infrared, except for the region near the Balmer break, $3600{\rm~\AA}<\lambda<3700{\rm~\AA}$, where we refine the bin width to 5--$10{\rm~\AA}$. The total number of wavelength bins is 122. The boundary conditions are the same as the gray simulations. We switch from adaptive to static mesh refinement as the temperature is now fixed. We prepare opacity tables for each wavelength group (see Section~\ref{subsec:opacity}) and initialize the opacities by linear interpolation (and nearest-neighbor extrapolation if needed), similar to the gray simulations. No update in opacity during the simulation is needed here because the gas density and temperature are fixed. We terminate the simulations when the fluxes at all locations and groups vary by less than one percent or less than ten times the machine precision limit per logarithmic time. The fluxes in the $z$ direction in the disk case or in the $r$ direction in the sphere case at the outer boundary constitute the desired spectrum.

\subsection{Opacity}
\label{subsec:opacity}
Wavelength-dependent opacities as a function of density and temperature are extensively used in this work. To prepare the opacity table, we use the code \textsc{FastChem} \citep[Version 2,][]{Stock2022} to evaluate the thermochemical equilibrium composition of an ideal gas of a given temperature, density, and elemental abundance. We assume a metallicity of $Z=0.1~Z_\odot$ and adopt the \citet{Asplund2009} solar composition, although the metallicities of LRDs are very uncertain. The chemical information is then passed to \textsc{optab} \citep{Hirose2022}, a public code to generate gas-phase Rosseland and Planck mean opacity tables given the density, temperature, and chemical composition. The original code only generates averaged opacities throughout the spectrum, but a slight adaptation allows us to calculate opacities averaged over arbitrary wavelength bins. \textsc{Optab} is designed to work with \textsc{FastChem}. Opacity sources implemented in \textsc{optab} include bremsstrahlung \citep{vanHoof2014,John1988,John1975}, photoionization \citep{Verner1995,Verner1996,Yan2001,Ohmura1960}, scattering \citep{Lee2005,Rohrmann2018,Tarafdar1973}, atomic line absorption \citep{kurucz_linelist}, and molecular line and collision-induced absorption (see Table~1 and Figure~1 in \citealt{Hirose2022}). However, in this work, we exclude molecular line absorption and collision-induced absorption, although molecules are accounted for when calculating the chemical composition. The optical effective temperature of most LRDs, $\sim5000{\rm~K}$, is expected to be sufficiently high to suppress the molecular population near the photosphere; moreover, spectrally confirmed LRDs do not seem to exhibit the strong molecular absorption features commonly seen in M-type stars. 

Now, we summarize the three use cases of opacities in this work. The first is the effective opacities $\kappa_{\rm eff}(3600{\rm~\AA})$ and $\kappa_{\rm eff}(4000{\rm~\AA})$ shown in Figure~\ref{fig:kappa_ratio} and $\kappa_{\rm eff}(5000{\rm~\AA})$ used in Figures~\ref{fig:sphere_teff} and \ref{fig:spheregray_m6}. They are obtained as the Planck mean of the quantity $\sqrt{3\kappa_{a,\nu}(\kappa_{a,\nu}+\kappa_{s,\nu})}$ over the wavelength ranges $3600{\rm~\AA}\leq\lambda\leq3620{\rm~\AA}$, $4000{\rm~\AA}\leq\lambda\leq4020{\rm~\AA}$, and $5000{\rm~\AA}\leq\lambda\leq5020{\rm~\AA}$ respectively. We use these opacities to diagnose the continuum, and thus only the continuum opacity sources are included. The second is the Rosseland and Planck mean opacity tables for our analytical calculations and gray simulations. These opacities include the line contributions; we also tested excluding them and found similar temperature structures in the gray simulations. The third is the multigroup opacities for spectral calculations. We only include the continuum opacity sources here because of our primary interest in the continuum and because of the computational cost of resolving individual lines, which is necessary for multigroup calculations to achieve spectral accuracy. With insufficient resolution, the line opacity would be smeared out to the entire wavelength bin that contained the line and appear as an unphysically enhanced continuum opacity. We caution that our treatment neglects metal line blocking. This effect tends to redden the continuum, although the extent depends on metallicity. Balmer lines also modify the spectrum on the red side of the Balmer limit, and thus realistic spectra would have a smoother increase in the range $3645{\rm~\AA}<\lambda\lesssim4000{\rm~\AA}$ than our calculations, especially in the presence of Doppler broadening. However, the strength of the Balmer break depends only on hydrogen photoionization, which is fully considered in our multigroup opacity calculations. 

\section{The disk scenario}
\label{sec:disk}
In this section, we consider the structure and continuum spectrum of a standard thin disk in the context of LRDs. The standard disk in the sub-Eddington regime has been found to produce a Balmer break if relatively cool annuli at $T_{\rm eff}\lesssim10^4{\rm~K}$ dominate the emission spectrum \citep[although they cautioned about uncertainties in the disk vertical structure]{Hubeny2000}. Bright UV continuum emission is theoretically expected to hide the break, which is consistent with the lack of a Balmer break in the spectrum of typical Type-I AGNs. However, LRDs are fainter in the UV and likely require cool emission components to dominate the optical continuum. This motivates us to revisit the thin-disk model and explore the parameter range that may reproduce the Balmer break and the optical redness of the LRDs.

In the remainder of this section, we follow the workflow outlined in Figure~\ref{fig:workflow}. Section~\ref{subsec:disk_analytical} examines the basic properties of the disk using a vertically integrated model, which constrains the parameter space relevant to LRDs and suggests a low-$T_{\rm eff}$ Balmer break at high accretion rates. We then evaluate the vertical density and temperature structures of the disk with analytical calculations and gray simulations, with details provided in Appendix~\ref{app:disk_analytical}. Finally, in Section~\ref{subsec:disk_spec}, we present the continuum spectra from multigroup simulations, which support the presence of a low-$T_{\rm eff}$ Balmer break but also highlight a fine-tuning problem inherent to the disk model.

\subsection{Analytical considerations}
\label{subsec:disk_analytical}
We consider the standard model \citep{Shakura1973} of a geometrically thin, optically thick, non-self-gravitating accretion disk around a black hole of mass $M_{\rm BH}$ and mass accretion rate $\dot{M}$. Assuming that the angular velocity at a given disk radius $r$ is Keplerian, the effective temperature,
\begin{equation}
    T_{\rm eff} = \left(\frac{3GM_{\rm BH}\dot{M}}{8\pi\sigma r^3}\right)^{1/4}\,,
    \label{eq:disk_Teff}
\end{equation}
is only a function of $M_{\rm BH}, \dot{M}$, and $r$. Here, $G$ and $\sigma$ are the gravitational constant and the Stefan-Boltzmann constant. We have ignored a factor $(1-\sqrt{R_{\rm ISCO}/r})$ in the bracket in Equation~(\ref{eq:disk_Teff}), where $R_{\rm ISCO}$ is the innermost stable circular orbit and has a value of several gravitational radii, $R_{\rm g}\equiv GM_{\rm BH}/c^2$. This is because we are interested in the cool outer disk, where $r\gg R_{\rm ISCO}$. 

The fixed power-law $T_{\rm eff} \propto r^{-3/4}$ implies that the disk continuum spectrum has three regimes. Let $R_{\rm in}$ and $R_{\rm out}$ $(R_{\rm ISCO} \ll R_{\rm in} < R_{\rm out})$ denote the disk's inner and outer truncation radii. For $\lambda < hc/k_BT_{\rm eff}(R_{\rm in})$, the continuum is dominated by emission from the innermost annulus. In the intermediate range, $hc/k_BT_{\rm eff}(R_{\rm in}) < \lambda < hc/k_BT_{\rm eff}(R_{\rm out})$, assuming blackbody emission, the spectrum follows $L_\lambda \propto \lambda^{-7/3}$. For $\lambda > hc/k_BT_{\rm eff}(R_{\rm out})$, the Rayleigh-Jeans tail dominates, with $L_\lambda \propto \lambda^{-4}$. The latter two regimes are too blue to match the optical continuum of LRDs, leaving only the innermost annulus as a plausible source. Reproducing the observed optical color temperature of $\sim 5000{\rm~K}$ thus requires $T_{\rm eff}(R_{\rm in}) \sim5000{\rm~K}$. As we shall see, such models face a significant fine-tuning issue, in that emission from disk annuli hotter than $\sim5000{\rm~K}$ needs to be excluded for the disk model to reproduce LRD spectra; we thus deem the disk scenario to be unlikely unless additional processes beyond our consideration set the truncation radius near this particular effective temperature. 

Notwithstanding this challenge, if we assume that the optical continuum of LRDs does emerge from a standard disk truncated at $T_{\rm eff,in}\equiv T_{\rm eff}(R_{\rm in})\sim 5000{\rm~K}$, then the total (one-sided) disk luminosity will be given by
\begin{align}
    L_{\rm disk}\simeq\frac{3GM_{\rm BH}\dot{M}}{4R_{\rm in}}&=8.4\times10^{43}{\rm~erg~s^{-1}} \left(\frac{T_{\rm eff,in}}{5000{\rm~K}}\right)^{4/3}\nonumber \\ &\times\left({\lambda_{\rm Edd}\over10^2}\right)^{2/3}\left(\frac{M_{\rm BH}}{10^7~M_\odot}\right)^{4/3}\,. \label{eq:disk_luminosity}
\end{align}
Here, we recast the accretion rate in terms of the Eddington ratio, $\lambda_{\rm Edd}\equiv \dot{M}/\dot{M}_{\rm Edd}$, where the Eddington accretion rate is $\dot{M}_{\rm Edd}\equiv L_{\rm Edd}/0.1c^2 = 4\pi GM_{\rm BH}/0.1c\kappa_{\rm es}$, assuming a conventional mass-to-radiation conversion efficiency of 0.1. We adopt the value of the Thomson electron scattering opacity $\kappa_{\rm es}=0.34{\rm~cm^2~g^{-1}}$. Currently known LRDs typically have optical continuum luminosities $L_{\rm opt}\sim10^{43}$--$10^{45}{\rm~erg~s^{-1}}$ (without dust correction; e.g., see the observed spectra in Figure~\ref{fig:diskspec_m7l2}). Equation~(\ref{eq:disk_luminosity}) then suggests that a super-Eddington thin accretion disk can match the LRD luminosities with a black hole mass of $M_{\rm BH}\sim10^7~M_\odot$. 

The super-Eddington accretion disk is vertically supported by radiation pressure in our region of interest \citep{Shakura1973}. Standard estimates of radiation-dominated accretion disks predict that the central density of the disk is given by
\begin{align}
    \rho_c &= 2\left(\frac{2}{3}\right)^{7/2}\left(\frac{0.1c^3}{\kappa_{\rm es}G}\right)^{3/2}\alpha^{-1}(\sigma T_{\rm eff}^4)^{-1/2}(\lambda_{\rm Edd}M_{\rm BH})^{-3/2} \nonumber \\
    &= 1.2\times10^{-11}{\rm~g~cm^{-3}} \nonumber \\ & \qquad\times\left(\frac{\alpha}{0.1}\right)^{-1}\left(\frac{T_{\rm eff}}{5000{\rm~K}}\right)^{-2}\left(\frac{\lambda_{\rm Edd}M_{\rm BH}}{10^9~M_\odot}\right)^{-3/2}\,, \label{eq:disk_midplane_density}
\end{align}
where we have expressed the density as a function of effective temperature rather than disk radius, and $\alpha$ is the standard viscous parameter. We have assumed that $\kappa_{\rm es}$ is the dominant opacity in the disk interior, which is consistent with the disk density we have obtained. We note that although the disks we consider have a super-Eddington accretion rate, the annuli $r\sim R_{\rm in}$ that produce the $T_{\rm eff}\sim5000$ K emission will locally remain consistent with the thin-disk assumption with negligible radial advective cooling compared to radiation,
\begin{align}
    {Q_{\rm adv}\over Q_{\rm rad}} &\simeq {\dot{M}c_s^2/2\pi r^2\over2\sigma T_{\rm eff}^4}=\left({\sqrt{3}GM_{\rm BH}\kappa_{\rm es}\sigma T_{\rm eff}^4\lambda_{\rm Edd}^2\over0.1^2\sqrt{2}c^5}\right)^{2/3} \nonumber\\
    &=0.0087\left(\frac{T_{\rm eff}}{5000{\rm~K}}\right)^{8/3}\left({M_{\rm BH}\over10^7~M_\odot}\right)^{2/3}\left({\lambda_{\rm Edd}\over10^2}\right)^{4/3}\,.
\end{align}
We will also directly show the validity of the geometric thinness after constructing the disk vertical profile (Appendix~\ref{app:disk_analytical}).

Equation~(\ref{eq:disk_midplane_density}) indicates that the midplane density at high accretion rates goes below the typical density of a main-sequence stellar atmosphere (see Figure~\ref{fig:kappa_ratio}). We note that the radiation-dominated disk is unstable \citep{Lightman1974,Jiang2013} and that models with magnetic support, which appear stable, have even lower central densities \citep{Jiang2025}. The photosphere density of the disk is expected to be still lower than the midplane density. Our argument in Section~\ref{sec:Balmer_break} then implies a Balmer break when $T_{\rm eff,in}\sim5000{\rm~K}$, as we will show with numerical continuum spectra in the next subsection. 

\subsection{Numerical continuum spectra}
\label{subsec:disk_spec}

\begin{figure*}
    \centering
    \includegraphics[width=0.49\textwidth]{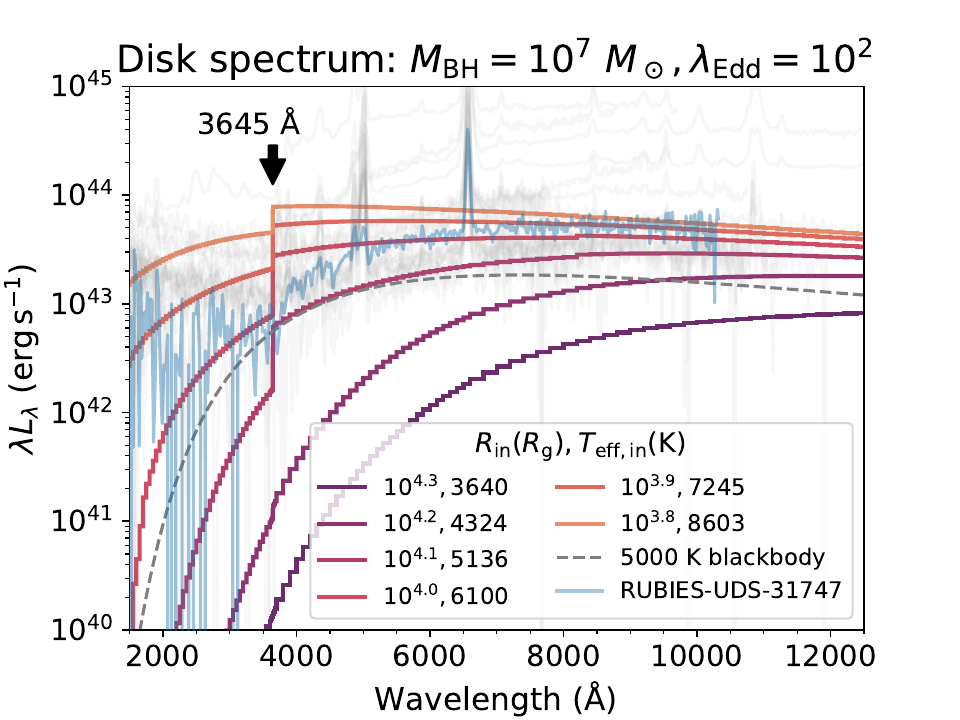}
    \includegraphics[width=0.49\textwidth]{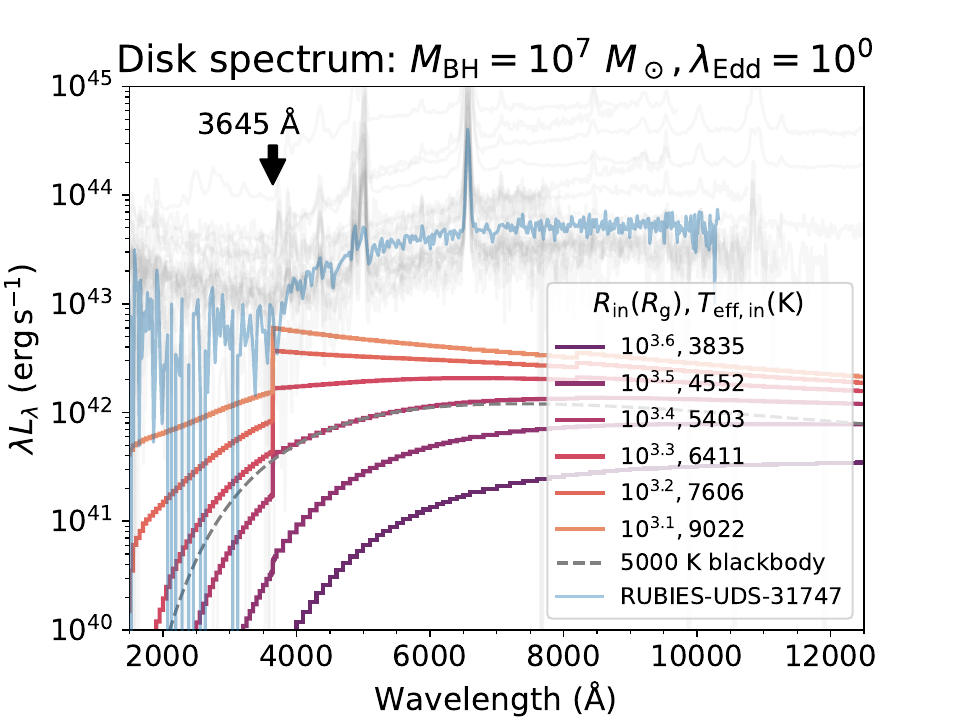}
    \caption{Model continuum spectra in the disk scenario. \textit{Left}: the $M_{\rm BH}=10^7~M_\odot, \lambda_{\rm Edd}=10^2$ case. Each curve shows the flux integrated over disk annuli from $R_{\rm in}$ to $R_{\rm out}=10^{4.3}~R_{\rm g}$. The legend marks the inner radius, $R_{\rm in}$, and the effective temperature at the inner radius, $T_{\rm eff,in}$, derived from Equation~(\ref{eq:disk_Teff}). A blackbody curve at $T=5000{\rm~K}$ and the spectrum of an LRD, RUBIES-UDS-31747, are shown for comparison. A collection of observed LRD spectra are shown in gray in the background. \textit{Right:} Same as left, but for the $M_{\rm BH}=10^7~M_\odot, \lambda_{\rm Edd}=10^0$ case, with $R_{\rm out}=10^{3.6}~R_{\rm g}$. }
    \label{fig:diskspec_m7l2}
\end{figure*}
Based on the estimates above, we calculate the vertical density and temperature profiles at each radius of the disk, elaborated in Appendix~\ref{app:disk_analytical}. We then use these profiles to calculate continuum spectra,  which we present in this subsection. 

We explore two cases in the disk model. The first case represents a disk around a black hole of mass $M_{\rm BH}=10^7~M_\odot$ and Eddington ratio $\lambda_{\rm Edd}=10^2$. The second case has the same black hole mass but a lower accretion rate, $\lambda_{\rm Edd}=1$. The main difference between the two cases is that a lower Eddington ratio corresponds to higher gas densities given the same effective temperature. For example, in the annulus where $T_{\rm eff}=5000{\rm~K}$, the photosphere densities of our vertical models (measured at the height where $T=T_{\rm eff}$) are $1.1\times10^{-11}{\rm~g~cm^{-3}}$ in the first case and $9.4\times10^{-10}{\rm~g~cm^{-3}}$ in the second case. We will show that the difference in density results in the Balmer break emerging at different effective temperatures. 

The results for the $M_{\rm BH}=10^7~M_\odot, \lambda_{\rm Edd}=10^2$ case are shown in the left panel of Figure~\ref{fig:diskspec_m7l2}. To account for the multiple annuli contributing to the emergent flux, we add the flux from each annulus at radius $r$ multiplied by the surface area $2\pi r^2\Delta\ln r$, where $\Delta\ln r=0.1\ln10$. The summation is performed from an outer radius where the disk is very cool ($R_{\rm out}=10^{4.3}~R_{\rm g}$, corresponding to $T_{\rm eff}=3640{\rm~K}$) to an inner radius indicated in the legend: for example, the spectrum marked with $R_{\rm in}=10^{4.1}~R_{\rm g}$ is the sum of the luminosities of three annuli at $r=10^{4.3}, 10^{4.2}, 10^{4.1}~R_{\rm g}$. For reference, we also mark in the legend the effective temperature of each innermost annulus, $T_{\rm eff,in}$, calculated from Equation~(\ref{eq:disk_Teff}). The model curves show a prominent Balmer break at $3645\rm~\AA$ when $R_{\rm in}\leq10^{4.1}~R_{\rm g}$. Note that the annulus at $10^{4.1}~R_{\rm g}$ has $T_{\rm eff}=5136{\rm~K}$, and the disk spectrum with $R_{\rm in}=10^{4.1}~R_{\rm g}$ appears slightly redder than a blackbody at $5000{\rm~K}$ in the optical spectrum due to the contribution of the cooler annuli. This illustrates the ability of a super-Eddington thin accretion disk to produce a Balmer break while having a red optical color, as we argued in Section~\ref{sec:Balmer_break}. As the truncation radius decreases and hotter annuli are included, the optical continuum becomes bluer, and the flux contrast at the Balmer break is diminished. The effective temperature at $10^{3.8}~R_{\rm g}$ is $8603{\rm~K}$, lower than that of a typical A-type main-sequence star, but the Balmer break is already weakened, as expected from Figure~\ref{fig:kappa_ratio} for low-density photospheres. 

We also plot in Figure~\ref{fig:diskspec_m7l2} a collection of JWST spectra of LRDs from the literature to represent the range of luminosities and spectral shapes from spectroscopically confirmed sources. The collection consists of all samples in \citet{Setton2024} that have a V-shaped continuum with significance above $2\sigma$. We highlight among them one source, RUBIES-UDS-31747, whose spectrum was taken as part of the RUBIES program \citep{deGraaff2024}. The spectrum has an optical continuum luminosity on the order of $3\times10^{43}{\rm~erg~s^{-1}}$, typical among the LRD samples plotted in the background, and visually shows a Balmer break. The model curve with $R_{\rm in}=10^{4.1}~R_{\rm g}$ gives an optical red color and a Balmer break strength similar to the data. The optical luminosity differs by a factor of two, but a slightly higher black hole mass and Eddington ratio will match the observation. Our intention here is not to fit the spectrum but to show the general behavior of disk spectra for parameters relevant to LRDs. 

To compare with the high-Eddington ratio case analyzed above, we show in the right panel of Figure~\ref{fig:diskspec_m7l2} the spectra of the $M_{\rm BH}=10^7~M_\odot, \lambda_{\rm Edd}=10^0$ case. The outer radius here is $R_{\rm out}=10^{3.6}~R_{\rm g}$, corresponding to $T_{\rm eff}=3835{\rm~K}$. Similarly to the previous case, different truncation radii result in a range of optical colors. However, the disk here will become too blue before reaching the optical luminosity of typical LRDs. The optical portion of the model spectrum with $R_{\rm in}=10^{3.4}~R_{\rm g}$ has a similar color to that of RUBIES-UDS-31747 but is 30 times dimmer. If we fix $R_{\rm in}$ at $T_{\rm eff}=5000{\rm~K}$, which in both panels produces an optical color similar to LRDs, then Equation~(\ref{eq:disk_Teff}) suggests $R_{\rm in}\propto(\lambda_{\rm Edd}M_{\rm BH}^2)^{1/3}$, and thus $L_{\rm disk}\sim3GM_{\rm BH}\dot{M}/4R_{\rm in}\propto (\lambda_{\rm Edd}M_{\rm BH}^2)^{2/3}$. This explains the low luminosity in the low-$\lambda_{\rm Edd}$ case.

The $M_{\rm BH}=10^7~M_\odot, \lambda_{\rm Edd}=10^0$ case is clearly disfavored. However, it helps to illustrate how the Balmer break is related to the disk photosphere condition. At an optical color comparable to that of a $T_{\rm eff}=5000{\rm~K}$ blackbody, the case with $\lambda_{\rm Edd}=10^0$ shows a weaker Balmer break than the high-$\lambda_{\rm Edd}$ counterpart (compare $R_{\rm in}=10^{3.4}~R_{\rm g}$ on the right to $R_{\rm in}=10^{4.1}~R_{\rm g}$ on the left). Moreover, in contrast to the $\lambda_{\rm Edd}=10^2$ case, the Balmer break here progressively strengthens when hotter and hotter annuli are included until the innermost annulus reaches $T_{\rm eff,in}=9022{\rm~K}$. Therefore, the Balmer break is optimally produced at a high effective temperature for a low Eddington ratio. We expect this behavior because the Eddington ratio anticorrelates with the photosphere density, which correlates with the optimal temperature, as shown in Figure~\ref{fig:kappa_ratio}. This supports the connection between the theoretical opacity law in Section~\ref{sec:Balmer_break} and the observable Balmer break feature. 

In summary, a truncated super-Eddington disk with a maximum effective temperature of $T_{\rm eff}\sim5000{\rm~K}$ produces a Balmer break, optical color, and optical luminosity similar to those of LRDs. The Balmer break and the cool effective temperature are reconciled at the low-density photosphere. Nevertheless, the model curves in either panel of Figure~\ref{fig:diskspec_m7l2} only span a factor of 3 in $R_{\rm in}$, but the spread in the optical color already exceeds that indicated by the collection of observed LRDs in the background. The abundance of these objects at similar optical colors requires a model insensitive to free parameters such as the truncation radius, which lies beyond the standard thin disk.

\section{The sphere scenario}
\label{sec:sphere}

We have seen that the disk model faces a fine-tuning challenge despite reproducing the observed Balmer break and the optical color of LRDs. Now, we consider an alternative scenario where the gas around the black hole follows a spherical geometry, as sketched in Figure~\ref{fig:cartoon}.   The broad motivation for this is that in super-Eddington accretion, the geometrical thickness of the flow approaches $H\sim r$, with optically thick inflows and/or outflows covering almost all of the solid angle 
\citep[e.g.,][]{Ohsuga2005,Jiang2014,McKinney2014,Sadowski2014,Jiang2019,Hu2022}.    In this initial exploratory calculation, we choose to approximate the true, more complicated geometry of such flows with a simple spherical model.

In this section, we construct spherical models and present their simulated spectra, following the workflow in Figure~\ref{fig:workflow}. In Section~\ref{subsec:sphere_photosphere}, we analytically evaluate the photosphere temperature and density, showing that the sphere scenario robustly yields a cool effective temperature. Section~\ref{subsec:sphere_gray} presents temperature profiles from gray simulations. Finally, in Section~\ref{subsec:sphere_multi}, we report our model continuum spectra and demonstrate their consistency with the observed optical colors and Balmer break features of LRDs.

\subsection{Estimating photosphere properties}
\label{subsec:sphere_photosphere}

We assume that a black hole of mass $M_{\rm BH}$ is surrounded by a steady, spherically symmetric, optically thick flow with density $\rho$, inward radial velocity $v$, and temperature $T$ as a function of radius $r$. The black hole accretes gas at a rate, $\dot{M} = 4\pi r^2\rho v$, independent of radius (this assumption is not critical, as we discuss below). We choose to specify the luminosity of the flow $L$ independent of $M_{\rm BH}$ or $\dot M$. In super-Eddington accretion, the luminosity is limited to $L_{\rm Edd}$ for a purely spherical laminar inflow but can exceed this limit with more realistic geometries. In this sense, our model represents black holes of mass $M_{\rm BH}\lesssim\kappa_{\rm es}L/(4\pi cG)$, where the equality holds if $L=L_{\rm Edd}$. Our simple model is agnostic to the exact radiative efficiency in that we vary $L$ and $\dot M$ independently and do not explicitly parameterize $M_{\rm BH}$. The luminosity relates the photosphere radius $R_{\rm ph}$ to the effective temperature $T_{\rm eff}$ by
\begin{equation}
    4\pi R_{\rm ph}^2\sigma T_{\rm eff}^4 = L\,.\label{eq:sphere_Ledd}
\end{equation}

In the analytical estimate in this section, we assume that the effective temperature is equal to the gas temperature at the photosphere (strictly speaking, the thermalization surface). We consider two estimates of the photosphere location. The conventional choice is to use the Rosseland mean opacity, $\kappa_R$, which captures the opacity over the full spectrum. However, when scattering dominates the Rosseland mean opacity (which is the case in our parameter space), the effective temperature calculated in this way may not accurately represent the spectral properties because the 
spectrum is set deeper in the atmosphere where the photons and gas can thermalize.  To account for this, we use the monochromatic effective opacity at $5000{\rm~\AA}$, $\kappa_{\rm eff}(5000{\rm~\AA})$, similar to Section~\ref{sec:Balmer_break}. The photosphere thus defined denotes the location where optical photons are thermalized and will better reflect the spectral color at $\sim5000{\rm~\AA}$. The ``effective temperature'' in this case is more properly understood as the color temperature at optical wavelengths,  but for simplicity, we will continue to refer to it as the effective temperature. We explore both definitions: 
\begin{equation}
    \rho_{\rm ph}\kappa(\rho_{\rm ph}, T_{\rm eff})H_{\rm ph} = \frac{2}{3}\,, \label{eq:sphere_photosphere}
\end{equation}
where $\kappa$ is either $\kappa_R$ or $\kappa_{\rm eff}(5000{\rm~\AA})$, $\rho_{\rm ph}=\rho(R_{\rm ph})$ is the gas density at the photosphere, and $H_{\rm ph}$ is the ``photosphere scale height", or the radius range where the optical depth is of order unity. This height (which we will clarify further in Section~\ref{subsec:sphere_gray}) will be determined self-consistently once we obtain the temperature profile in the numerical calculations, but we treat $H_{\rm ph}/R_{\rm ph}$ as a free parameter in this subsection. Then, the photosphere conditions will be determined if the photosphere density is known.

The photosphere density, or more generally the entire density profile, is related to $\dot{M}$ and $v$. We assume that the inflow velocity at a given radius is a fixed fraction of the Keplerian velocity:
\begin{equation}
    v = \delta \sqrt{\frac{GM_{\rm BH}}{r}}\,. \label{eq:sphere_velocity}
\end{equation}
Note that this velocity only reflects the net flow, but turbulent gas motion could be faster in principle.  In addition, if there is an escaping outflow with a mass outflow rate comparable to $\dot M$, it could have a significantly higher velocity (e.g., \citealt{Shakura1973}; we will return to this in Section~\ref{sec:conclusions}). From Equation~(\ref{eq:sphere_velocity}), for a given system, the density profile will follow a power law, $\rho=\dot{M}/4\pi r^2v\propto r^{-3/2}$. To set a normalization value of this profile, we introduce a free parameter, $\rho_{\rm ref}$, a reference density at a fixed radius, $R_{\rm ref}\equiv1.477\times10^{16}{\rm~cm}$, which is equal to $10^5~R_{\rm g}$ for a black hole mass of $10^6~M_\odot$. Then, 
\begin{equation}
    \rho_{\rm ph}R_{\rm ph}^{3/2} = \rho_{\rm ref}R_{\rm ref}^{3/2}\,.\label{eq:sphere_density}
\end{equation}

The reference radius is chosen to be comparable to the photosphere radius given a cool effective temperature. In comparison, if $T_{\rm eff}=5000{\rm~K}$, then Equation~(\ref{eq:sphere_Ledd}) gives $R_{\rm ph}=5.75\times10^{15}{\rm~cm}$ for $L=L_{\rm Edd}(10^5~M_\odot)$, $R_{\rm ph}=1.82\times10^{16}{\rm~cm}$ for $L=L_{\rm Edd}(10^6~M_\odot)$, and $R_{\rm ph}=5.75\times10^{16}{\rm~cm}$ for $L=L_{\rm Edd}(10^7~M_\odot)$. The reference density thus resembles the midplane density in the disk model in the sense of indicating the order of magnitude of the photosphere density. 

The reference density is related to the Eddington ratio as
\begin{align}
    \lambda_{\rm Edd}&=\frac{4\pi r^2\rho v}{\dot{M}_{\rm Edd}}\notag\\ &= 1.59\times10^3\frac{\delta}{10^{-2}}\left(\frac{M_{\rm BH}}{10^6M_\odot}\right)^{-1/2}\frac{\rho_{\rm ref}}{10^{-12}{\rm~g~cm^{-3}}}\,. \label{eq:sphere_Eddington_ratio}
\end{align}
Here, $\delta$ depends on the turbulent viscosity operating in the sphere and is uncertain. The fiducial value of $\delta=10^{-2}$ is consistent with an $\alpha$-disk model with $\alpha=0.1$ due to the magnetorotational instability \citep{Balbus1991} and a dimensionless scale height of $\sim0.3$. Note that the Eddington ratio does not directly appear in our spherical models: the free parameters are $L$ and $\rho_{\rm ref}$, and Equation~(\ref{eq:sphere_Eddington_ratio}) only serves to provide some intuition on how to translate $\rho_{\rm ref}$ to $\lambda_{\rm Edd}$. We also note that we have assumed a net mass accretion rate independent of radius, although outflows can significantly reduce the angle-averaged net accretion rate close to the black hole \citep[e.g.,][]{Jiang2019,Hu2022}. Thus, the Eddington ratio in Equation~(\ref{eq:sphere_Eddington_ratio}) should be interpreted as the mass inflow rate at a relatively large distance, with the net accretion rate onto the black hole being possibly significantly lower. A varying net accretion rate with radius would also modify the power law of the density profile in the accretion flow. However, we expect details of the density-radius relation to have a small effect on the emergent flux compared to the photosphere temperature and density because the sphere interior is hidden in the optically thick region.

We also calculate the minimum mass needed to sustain such a photosphere, 
\begin{align}
    M_{\rm sph,min}={8\pi\over3}\rho_{\rm ph}R_{\rm ph}^3=2.5\times10^4~M_\odot \nonumber\\
    \times{\rho_{\rm ph}\over10^{-12}{\rm~g~cm^{-3}}}\left(L\over L_{\rm Edd}(10^6~M_\odot)\right)^{3/2}\left({T_{\rm eff}\over5000{\rm~K}}\right)^{-6}\,.
\end{align}
This mass is small compared to the black hole mass ($M_{\rm BH}\sim10^5-10^7~M_\odot$ if we assume $L=L_{\rm Edd}$). How to fuel this sphere from larger scales is an interesting question but lies beyond the scope of this paper.

\begin{figure}
    \centering
    \includegraphics[width=0.49\textwidth]{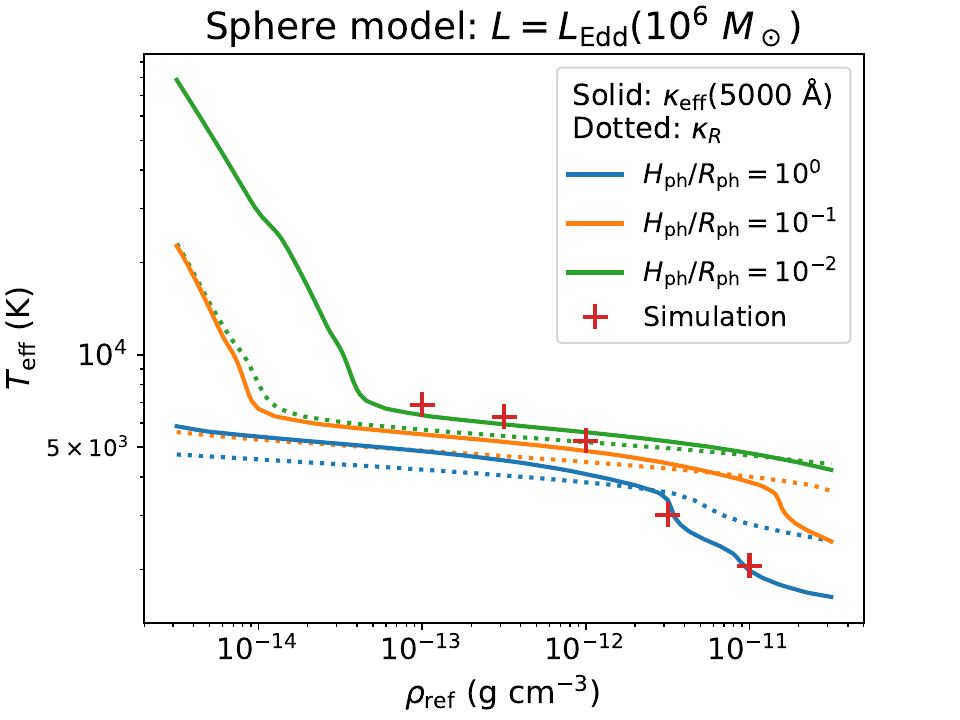}
    \caption{Effective temperature of the sphere model with $L=L_{\rm Edd}(10^6~M_\odot)$. The horizontal axis is the reference density, $\rho_{\rm ref}$, a proxy for the Eddington ratio (Equation~(\ref{eq:sphere_Eddington_ratio})) and the photosphere density ($\rho_{\rm ref}\sim\rho_{\rm ph}$ if $T_{\rm eff}\sim5000{\rm~K}$). Curves show analytical estimates using the effective opacity at $5000\rm~\AA$ (solid) or the Rosseland mean opacity (dotted). These estimates depend on the dimensionless photosphere scale height, $H_{\rm ph}/R_{\rm ph}$, as a free parameter. Plus signs mark measurements of the gas temperature at $\tau_{\rm eff}(5000{\rm~\AA})=1/2$ from the gray simulations, where the photosphere scale height is determined self-consistently by radiation energy transport. The analytical effective temperature is very insensitive to the reference density for a given luminosity and $H_{\rm ph}/R_{\rm ph}$, although numerical measurements suggest a somewhat steeper $T_{\rm eff}-\rho_{\rm ref}$ relation.} 
    \label{fig:sphere_teff}
\end{figure}

We now solve for $T_{\rm eff}$ from Equations~(\ref{eq:sphere_Ledd})(\ref{eq:sphere_photosphere})(\ref{eq:sphere_density}). Figure~\ref{fig:sphere_teff} shows the effective temperature for $L=L_{\rm Edd}(10^6~M_\odot)=1.47\times10^{44}{\rm~erg~s^{-1}}$. We first discuss the effective temperature calculated with $\kappa_{\rm eff}(5000\rm~\AA)$ (solid curves). For a given $H_{\rm ph}/R_{\rm ph}$, the effective temperature decreases with the reference density because a denser sphere tends to increase the optical depth and move the photosphere further out. However, $T_{\rm eff}$ is very insensitive to $\rho_{\rm ref}$ in the range $4000{\rm~K}\lesssim T_{\rm eff}\lesssim6000{\rm~K}$. This implies that a wide range of accretion rates tend to produce very similar $T_{\rm eff}$. For example, if we set $\delta=0.01$ and $M_{\rm BH}=10^6~M_\odot$ in Equation~(\ref{eq:sphere_Eddington_ratio}), then $\rho_{\rm ref}=6.3\times10^{-16}\lambda_{\rm Edd}{\rm~g~cm^{-3}}$. Thus, on the curve of $H_{\rm ph}/R_{\rm ph}=10^0$, a relatively small Eddington ratio of $\lambda_{\rm Edd}=10$ will result in $T_{\rm eff}=5500{\rm~K}$, while increasing the Eddington ratio by two orders of magnitude only shifts the effective temperature by a small amount, to $T_{\rm eff}=4300{\rm~K}$. Our argument has assumed a fixed luminosity, and if the luminosity increases with the accretion rate (and hence $\rho_{\rm ref}$), as found in many super-Eddington accretion simulations \citep{Ohsuga2005,Jiang2014,Sadowski2015,Jiang2019}, this would make the effective temperature an even weaker function of $\rho_{\rm ref}$ than the curves in Figure~\ref{fig:sphere_teff}. The effective temperature calculated with $\kappa_R$ (dotted curves) shows similar trends, with an even flatter $T_{\rm eff}-\rho_{\rm ref}$ relation at $4000{\rm~K}\lesssim T_{\rm eff}\lesssim6000{\rm~K}$, although we again caution that the effective temperature here does not accurately predict optical colors. We will focus on the solid curves hereafter. 

The behavior in Figure~\ref{fig:sphere_teff} is due to the sharp increase of opacity with temperature at $T_{\rm eff}\sim5000{\rm~K}$. If we parameterize the opacity as $\kappa_{\rm eff}\propto \rho_{\rm ph}^\alpha T_{\rm eff}^\beta$, then, from Equations~(\ref{eq:sphere_Ledd})(\ref{eq:sphere_photosphere})(\ref{eq:sphere_density}), $T_{\rm eff}\propto\rho_{\rm ref}^{-(\alpha+1)/(3\alpha+\beta+1)}$. At $T_{\rm eff}\gtrsim10^4{\rm~K}$, the opacity weakly depends on the density and temperature. We empirically obtain $\alpha\simeq0.5$, $\beta\simeq-0.6$ from the opacity table, and thus $T_{\rm eff}\propto\rho_{\rm ref}^{-0.8}$, in agreement with the leftmost section of the curves with $H_{\rm ph}/R_{\rm ph}=10^{-1}$ and $H_{\rm ph}/R_{\rm ph}=10^{-2}$. At lower temperatures, the density dependence remains modest, $-1\leq\alpha\leq1$, but $\beta\simeq12-16$. This steepness results from electron recombination with hydrogen, which exponentially limits all important opacity sources in this regime: the electron scattering opacity, the $\rm H^-$ photoionization opacity (electrons are required to form the anions), and the metal bremsstrahlung opacity. This leads to the flat segments in Figure~\ref{fig:sphere_teff}. At even lower temperatures, $T_{\rm eff}<3000{\rm~K}$, the completion of electron recombination ends the steep dependence of opacity on temperature, and thus the $T_{\rm eff}-\rho_{\rm ref}$ curves again have a larger negative gradient at low $T_{\rm eff}$  (high $\rho_{\rm ref}$).

The $T_{\rm eff}-\rho_{\rm ref}$ relation also depends on $H_{\rm ph}/R_{\rm ph}$. For a given effective temperature, if the photosphere scale height is small, a higher density will be needed to provide the same optical depth, and thus the curves with lower $H_{\rm ph}/R_{\rm ph}$ are located higher in the figure. We will return to this point when we discuss the simulation results in Section~\ref{subsec:sphere_gray}.

\subsection{Numerical temperature profile}
\label{subsec:sphere_gray}
\begin{figure*}
    \centering
    \includegraphics[width=0.98\textwidth]{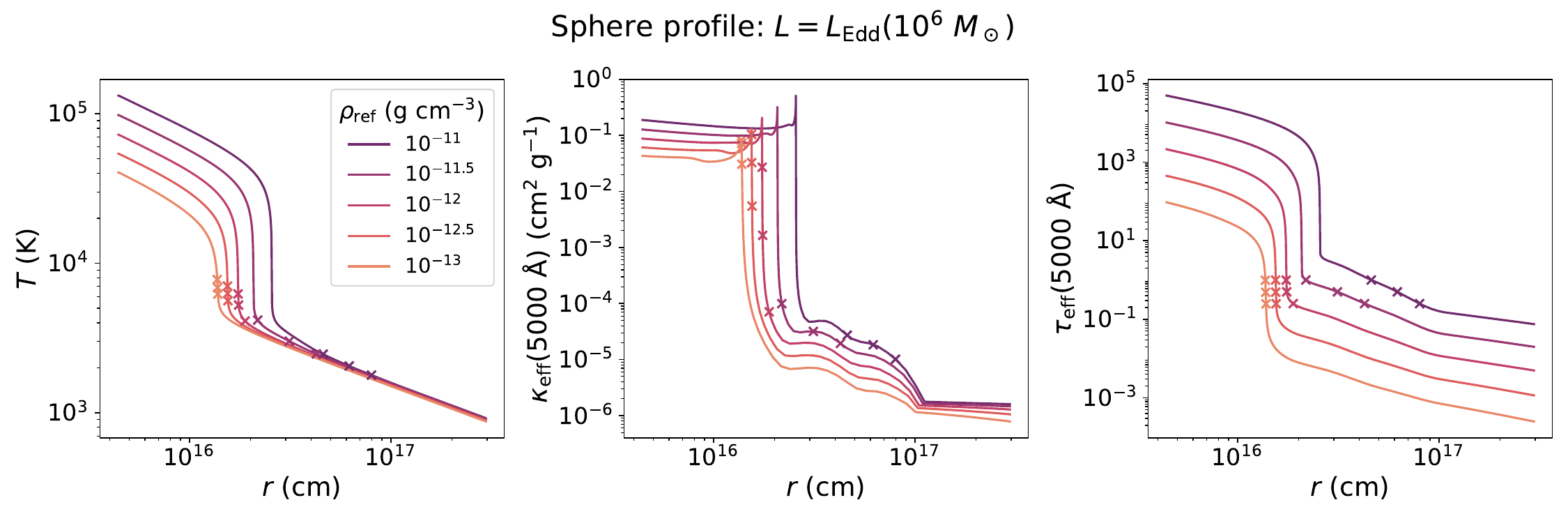}
    \caption{Temperature (left) from gray simulations of the sphere model with $L=L_{\rm Edd}(10^6~M_\odot)$. The effective opacity and optical depth at $5000{\rm~\AA}$ (mid, right) are calculated using the temperature profiles. Dark colors indicate high $\rho_{\rm ref}$ and hence high accretion rate. The three crosses on each curve mark the locations where $\tau_{\rm eff}(5000{\rm~\AA})=1/4, 1/2$, and $1$ to indicate the photosphere location and its  scale height. }
    \label{fig:spheregray_m6}
\end{figure*}

We conduct a grid of simulations with $L=1.47\times(10^{43},10^{44},10^{45}){\rm~erg~s^{-1}}$ and a range of $\rho_{\rm ref}$ to cover the luminosities and colors of LRDs. The chosen values of luminosity correspond to the Eddington luminosity of a black hole mass of $10^5$--$10^7~M_\odot$, though we reiterate that our model does not stipulate or require $L = L_{\rm Edd}$.

For each parameter set, we calculate the initial temperature using the radiative diffusion equation dominated by the electron scattering opacity and the power-law density profile in Equation~(\ref{eq:sphere_density}). We obtain
\begin{equation}
    T^4(r)\simeq \frac{3\kappa_{\rm es}L}{10\pi ac}\rho_{\rm ref}R_{\rm ref}^{3/2}r^{-5/2}\,. \label{eq:sphere_T_inner}
\end{equation}
This is accurate when $r\ll R_{\rm ph}$ and $T\gg T_{\rm eff}$. We choose the inner boundary, $R_{\rm inner}$, to be sufficiently optically thick (Rosseland mean optical depth greater than 100), and use Equation~(\ref{eq:sphere_T_inner}) to set the inner boundary condition of the radiation that has an energy density equal to $aT^4(R_{\rm inner})$ and flux equal to $L/(4\pi R_{\rm inner}^2)$.

In the remainder of this subsection, we describe the results from the gray simulations. The left panel of Figure~\ref{fig:spheregray_m6} shows the temperature profiles for $L=L_{\rm Edd}(10^6~M_\odot)$. We also show the effective opacity $\kappa_{\rm eff}$ at $5000{\rm~\AA}$ calculated from the temperature profiles and the corresponding optical depth $\tau_{\rm eff}(5000{\rm~\AA})$ in the middle and right panels for diagnostic purposes, although wavelength-dependent opacities are not used in the gray simulations.  The temperature decreases with radius, converging at large radii where the gas becomes optically thin and $E_{\rm rad}=aT^4\simeq L/4\pi cr^2$. These curves show a steep drop at $T\sim8000{\rm~K}$. This occurs where the effective opacity reaches a maximum, as expected from radiative transport. Simulations with higher $\rho_{\rm ref}$ have higher temperature in the optically thick region at a given radius due to the higher optical depth. 

Now, we focus on the photosphere properties indicated by the gray simulations and compare them with analytical estimates. On each curve, we find the locations where $\tau_{\rm eff}(5000{\rm~\AA})=1/4, 1/2,$ and 1 and mark them with crosses in Figure~\ref{fig:spheregray_m6}. The radial distance between the first and last markers roughly indicates the photosphere scale height $H_{\rm ph}$ (most clearly seen in the right panel). We use the gas temperature at $\tau_{\rm eff}(5000{\rm~\AA})=1/2$ to represent the photosphere temperature.\footnote{Empirically, we find that the optical continuum is better approximated by a blackbody of the temperature at $\tau_{\rm eff}(5000{\rm~\AA})=1/2$ instead of 2/3. } We observe two trends. First, the photosphere temperature decreases with the reference density, as expected, and as in Figure \ref{fig:sphere_teff}. Second, while the photosphere scale height is comparable to the photosphere radius for $\rho_{\rm ref}\geq10^{-11.5}{\rm~g~cm^{-3}}$, $H_{\rm ph}$ becomes significantly smaller for lower $\rho_{\rm ref}$. This shows that $H_{\rm ph}/R_{\rm ph}$ changes with $\rho_{\rm ref}$, a complication not considered in our analytical estimates in Section~\ref{subsec:sphere_photosphere}. 

We show the temperature at $\tau_{\rm eff}(5000{\rm~\AA})=1/2$ in Figure~\ref{fig:sphere_teff} to compare with our analytic estimates. The measured temperature decreases faster with $\rho_{\rm ref}$ than suggested by individual analytical curves. This is due to the change in $H_{\rm ph}/R_{\rm ph}$ in the simulations. The location of the measured points relative to the solid curves indicates that $H_{\rm ph}/R_{\rm ph}\sim10^0$ for $\rho_{\rm ref}\geq10^{-11.5}{\rm~g~cm^{-3}}$ and $H_{\rm ph}/R_{\rm ph}<10^{-1}$ otherwise, in agreement with Figure~\ref{fig:spheregray_m6}. 

Although the photosphere temperature is empirically more sensitive to the accretion rate than analytical expectations, the simulations assume that the sphere is laminar with purely radiative energy transport. In reality, turbulent mixing can significantly boost $H_{\rm ph}/R_{\rm ph}$ to order unity for the cases with small photosphere scale heights. Turbulent energy dissipation in the optically thin region could do the same as well, by smoothing out the temperature profile. Simulations considering these additional effects are likely to achieve better agreement with the analytical curve of $H_{\rm ph}/R_{\rm ph}\sim 1$. 

\subsection{Numerical continuum spectra and comparison to observation}
\label{subsec:sphere_multi}
\begin{figure}
    \centering
    \includegraphics[width=0.49\textwidth]{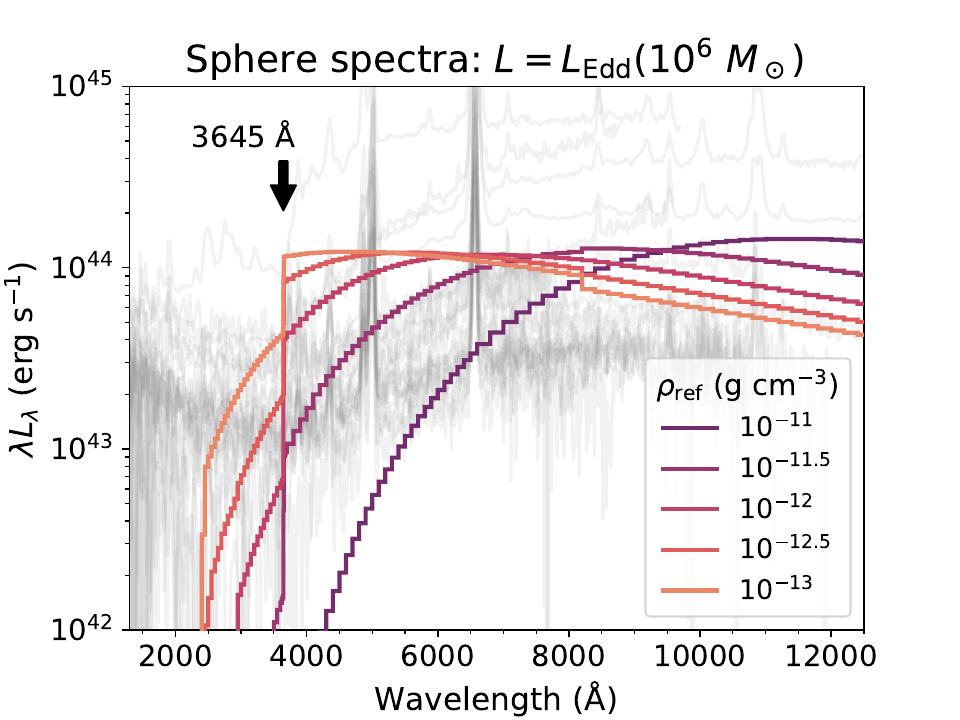}
    \caption{Continuum spectra from sphere models at a total luminosity $L=L_{\rm Edd}(10^6~M_\odot)=1.47\times10^{44}{\rm~erg~s^{-1}}$. Model spectra are color-coded by the reference density, $\rho_{\rm ref}$, a proxy for the Eddington ratio (Equation~(\ref{eq:sphere_Eddington_ratio})) and the photosphere density ($\rho_{\rm ref}\sim\rho_{\rm ph}$ if $T_{\rm eff}\sim5000{\rm~K}$). A collection of observed LRD spectra are shown in the background for comparison. }
    \label{fig:spherespec_m6}
\end{figure}

\begin{figure}
    \centering
    \includegraphics[width=0.49\textwidth]{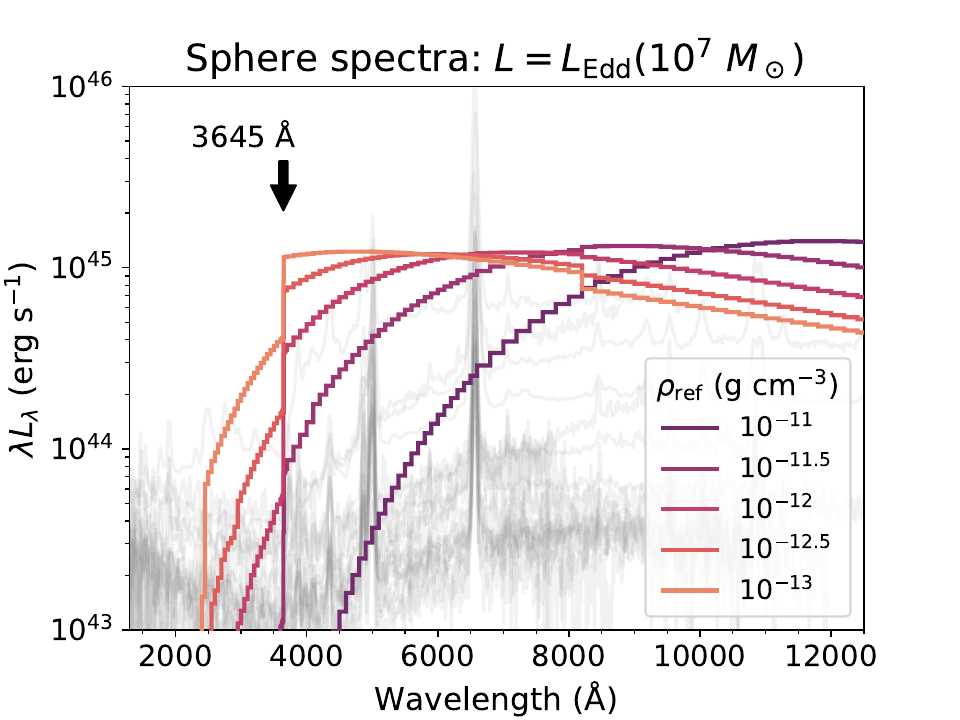}
    \caption{Same as Figure~\ref{fig:spherespec_m6}, but for the models with $L=L_{\rm Edd}(10^7~M_\odot)$.}
    \label{fig:spherespec_m7}
\end{figure}

As we described in Section~\ref{subsec:sphere_photosphere}, the free parameters in our sphere scenario are the total luminosity $L$, defined in Equation~(\ref{eq:sphere_Ledd}), and a reference density $\rho_{\rm ref}$, whose physical meaning is defined in Equation~(\ref{eq:sphere_density}) and discussed in the text above and below the equation. Figures~\ref{fig:spherespec_m6} and \ref{fig:spherespec_m7} show numerical continuum spectra from the multigroup simulations for a range of $L$ and $\rho_{\rm ref}$. In the background, we plot the same collection of LRD spectra as in Figure~\ref{fig:diskspec_m7l2}. The optical continuum luminosity of these LRDs ranges from $10^{43}$ to $10^{45}{\rm~erg~s^{-1}}$ (no dust attenuation is included here), approximately corresponding to the Eddington luminosity of black holes of mass from $10^5$ to $10^7~M_\odot$.

Our model curves in each figure exhibit the same bolometric luminosity by construction. The optical color reddens as $\rho_{\rm ref}$ increases, in agreement with the decreasing photosphere temperature measured in Section~\ref{subsec:sphere_gray}. At $3645\rm~\AA$, almost all model spectra show a strong Balmer break with flux contrast on the order of a few. This is expected from the photosphere density of the spherical models (see the horizontal axis range of Figure~\ref{fig:sphere_teff}, where $\rho_{\rm ref}\sim \rho_{\rm ph}$ if $T_{\rm eff}\sim5000{\rm~K}$). Some other discontinuities are present at shorter wavelengths as a result of metal photoionization. The jump in some spectra at $8200{\rm~\AA}$ corresponds to the Paschen limit and is physical; this is possible in the presence of scattering opacity \citep{Schuster1905,Underhill1949}. The model spectra of $L=L_{\rm Edd}(10^5~M_\odot)$ appear similar to those of higher masses except for their lower bolometric luminosities, and we do not plot them for brevity.

\begin{figure}
    \centering
    \includegraphics[width=0.49\textwidth]{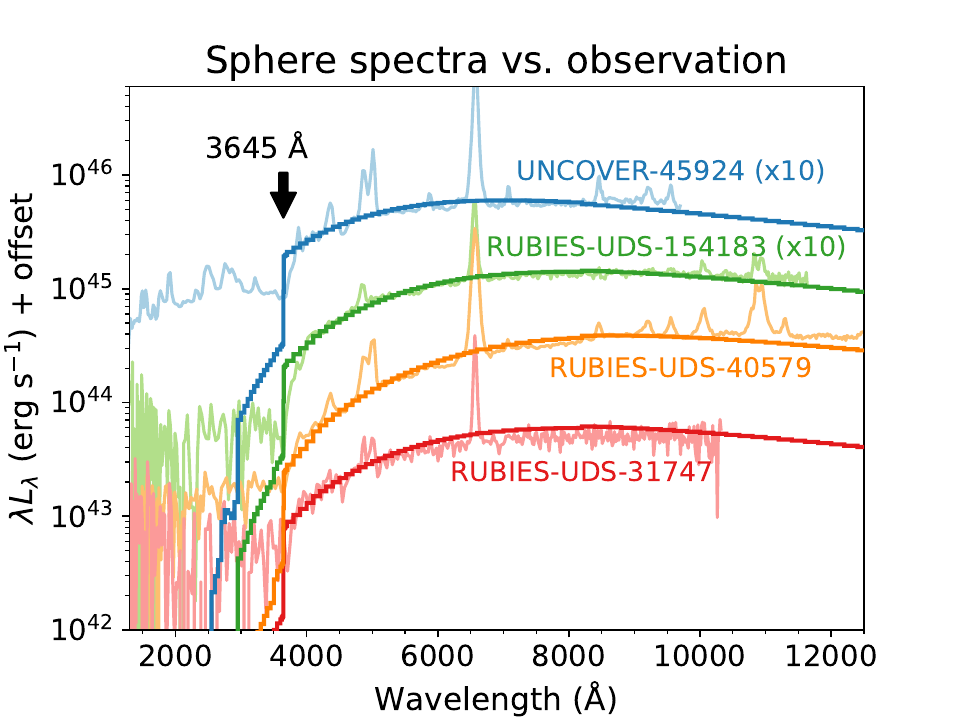}
    \caption{Continuum spectra from spherical models compared to four LRD spectra showing strong Balmer breaks. The names of the observed objects are annotated. The model and observed spectra of UNCOVER-45924 and RUBIES-UDS-154183 are shifted up by 10 for visual clarity. The paramters for the four model curves are $(L, \rho_{\rm ref})=(7.4\times10^{44}$, $2.0\times10^{-12})$, $ (1.8\times10^{44},2.1\times10^{-12})$, $ (4.4\times10^{44},5.5\times10^{-12})$, and $(7.4\times10^{43},1.6\times10^{-12})$ (in CGS units) respectively, from top to bottom. }
    \label{fig:spherespec_obs}
\end{figure}

To compare with observation in more detail, in Figure~\ref{fig:spherespec_obs}, we select four sources with strong Balmer breaks, UNCOVER-45924, RUBIES-UDS-154183, RUBIES-UDS-40579, and RUBIES-UDS-31747. The first source is discovered in the UNCOVER survey \citep{Bezanson2024}, photometrically selected \citep{Labbe2025}, and studied in detail with high-signal-to-noise-ratio NIRSpec/PRISM spectroscopy \citep{Labbe2024}, known as one of the most optically luminous LRDs observed so far. The latter three sources are observed as part of the RUBIES program \citep{deGraaff2024}. RUBIES-UDS-154183 shows an outstanding Balmer break strength exceeding the limit expected from any stellar population \citep{deGraaff2025}. RUBIES-UDS-40579 is known for its extreme red color and has been studied individually in \citet{Wang2025}. RUBIES-UDS-31747 is the same object that we highlighted in Figure~\ref{fig:diskspec_m7l2}. These sources represent LRDs with strong Balmer breaks at different luminosities and colors. On top of them are the spectra interpolated from our parameter grid: we choose $L$ to match the peak luminosities of each object and then choose $\rho_{\rm ref}$ to match the optical color. These are not intended as rigorous fits of the observations, but the visual agreement demonstrates that the spherical model can reproduce a strong break and a red optical color simultaneously without invoking stellar populations or external gas or dust absorption. The model spectra have a discontinuity at the Balmer break instead of a smooth rollover as in the observed spectra because we do not consider the Balmer series near the break, which will blend and form quasi-continuum absorption in the presence of turbulent gas motion. At longer wavelengths, the model curves become somewhat bluer than the observed spectra\footnote{Our model spectra, especially the reddest ones, have larger curvature than a blackbody in the optical \citep[see also][]{Lin2025b}. When the gas is mostly neutral (e.g., $T\lesssim5000{\rm~K}$ at $\rho=10^{-12}{\rm~g~cm^{-3}}$), Rayleigh scattering will become the dominant opacity source at optical wavelengths \citep[see][although they studied zero-metallicity gas]{Lenzuni1991}. It preferentially suppresses short-wavelength spectral flux, similar to but more manifest than its role in metal-poor giant stars \citep{Cayrel2004}. }. This discrepancy can be alleviated by extensions of our simplistic model, e.g., by allowing for a distribution of $\rho_{\rm ref}$. This will produce a range of effective temperatures at the photosphere, which tends to boost the infrared flux and smooth out the inverse Paschen break (which disappears at high $\rho_{\rm ref}$ in our models) while maintaining the Balmer break. This will arise naturally if rotation breaks spherical symmetry and the infalling gas has a more oval shape.

Assuming $L=L_{\rm Edd}$, the models in Figure~\ref{fig:spherespec_obs} correspond to black hole masses on the order of $10^5$--$10^7~M_\odot$. However, we reiterate that models of super-Eddington accretion can have $L>L_{\rm Edd}$, in which case the BH masses inferred for the LRDs would be yet smaller. 
These black hole masses are significantly lower than previous estimates of the population based on the broad H$\alpha$ line \citep[e.g.,][]{Maiolino2024,Greene2024}. However, these works usually included substantial dust correction to the broad-line luminosity, which likely overestimated the black hole mass in light of the absence of dust re-emission. These measurements also assumed that the line was Doppler-broadened by virial motion of gas in the broad line region, whose size was calibrated with local Type-I AGNs, but it remains unclear whether the gas in the broad line region gas is virialized \citep{King2024}, whether the broad line region of LRDs follows the same radius--luminosity relation as that of Type-I AGNs \citep{Lupi2024}, and whether the lines are alternatively scattering-broadened (\citealt{Rusakov2025,Naidu2025,Chang2025}; but see \citealt{Juodzbalis2024,Brazzini2025}). In addition, the modest black hole masses suggested here may align better with the clustering measurements of LRDs \citep{Matthee2024b,Lin2025,Carranza-Escudero2025} and with the host galaxy masses constrained from extended emission \citep{Chen2025a,Chen2025b,Li2025} and narrow line dispersion \citep{Ubler2023,Maiolino2024,Wang2025,Ji2025}. 

The reference densities are on the order of $10^{-12}$--$10^{-11}{\rm~g~cm^{-3}}$, with the photosphere densities spanning $10^{-13}$--$10^{-11}{\rm~g~cm^{-3}}$. The range would likely widen in more realistic models that consider turbulence (see Section~\ref{subsec:sphere_gray}). These values are chosen to match the optical color of LRDs regardless of their Balmer break strength, but the strong Balmer break is ubiquitous in our models, as expected from Section~\ref{sec:Balmer_break}. However, our spherical model does not account for the UV continuum flux of LRDs, whose origin is still debated \citep[e.g.,][]{Greene2024,Barro2024,Naidu2025,Chen2025a,Rinaldi2024,Chen2025b}. Assuming that this component does not produce a stronger Balmer break than the model spectra here, the total flux will show a partially or completely filled Balmer break. In this case, the optical spectrum shown here represents the upper limit of the break strength generated by the spherical model. This also naturally places the inflection point of the V-shaped spectrum near the Balmer limit over a wide range of UV-to-optical luminosity ratios, suggesting a robust explanation for its observed location \citep{Setton2024}. 

\begin{figure}
    \centering
    \includegraphics[width=0.48\textwidth]{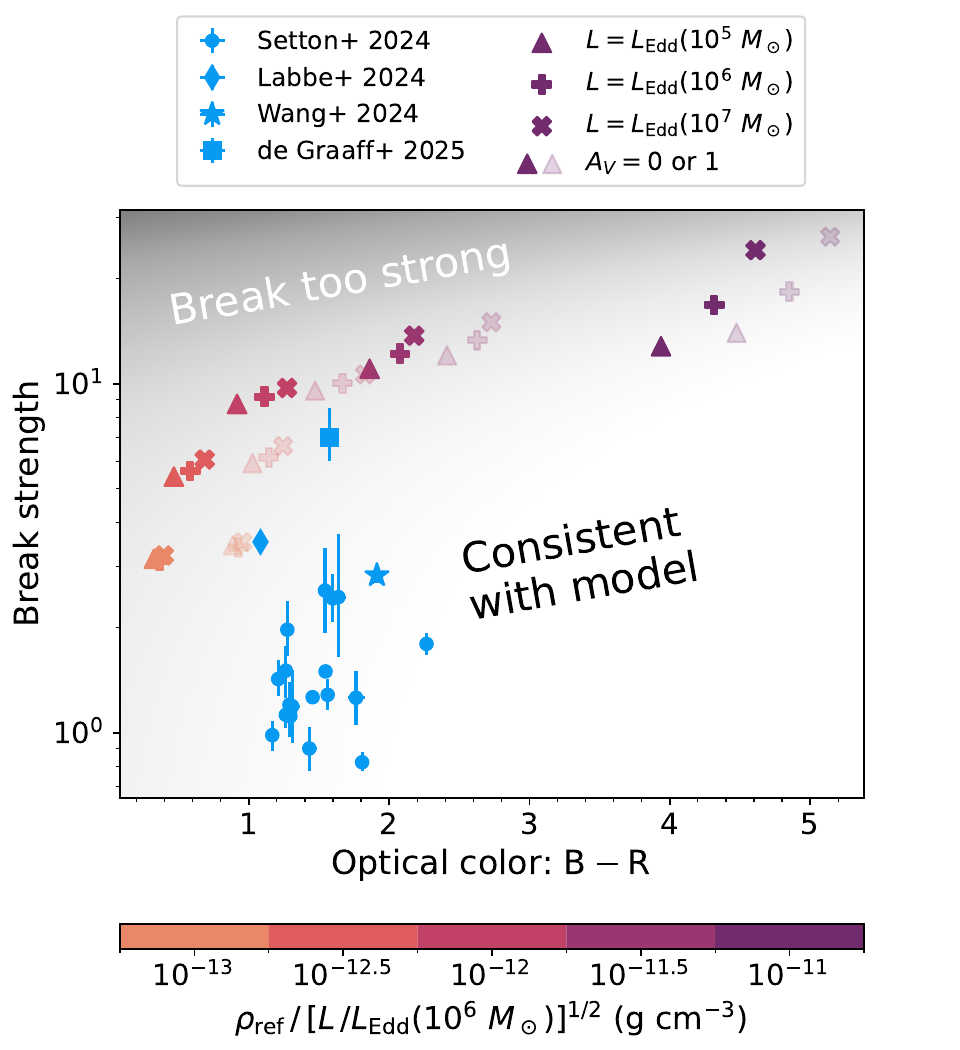}
    \caption{Balmer break strength and optical color of the sphere models (color-coded proportional to $\rho_{\rm ref}L^{-1/2}$; points with high transparency represent spectra extincted by a Small Magellanic Cloud-average dust law with $A_V=1$) compared to observation (blue points with error bars). As qualitatively indicated by the light shade in the background, the spherical model can explain the region below the model data points (due to filling by the UV component), consistent with the distribution of known LRDs. }
    \label{fig:color_break}
\end{figure}

In Figure~\ref{fig:color_break}, we further quantify the model Balmer break strength and compare it to observation, including the same collection of observed LRDs that we showed in Figure~\ref{fig:diskspec_m7l2} and RUBIES-UDS-154183. We follow \citet{deGraaff2025} to evaluate the observed Balmer break strengths and uncertainties with the flux density ($f_\nu$) ratio at two bands, $[4000,4100]{\rm~\AA}$ and $[3620,3670]{\rm~\AA}$. To characterize the Balmer break from the models, we use the luminosity ($L_\nu$) ratio in two wavelength bins, $[4000,4100]{\rm~\AA}$ and $[3640,3645]{\rm~\AA}$. The blue band is narrower than the one used for the observed LRDs in order not to traverse the theoretical Balmer limit. We also quantify the optical color of the model and observed spectra. We follow \citet{Setton2024} and evaluate the model and observed flux densities within two top-hat filters, ``B'' in $[4000,4700]{\rm~\AA}$ and ``R'' in $[6700,7400]{\rm~\AA}$. We exclude sources at $z>6$ as their ``R'' band is not fully covered by JWST/NIRSpec. Our break strength and color measurements are broadly consistent with those reported in \citet{deGraaff2025} and \citet{Setton2024}.

We have so far neglected dust attenuation. For the brightest LRDs, rest-frame mid- to far-infrared non-detections put upper limits at $A_V\lesssim1$, depending on the dust extinction curve assumed \citep{Setton2025,Chen2025}. We investigate this effect by applying an extinction of $A_V=1$ to the model continuum spectra using a dust extinction curve averaged over the Small Magellanic Cloud \citep{Gordon2024}. The measured colors and break strengths of these extincted models are shown as translucent markers in Figure~\ref{fig:color_break}. The model points with $A_V=0$ and $A_V=1$ represent the observational uncertainty in dust extinction. 

The model points without dust extinction are distributed roughly around a break strength of 10 in Figure~\ref{fig:color_break}, with redder optical colors correlating with stronger breaks. This trend persists even when $\rm B-R>2$, where the effective temperature falls below the optimal temperature in Figure~\ref{fig:kappa_ratio}. This is presumably because the atmosphere layer where $T\sim5000{\rm~K}$ is not very optically thick even for the coolest models (see the right panel in Figure~\ref{fig:spheregray_m6}), so a Balmer break formed there can still manifest itself. At a given optical color, the break strength only weakly depends on luminosity. Dust extinction of $A_V=1$ shifts each model point redward by around 0.6, while only slightly enhancing the break strength. 

We broadly term the area at or below the stripe of the model points as consistent with the models (when complemented with a UV component, as discussed above). This is qualitatively indicated by the light shade in the background of Figure~\ref{fig:color_break}. The boundary of this area is slightly more stringent when assuming $A_V=1$, especially for relatively blue optical colors. Regardless of the uncertainty in dust extinction, the points denoting observed objects generally fall into this area, even for RUBIES-UDS-154183 \citep{deGraaff2025}; see also Figure~\ref{fig:spherespec_obs}. In the current literature, this source is unmatched in the break strength except for ``MoM-BH*-1'' \citep{Naidu2025}, an LRD at $z=7.76$ with a very similar Balmer break and continuum up to $\lambda_{\rm rest}=6000{\rm~\AA}$, the red edge of the available spectrum. Therefore, we find no significant inconsistency between the model and most, if not all, known LRDs in terms of the Balmer break strength.

\section{Discussion and summary}

\label{sec:conclusions}

This work aims to connect the observed Balmer break and optical redness of LRDs with the theoretical picture of super-Eddington accretion onto black holes. To this end, we have constructed idealized atmosphere models of super-Eddington accretion systems in a disk or spherical geometry (Figure~\ref{fig:cartoon}) and conducted radiation transport calculations to study their continuum spectral properties. Our study contributes to a physically motivated interpretation of LRDs. 

Our first conclusion is that a photosphere at $T\sim5000{\rm~K}$ and $\rho<10^{-9}{\rm~g~cm^{-3}}$ produces a Balmer break. The underlying physics is the same as the well-known break feature in early-type stars, namely a discontinuity in opacity at the Balmer limit.   The optimal photosphere temperature to produce the Balmer break increases with density as indicated by the opacity law (Figure~\ref{fig:kappa_ratio}).   This is why our models produce strong Balmer breaks at lower effective temperatures than stellar atmospheres.

We further find that super-Eddington accretion systems naturally have $\rho<10^{-9}{\rm~g~cm^{-3}}$ at the photosphere. This is true for both the disk (Equation~(\ref{eq:disk_midplane_density})) and the sphere (Figure~\ref{fig:sphere_teff}) scenarios. We conclude that a super-Eddington system will robustly produce a Balmer break if the photosphere is cool (Figures~\ref{fig:diskspec_m7l2}, \ref{fig:spherespec_m6}--\ref{fig:color_break}). 

The remaining question is the ubiquity of the cool photosphere, i.e. the optical redness of LRDs. We find that the sphere scenario gives an effective temperature of $4000{\rm~K}\lesssim T_{\rm eff}\lesssim6000{\rm~K}$ over a wide range of the reference densities, i.e., accretion rates (Figure~\ref{fig:sphere_teff}). This is due to electron recombination producing a strong temperature dependence of the opacity, and is analogous to the origin of the Hayashi line in stellar models. Together, the sphere scenario naturally reproduces the optical color and the Balmer break of LRDs (Figure~\ref{fig:spherespec_obs}).  By contrast, a standard thin disk model (which can be present in the outer parts of super-Eddington accretion flows) would explain the optical redness of LRDs only if the disk is truncated at the radius where the effective temperature is $\sim 5000$ K. We are unsure why this should generically be the case.   

\subsection{Comparison to prior work}

Our explanation of the Balmer break is related to the work of \citet{Inayoshi2025}, who proposed that a dense gas shell, where neutral hydrogen is collisionally excited to the $n=2$ state, absorbs photons bluer than the Balmer limit and thus imprints a Balmer break on an otherwise featureless continuum spectrum. They inferred the properties of the proposed gas shell using \textsc{Cloudy} models, a method subsequently adopted in various case studies \citep{Ji2025,Naidu2025,deGraaff2025,Taylor2025}. The photospheric Balmer break in our models has a similar microphysical origin and does not contradict the absorbing gas shell picture. However, in the language of \textsc{Cloudy}, our work does not need to assume an ``incident'' radiation spectrum: the continuum flux and the Balmer break  arise self-consistently from radiation transport calculations given a physically motivated model of the gas distribution. This allows us to predict the continuum color from atmosphere models. We also note that our method can handle large scattering optical depths expected in super-Eddington accretion flows, which \textsc{Cloudy} presently cannot model \citep[see Section 10.8 in the documentation ``Hazy1'',][]{Ferland2023}.

When this work was close to completion, \citet{Kido2025} proposed that an optically thick envelope around the black hole produces an effective temperature of $T_{\rm eff}\sim 5000$--$7000{\rm~K}$, close to the Hayashi limit \citep{Hayashi1961}. Both \citet{Kido2025} and we argue that for an optically thick spherical system, hydrogen recombination produces a large change in opacity with temperature and thus sets the effective temperature observed.  However, our work differs from \citet{Kido2025} in that we carry out detailed radiation transport calculations to predict spectra that are similar to those observed. In addition, we argue that the Balmer break and the continuum emission can be produced roughly at the same location in the accretion flow photosphere, rather than requiring a separate dynamical component at larger radii to produce the Balmer break.  The importance of the change in opacity at hydrogen recombination emphasized here and in \citet{Kido2025} in the context of LRDs was in fact predicted in the ``quasi-star'' model \citep{Begelman2008} for the early growth of supermassive black hole seeds, whose late stage may be related to the LRDs themselves \citep{Begelman2025}.  \citet{Begelman2008} also highlighted the analogy to the Hayashi line in stellar models that we have emphasized \citep[see also][]{Hosokawa2013}.

\subsection{Caveats and future directions}

All models in this work have assumed LTE, and our spectral calculations do not account for lines or gas motion. Detailed spectral predictions, including the emission and absorption lines observed from LRDs, require non-LTE radiation transport and likely depend on the atmosphere structure. However, the physical analogy between the photosphere of spherical super-Eddington accretion flows and that of stars suggests possible observational similarity in their spectral features, of which the Balmer break is just one example. We anticipate that future high-resolution spectroscopy can probe the density, temperature, metallicity, and kinematics of the LRD photosphere, which will test our scenarios directly \citep[see an interesting observational case in][]{Lin2025b}. 

Despite the success of our model in reproducing aspects of the observed spectra of LRDs, many open questions remain. Returning to the cartoon in Figure~\ref{fig:cartoon}, while we can reproduce a Balmer break from a disk geometry, it is unclear how to truncate the disk at $T_{\rm eff}\sim5000{\rm~K}$, which is required to reproduce the optical redness of the LRDs (Figure~\ref{fig:diskspec_m7l2}). A structural change of the disk, due to the opacity or other properties of the gas at this particular effective temperature, might generate such a truncation. Specifically, the disk could potentially become quasi-spherical at small radii, leading to a composite disk-sphere scenario.   However, we have not explored at what radius this transition should occur and whether it is connected to an effective temperature of $\sim 5000$ K. In the sphere scenario, we have remained agnostic about the exact cause of the geometric thickness. Radiation pressure alone can only support a geometrically thick disk within a radius $\sim10\lambda_{\rm Edd}R_{\rm g}$, which is generally smaller than the photospheric radius in our spherical models.  If, however, dense gas is fueled from larger scales with a wide range of angular momenta, or if optically thick winds from small radii modify the dynamical and emission properties of the gas at large radii, the spherical structure may be maintained at the scale of the $\sim5000{\rm~K}$ photosphere. 

We have yet to account for the origin of the broad emission lines and the blue UV continua of LRDs. Super-Eddington accretion theory and simulations generically predict rapid outflows with a much lower density than other regions of the flow \citep[e.g.,][]{Shakura1973,Jiang2019}; these may also be quite clumpy \citep{Kobayashi2018}, as is the case for radiation-driven stellar winds \citep[e.g.,][]{Owocki2004}.  Such lower-density regions may expose hot gas at smaller radii, whose thermal emission may contribute to the blue UV continuum and emit ionizing photons that power the broad lines. Indeed, our spherical models with lower reference density (which could correspond to a higher-speed component at a given mass flow rate) have significantly higher effective temperatures (Figures~\ref{fig:sphere_teff} and \ref{fig:spherespec_m6}).   An appropriate superposition of a range of flow velocities and densities could plausibly explain the full spectra of LRDs from the UV to the near-infrared, including the broad emission lines. Still, the UV continuum may additionally contain starlight from the host galaxy \citep{Naidu2025,Rinaldi2024,Chen2025a,Lin2025b}, nebular emission \citep{Chen2025b}, or a mixture of these contributions. If so, a complete model of LRDs will need to extend beyond the accreting black hole towards the galactic scale. 

The gas density and velocity distributions are undoubtedly much more complex than our simple models, which may be responsible for the complex broad line profiles, including Balmer absorption, of many LRDs \citep[e.g.,][]{Maiolino2025,Kocevski2024,Juodzbalis2024}. In addition, super-Eddington accretion flows may explain the X-ray weakness of LRDs \citep{Pacucci2024,Lambrides2024,Inayoshi2024,Madau2024,Trinca2024}; while this was recently questioned in \citet{Sacchi2025}, we note that quantitative X-ray predictions depend on computational methods, the assumed gas geometry, the bolometric correction, and physical parameters such as the Eddington ratio. Ultimately, the simple super-Eddington models presented here will need to be supplemented by realistic three-dimensional simulations tailored for LRDs---those with physical parameters relevant to supermassive black holes and enough spatial dynamic range to include the regions responsible for emission from the X-ray (if present) to the optical-infrared (we expect that the optical-infrared emission arises at large radii where current simulations are not in statistically steady state).   The results of super-Eddington simulations may be particularly sensitive to the geometry of the accreting gas.  The gas is possibly much less ordered than standard rotating initial conditions used in most simulations, which could suppress the X-ray emission and the viewing angle dependence of the spectra.  Drawing initial conditions from simulations of high-redshift galaxies could inform more physical models of super-Eddington accretion in LRDs.

The interpretation favored in this paper is one in which LRDs are powered by relatively low-mass black holes that are highly super-Eddington. This suggests an important phase of black hole growth, with potential evolutionary links to the initial seeding stage (\citealt{Inayoshi2025b,Inayoshi2020,Volonteri2021}; see also the quasi-star scenario in \citealt{Begelman2008,Coughlin2024}). Although significant work remains, our work points to LRDs as an exciting and unique new probe of black hole evolution and demographics.

\begin{acknowledgments}
We thank the anonymous referee for constructive comments that helped to improve this work. H. L. thanks Lizhong Zhang, Minghao Guo, Xiurui Zhao, and Bertrand Plez for constructive discussions. The simulations presented in this article were performed on computational resources managed and supported by Princeton Research Computing, a consortium of groups including the Princeton Institute for Computational Science and Engineering (PICSciE) and the Office of Information Technology's High Performance Computing Center and Visualization Laboratory at Princeton University. The Center for Computational Astrophysics at the Flatiron Institute is supported by the Simons Foundation.
\end{acknowledgments}

The spherical model spectra are available on Zenodo under an open-source Creative Commons Attribution license: \dataset[10.5281/zenodo.17204644]{https://doi.org/10.5281/zenodo.17204644}.

\software{Astropy \citep{astropy2013,astropy2018,astropy2022},  Athena++ \citep{Stone2020}, Dust\_extincton \citep{Gordon2024JOSS}, FastChem \citep{Stock2022},  Matplotlib \citep{Hunter2007}, Numpy \citep{Harris2020}, Optab \citep{Hirose2022}, Pyphot \citep{zenodopyphot}, Scipy \citep{Scipy2020}, Seaborn \citep{Waskom2021} }

\appendix
\section{Vertical structure of the disk model}
\label{app:disk_analytical}
We consider a radiation-dominated thin disk. At each annulus at radius $r$, the equations include hydrostatic equilibrium,
\begin{equation}
    \frac{1}{3}\frac{dE_{\rm rad}}{dz} = -\Omega^2\rho z\,, \label{eq:disk_hydro}
\end{equation}
definition of the optical depth,
\begin{equation}
    \frac{d\tau}{dz} = -\kappa_R\rho\,, \label{eq:disk_opacity}
\end{equation}
radiative energy transport,
\begin{equation}
    \frac{1}{3}\frac{dE_{\rm rad}}{d\tau} = \frac{F}{c}\,, \label{eq:disk_radiation}
\end{equation}
and thermal equilibrium,
\begin{equation}
    \frac{dF}{d\tau} = -\frac{Q}{\kappa_R\rho}\,. \label{eq:disk_thermal_equilibrium}
\end{equation}
Here, $z$ is the vertical coordinate, with $z=0$ being the midplane location, $E_{\rm rad}\equiv aT^4$ is the radiation energy density, $\Omega\equiv\sqrt{GM_{\rm BH}/r^3}$ is the Keplerian angular velocity, $F$ is the vertical radiation flux, $\kappa_R$ is the Rosseland-mean opacity, and $Q$ is the heating rate per unit volume due to accretion. Consistent with previous radiation hydrodynamic simulations of radiation-dominated accretion disks \citep{Hirose2009}, we assume $Q\propto \rho\tau^{-1/2}$. The boundary conditions are $\{\rho=0$, $F=\sigma T_{\rm eff}^4$, $E=aT_{\rm eff}^4/2\}$ at $\tau=0$, where $T_{\rm eff}$ is given by Equation~(\ref{eq:disk_Teff}), and $\{z=0$, $F=0\}$ at $\tau=\tau_m$, where $\tau_m\gg1$ is the midplane optical depth to be determined later. We have five boundary conditions for the four equations; the one extra condition serves to fix the normalization factor of $Q$. 

The above system of equations is challenging to solve even numerically due to the complex temperature and density dependence of the opacity. Moreover, we expect the solution to involve dynamically unstable density inversion (Appendix~\ref{subapp:disk_inversion}). Therefore, we adopt an approximate solution as follows. We first assume that $\kappa_R=\kappa_{\rm es}$, which is accurate in the ionized disk interior. This allows us to solve Equation~(\ref{eq:disk_thermal_equilibrium}) and then Equation~(\ref{eq:disk_radiation}) analytically:
\begin{gather}
    F(\tau) = \sigma T_{\rm eff}^4\left(1-\sqrt{\frac{\tau}{\tau_m}}\right)\,, \label{eq:disk_F_sol}\\
    E_{\rm rad}(\tau) = \frac{1}{2}aT_{\rm eff}^4\left(1+\frac{3}{2}\tau-\sqrt{\frac{\tau}{\tau_m}}\tau\right)\,,\label{eq:disk_Erad_sol}
\end{gather} 
and then Equations~(\ref{eq:disk_hydro})(\ref{eq:disk_opacity}) imply
\begin{equation}
    \rho(z) = \frac{2\tau_m}{\kappa_{\rm es}H}\left(1-\frac{z}{H}\right)\,, \label{eq:disk_density}
\end{equation}
where $0\leq z \leq H$, and
\begin{equation}
    H = \frac{\kappa_{\rm es}\sigma T_{\rm eff}^4}{\Omega^2c} \label{eq:disk_height}
\end{equation}
is the height at which the density drops to zero. Note that $\rho$ monotonically decreases with $z$, so this solution excludes any density inversion. The temperature profile follows from $E_{\rm rad}=aT^4$. Next, we partially relax the assumption of $\kappa_R=\kappa_{\rm es}$. Using the density profile, Equation~(\ref{eq:disk_density}), we calculate $\tau(z)$ from Equation~(\ref{eq:disk_opacity}) using a realistic table of the Rosseland-mean opacity $\kappa_R(\rho,T)$. The optical depth profile then gives $F(z)$ and $E_{\rm rad}(z)$ from Equation~(\ref{eq:disk_F_sol})(\ref{eq:disk_Erad_sol}). In effect, if we assume $Q\propto\kappa_R\rho\tau^{-1/2}$ instead of $Q\propto\rho\tau^{-1/2}$ (being still consistent with \citealt{Hirose2009}, where $\kappa_R$ is dominated by $\kappa_{\rm es}$), then our approximate solution will strictly satisfy Equations~(\ref{eq:disk_opacity})(\ref{eq:disk_radiation})(\ref{eq:disk_thermal_equilibrium}) with realistic opacities, agree with Equation~(\ref{eq:disk_hydro}) in the disk interior, and avoid density inversion near the disk surface when $T<10^4{\rm~K}$. 

The remaining undetermined quantity is $\tau_m$. We use the $\alpha$-disk prescription, requiring
\begin{equation}
    \frac{\alpha}{\Omega}\int_0^H \frac{1}{3}E_{\rm rad}dz = \frac{\dot{M}}{6\pi}\,,\label{eq:disk_alpha}
\end{equation}
which is generalized from the standard, vertically integrated $\alpha$-disk model. We fix $\alpha=0.1$ in this work. This additional equation determines $\tau_m$ and completes our analytical model. In practice, $\tau_m$ is determined iteratively. For each set of independent parameters ($M, \lambda_{\rm Edd}$), we explore a range of radii that give effective temperatures around $5000{\rm~K}$. For each $r$, we make an initial guess for $\tau_m$, obtain $F(\tau), E_{\rm rad}(\tau), \rho(z), $ and $\tau(z)$, then compare the two sides of Equation~(\ref{eq:disk_alpha}) and update $\tau_m$ until this equation is satisfied. Finally, we collect $\rho(z), Q(z), T(z)$, and $F(z)$ and apply them to the simulations. The density profile is fixed. The heating rate is implemented as a fixed explicit energy source term in \textsc{Athena++}. The last two terms serve as initial conditions, as described in Section~\ref{subsec:method_Athena}.

We finally note that the disk vertical profile remains geometrically thin in our parameter range of interest. Combining Equations~(\ref{eq:disk_Teff})(\ref{eq:disk_height}), we obtain the dimensionless scale height
\begin{align}
    {H\over r} &= {\kappa_{\rm es}\over c}\left({9\dot{M}^2\sigma T_{\rm eff}^4\over64\pi^2GM_{\rm BH}}\right)^{1/3} \nonumber\\
    &=0.114\left({\lambda_{\rm Edd}\over10^2}\right)^{2/3}\left({M_{\rm BH}\over 10^7~M_\odot}\right)^{1/3}\left({T_{\rm eff}\over5000\rm~K}\right)^{4/3}\,.
\end{align}
This further supports the thin-disk assumption, which we have justified in Section~\ref{subsec:disk_analytical} by comparing advective and radiative cooling of the disk.

\subsection{A note on density inversion}
\label{subapp:disk_inversion}
\begin{figure}
    \centering
    \includegraphics[width=0.49\textwidth]{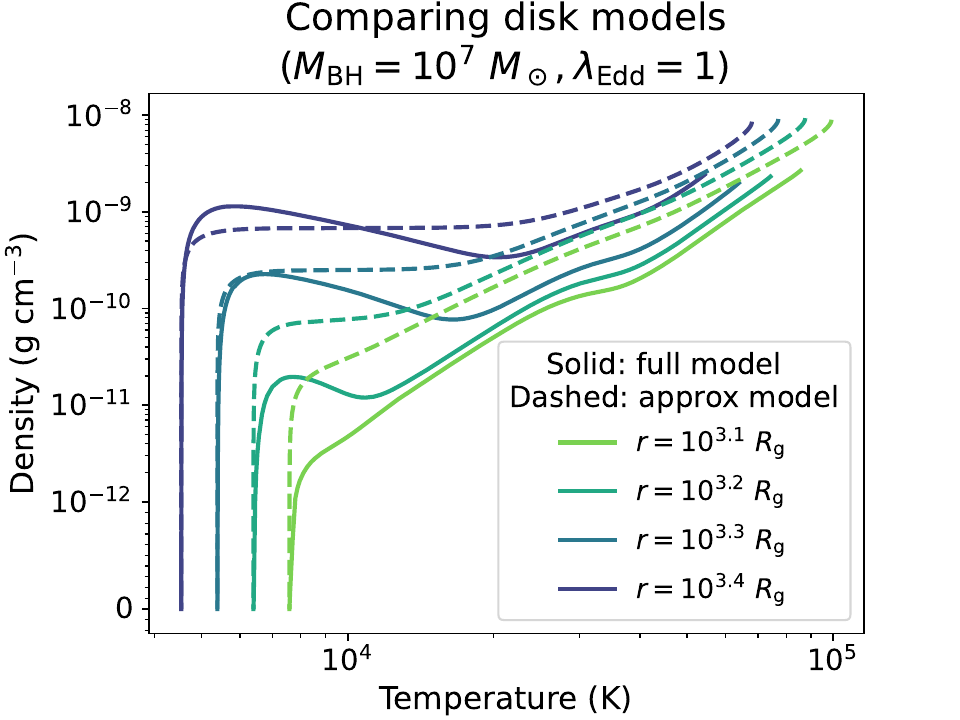}
    \caption{Comparison of model disk vertical profiles for the case $M_{\rm BH}=10^7~M_\odot, \lambda_{\rm Edd}=1$. The vertical axis is scaled linearly below $10^{-12}{\rm~g~cm^{-3}}$ and logarithmically above. Solid curves denote the full solution described in Appendix~\ref{subapp:disk_inversion}. Dashed curves denote the approximation described in Appendix~\ref{app:disk_analytical} and applied in the main text. The full model shows density inversion at radii $r\geq10^{3.2}~R_{\rm g}$, which is avoided in the approximate model. }
    \label{fig:disk_compare}
\end{figure}
We have described the equations satisfied by a thin disk annulus under hydrostatic equilibrium and radiative energy transport but did not solve them exactly. The choice is technical and physical.

The main challenge in solving the equations is the need to obtain $\rho$ implicitly from $\kappa_R$: we cannot directly update $\rho$ because it does not appear on the left-hand side of Equations~(\ref{eq:disk_hydro})--(\ref{eq:disk_thermal_equilibrium}). This requires mapping $\kappa_R$ to $\rho$ at a given temperature. But this is not always possible: sometimes, no value of $\rho$ in the entire opacity table will give the desired $\kappa_R$.  

This issue will be partially circumvented if we include the gas pressure in the hydrostatic equibiliurm equation, i.e., if we replace Equation~(\ref{eq:disk_hydro}) with
\begin{equation}
    \frac{d}{dz}\left(\frac{E_{\rm rad}}{3}+\frac{\rho k_BT}{\mu m_p}\right) = -\Omega^2\rho z\,. \label{eq:disk_hydro_gas}
\end{equation}
This introduces $\rho$ to the derivative term, which allows us to find $\rho$ using standard techniques of initial value problems and eliminates the need to infer $\rho$ from $\kappa_R$. However, Equation~(\ref{eq:disk_hydro_gas}) will become stiff in the super-Eddington regime, where $E_{\rm rad}\gg \rho k_BT/\mu m_p$, and evaluating the derivative of $\rho$ will result in catastrophic cancellation. In practice, we have successfully solved the system of Equations~(\ref{eq:disk_hydro_gas})(\ref{eq:disk_opacity})(\ref{eq:disk_radiation})(\ref{eq:disk_thermal_equilibrium}) when $M_{\rm BH}=10^7~M_\odot,\lambda_{\rm Edd}=1$ but not when $M_{\rm BH}=10^7~M_\odot,\lambda_{\rm Edd}=10^2$.

Even without these technical difficulties, the solution may show an unstable density inversion. We show this by solving Equations~(\ref{eq:disk_hydro_gas})(\ref{eq:disk_opacity})(\ref{eq:disk_radiation})(\ref{eq:disk_thermal_equilibrium}) numerically in the case $M_{\rm BH}=10^7~M_\odot, \lambda_{\rm Edd}=1$. Figure~\ref{fig:disk_compare} presents the solution for a range of radii in solid curves. For comparison, the dashed curves show the approximate solutions that we developed above in Appendix~\ref{app:disk_analytical} and applied to calculate the spectra in Figure~\ref{fig:diskspec_m7l2}. At $r\geq10^{3.2}~R_{\rm g}$, or $T_{\rm eff}\leq7606{\rm~K}$, a local maximum in density at $T<10^4{\rm~K}$ appears in the full solution, implying a disk surface layer that is denser than the interior. Previous work on sub-Eddington disks and primordial supermassive stars has noted similar behavior \citep{Hubeny2000,Hosokawa2013}. This configuration will unlikely be sustained in a realistic disk due to the Rayleigh-Taylor instability. On the other hand, in the approximate solutions, the density monotonically increases with temperature and hence decreases with height throughout each annulus; see Equation~(\ref{eq:disk_density}).

In summary, we adopt the approximate model because it is computationally tractable in the super-Eddington regime and because it does not show any physically unstable density inversion. As a caveat, we do not expect the model to represent a realistic annulus structure, which requires dedicated simulations especially in the presence of turbulence. 

\begin{figure}
    \centering
    \includegraphics[width=0.49\textwidth]{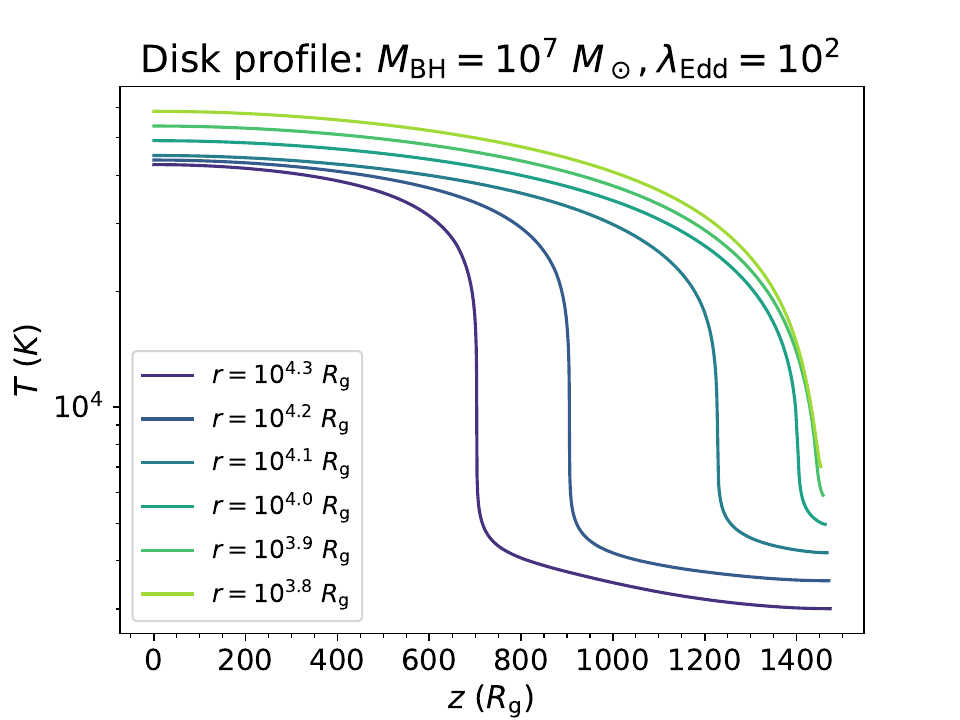}
    \caption{Disk vertical temperature profile from the gray simulations. The disk has $M_{\rm BH}=10^7~M_\odot$ and $\lambda_{\rm Edd}=10^2$. Each curve represents an individual simulation of an annulus. }
    \label{fig:disktemp_m7l2}
\end{figure}

\subsection{Temperature structure from gray simulations}

For completeness, this subsection reports the temperature profiles from gray simulations in the disk models. Figure~\ref{fig:disktemp_m7l2} shows the temperature of a range of annulus radii in the case with $M_{\rm BH}=10^7~M_\odot, \lambda_{\rm Edd}=10^2$. For each annulus, the temperature decreases with height, which is required for radiative energy transport. The right end of each curve reaches $T(\tau=0)=2^{-1/4}T_{\rm eff}$ (relative error $<4\%$), as expected from Equation~(\ref{eq:disk_Erad_sol}). For those annuli with low $T_{\rm eff}$, the temperature decreases dramatically with $z$ at $T\sim8000{\rm~K}$ and then flattens due to variations in the Rosseland- and Planck-mean opacities with temperature. 

\bibliography{bib}{}

\begin{thebibliography}{}
\expandafter\ifx\csname natexlab\endcsname\relax\def\natexlab#1{#1}\fi
\providecommand{\url}[1]{\href{#1}{#1}}
\providecommand{\dodoi}[1]{doi:~\href{http://doi.org/#1}{\nolinkurl{#1}}}
\providecommand{\doeprint}[1]{\href{http://ascl.net/#1}{\nolinkurl{http://ascl.net/#1}}}
\providecommand{\doarXiv}[1]{\href{https://arxiv.org/abs/#1}{\nolinkurl{https://arxiv.org/abs/#1}}}

\bibitem[{H.~B. {Akins} {et~al.}(2024){Akins}, {Casey}, {Lambrides}, {Allen}, {Andika}, {Brinch}, {Champagne}, {Cooper}, {Ding}, {Drakos}, {Faisst}, {Finkelstein}, {Franco}, {Fujimoto}, {Gentile}, {Gillman}, {Gozaliasl}, {Harish}, {Hayward}, {Hirschmann}, {Ilbert}, {Kartaltepe}, {Kocevski}, {Koekemoer}, {Kokorev}, {Liu}, {Long}, {McCracken}, {McKinney}, {Onoue}, {Paquereau}, {Renzini}, {Rhodes}, {Robertson}, {Shuntov}, {Silverman}, {Tanaka}, {Toft}, {Trakhtenbrot}, {Valentino}, \& {Zavala}}]{Akins2024}
{Akins}, H.~B., {Casey}, C.~M., {Lambrides}, E., {et~al.} 2024, \bibinfo{title}{{COSMOS-Web: The over-abundance and physical nature of ``little red dots''--Implications for early galaxy and SMBH assembly},} arXiv e-prints, arXiv:2406.10341, \dodoi{10.48550/arXiv.2406.10341}

\bibitem[{T.~T. {Ananna} {et~al.}(2024){Ananna}, {Bogd{\'a}n}, {Kov{\'a}cs}, {Natarajan}, \& {Hickox}}]{Ananna2024}
{Ananna}, T.~T., {Bogd{\'a}n}, {\'A}., {Kov{\'a}cs}, O.~E., {Natarajan}, P., \& {Hickox}, R.~C. 2024, \bibinfo{title}{{X-Ray View of Little Red Dots: Do They Host Supermassive Black Holes?},} \apjl, 969, L18, \dodoi{10.3847/2041-8213/ad5669}

\bibitem[{M. {Asplund} {et~al.}(2009){Asplund}, {Grevesse}, {Sauval}, \& {Scott}}]{Asplund2009}
{Asplund}, M., {Grevesse}, N., {Sauval}, A.~J., \& {Scott}, P. 2009, \bibinfo{title}{{The Chemical Composition of the Sun},} \araa, 47, 481, \dodoi{10.1146/annurev.astro.46.060407.145222}

\bibitem[{ {Astropy Collaboration} {et~al.}(2013){Astropy Collaboration}, {Robitaille}, {Tollerud}, {Greenfield}, {Droettboom}, {Bray}, {Aldcroft}, {Davis}, {Ginsburg}, {Price-Whelan}, {Kerzendorf}, {Conley}, {Crighton}, {Barbary}, {Muna}, {Ferguson}, {Grollier}, {Parikh}, {Nair}, {Unther}, {Deil}, {Woillez}, {Conseil}, {Kramer}, {Turner}, {Singer}, {Fox}, {Weaver}, {Zabalza}, {Edwards}, {Azalee Bostroem}, {Burke}, {Casey}, {Crawford}, {Dencheva}, {Ely}, {Jenness}, {Labrie}, {Lim}, {Pierfederici}, {Pontzen}, {Ptak}, {Refsdal}, {Servillat}, \& {Streicher}}]{astropy2013}
{Astropy Collaboration}, {Robitaille}, T.~P., {Tollerud}, E.~J., {et~al.} 2013, \bibinfo{title}{{Astropy: A community Python package for astronomy},} \aap, 558, A33, \dodoi{10.1051/0004-6361/201322068}

\bibitem[{ {Astropy Collaboration} {et~al.}(2018){Astropy Collaboration}, {Price-Whelan}, {Sip{\H{o}}cz}, {G{\"u}nther}, {Lim}, {Crawford}, {Conseil}, {Shupe}, {Craig}, {Dencheva}, {Ginsburg}, {VanderPlas}, {Bradley}, {P{\'e}rez-Su{\'a}rez}, {de Val-Borro}, {Aldcroft}, {Cruz}, {Robitaille}, {Tollerud}, {Ardelean}, {Babej}, {Bach}, {Bachetti}, {Bakanov}, {Bamford}, {Barentsen}, {Barmby}, {Baumbach}, {Berry}, {Biscani}, {Boquien}, {Bostroem}, {Bouma}, {Brammer}, {Bray}, {Breytenbach}, {Buddelmeijer}, {Burke}, {Calderone}, {Cano Rodr{\'\i}guez}, {Cara}, {Cardoso}, {Cheedella}, {Copin}, {Corrales}, {Crichton}, {D'Avella}, {Deil}, {Depagne}, {Dietrich}, {Donath}, {Droettboom}, {Earl}, {Erben}, {Fabbro}, {Ferreira}, {Finethy}, {Fox}, {Garrison}, {Gibbons}, {Goldstein}, {Gommers}, {Greco}, {Greenfield}, {Groener}, {Grollier}, {Hagen}, {Hirst}, {Homeier}, {Horton}, {Hosseinzadeh}, {Hu}, {Hunkeler}, {Ivezi{\'c}}, {Jain}, {Jenness}, {Kanarek}, {Kendrew}, {Kern}, {Kerzendorf}, {Khvalko}, {King}, {Kirkby}, {Kulkarni}, {Kumar}, {Lee}, {Lenz}, {Littlefair}, {Ma}, {Macleod}, {Mastropietro}, {McCully}, {Montagnac}, {Morris}, {Mueller}, {Mumford}, {Muna}, {Murphy}, {Nelson}, {Nguyen}, {Ninan}, {N{\"o}the}, {Ogaz}, {Oh}, {Parejko}, {Parley}, {Pascual}, {Patil}, {Patil}, {Plunkett}, {Prochaska}, {Rastogi}, {Reddy Janga}, {Sabater}, {Sakurikar}, {Seifert}, {Sherbert}, {Sherwood-Taylor}, {Shih}, {Sick}, {Silbiger}, {Singanamalla}, {Singer}, {Sladen}, {Sooley}, {Sornarajah}, {Streicher}, {Teuben}, {Thomas}, {Tremblay}, {Turner}, {Terr{\'o}n}, {van Kerkwijk}, {de la Vega}, {Watkins}, {Weaver}, {Whitmore}, {Woillez}, {Zabalza}, \& {Astropy Contributors}}]{astropy2018}
{Astropy Collaboration}, {Price-Whelan}, A.~M., {Sip{\H{o}}cz}, B.~M., {et~al.} 2018, \bibinfo{title}{{The Astropy Project: Building an Open-science Project and Status of the v2.0 Core Package},} \aj, 156, 123, \dodoi{10.3847/1538-3881/aabc4f}

\bibitem[{ {Astropy Collaboration} {et~al.}(2022){Astropy Collaboration}, {Price-Whelan}, {Lim}, {Earl}, {Starkman}, {Bradley}, {Shupe}, {Patil}, {Corrales}, {Brasseur}, {N{\"o}the}, {Donath}, {Tollerud}, {Morris}, {Ginsburg}, {Vaher}, {Weaver}, {Tocknell}, {Jamieson}, {van Kerkwijk}, {Robitaille}, {Merry}, {Bachetti}, {G{\"u}nther}, {Aldcroft}, {Alvarado-Montes}, {Archibald}, {B{\'o}di}, {Bapat}, {Barentsen}, {Baz{\'a}n}, {Biswas}, {Boquien}, {Burke}, {Cara}, {Cara}, {Conroy}, {Conseil}, {Craig}, {Cross}, {Cruz}, {D'Eugenio}, {Dencheva}, {Devillepoix}, {Dietrich}, {Eigenbrot}, {Erben}, {Ferreira}, {Foreman-Mackey}, {Fox}, {Freij}, {Garg}, {Geda}, {Glattly}, {Gondhalekar}, {Gordon}, {Grant}, {Greenfield}, {Groener}, {Guest}, {Gurovich}, {Handberg}, {Hart}, {Hatfield-Dodds}, {Homeier}, {Hosseinzadeh}, {Jenness}, {Jones}, {Joseph}, {Kalmbach}, {Karamehmetoglu}, {Ka{\l}uszy{\'n}ski}, {Kelley}, {Kern}, {Kerzendorf}, {Koch}, {Kulumani}, {Lee}, {Ly}, {Ma}, {MacBride}, {Maljaars}, {Muna}, {Murphy}, {Norman}, {O'Steen}, {Oman}, {Pacifici}, {Pascual}, {Pascual-Granado}, {Patil}, {Perren}, {Pickering}, {Rastogi}, {Roulston}, {Ryan}, {Rykoff}, {Sabater}, {Sakurikar}, {Salgado}, {Sanghi}, {Saunders}, {Savchenko}, {Schwardt}, {Seifert-Eckert}, {Shih}, {Jain}, {Shukla}, {Sick}, {Simpson}, {Singanamalla}, {Singer}, {Singhal}, {Sinha}, {Sip{\H{o}}cz}, {Spitler}, {Stansby}, {Streicher}, {{\v{S}}umak}, {Swinbank}, {Taranu}, {Tewary}, {Tremblay}, {de Val-Borro}, {Van Kooten}, {Vasovi{\'c}}, {Verma}, {de Miranda Cardoso}, {Williams}, {Wilson}, {Winkel}, {Wood-Vasey}, {Xue}, {Yoachim}, {Zhang}, {Zonca}, \& {Astropy Project Contributors}}]{astropy2022}
{Astropy Collaboration}, {Price-Whelan}, A.~M., {Lim}, P.~L., {et~al.} 2022, \bibinfo{title}{{The Astropy Project: Sustaining and Growing a Community-oriented Open-source Project and the Latest Major Release (v5.0) of the Core Package},} \apj, 935, 167, \dodoi{10.3847/1538-4357/ac7c74}

\bibitem[{J.~F.~W. {Baggen} {et~al.}(2024){Baggen}, {van Dokkum}, {Brammer}, {de Graaff}, {Franx}, {Greene}, {Labb{\'e}}, {Leja}, {Maseda}, {Nelson}, {Rix}, {Wang}, \& {Weibel}}]{Baggen2024}
{Baggen}, J. F.~W., {van Dokkum}, P., {Brammer}, G., {et~al.} 2024, \bibinfo{title}{{The Small Sizes and High Implied Densities of ``Little Red Dots'' with Balmer Breaks Could Explain Their Broad Emission Lines without an Active Galactic Nucleus},} \apjl, 977, L13, \dodoi{10.3847/2041-8213/ad90b8}

\bibitem[{S.~A. {Balbus} \& J.~F. {Hawley}(1991){Balbus} \& {Hawley}}]{Balbus1991}
{Balbus}, S.~A., \& {Hawley}, J.~F. 1991, \bibinfo{title}{{A Powerful Local Shear Instability in Weakly Magnetized Disks. I. Linear Analysis},} \apj, 376, 214, \dodoi{10.1086/170270}

\bibitem[{G. {Barro} {et~al.}(2024){Barro}, {P{\'e}rez-Gonz{\'a}lez}, {Kocevski}, {McGrath}, {Trump}, {Simons}, {Somerville}, {Yung}, {Arrabal Haro}, {Akins}, {Bagley}, {Cleri}, {Costantin}, {Davis}, {Dickinson}, {Finkelstein}, {Giavalisco}, {G{\'o}mez-Guijarro}, {Hathi}, {Hirschmann}, {Holwerda}, {Huertas-Company}, {Kartaltepe}, {Koekemoer}, {Lucas}, {Papovich}, {Pirzkal}, {Seill{\'e}}, {Tacchella}, {Wuyts}, {Wilkins}, {de la Vega}, {Yang}, \& {Zavala}}]{Barro2024}
{Barro}, G., {P{\'e}rez-Gonz{\'a}lez}, P.~G., {Kocevski}, D.~D., {et~al.} 2024, \bibinfo{title}{{Extremely Red Galaxies at z = 5{\textendash}9 with MIRI and NIRSpec: Dusty Galaxies or Obscured Active Galactic Nuclei?},} \apj, 963, 128, \dodoi{10.3847/1538-4357/ad167e}

\bibitem[{M.~C. {Begelman} \& J. {Dexter}(2025){Begelman} \& {Dexter}}]{Begelman2025}
{Begelman}, M.~C., \& {Dexter}, J. 2025, \bibinfo{title}{{Little Red Dots As Late-stage Quasi-stars},} arXiv e-prints, arXiv:2507.09085, \dodoi{10.48550/arXiv.2507.09085}

\bibitem[{M.~C. {Begelman} {et~al.}(2008){Begelman}, {Rossi}, \& {Armitage}}]{Begelman2008}
{Begelman}, M.~C., {Rossi}, E.~M., \& {Armitage}, P.~J. 2008, \bibinfo{title}{{Quasi-stars: accreting black holes inside massive envelopes},} \mnras, 387, 1649, \dodoi{10.1111/j.1365-2966.2008.13344.x}

\bibitem[{R. {Bezanson} {et~al.}(2024){Bezanson}, {Labbe}, {Whitaker}, {Leja}, {Price}, {Franx}, {Brammer}, {Marchesini}, {Zitrin}, {Wang}, {Weaver}, {Furtak}, {Atek}, {Coe}, {Cutler}, {Dayal}, {van Dokkum}, {Feldmann}, {F{\"o}rster Schreiber}, {Fujimoto}, {Geha}, {Glazebrook}, {de Graaff}, {Greene}, {Juneau}, {Kassin}, {Kriek}, {Khullar}, {Maseda}, {Mowla}, {Muzzin}, {Nanayakkara}, {Nelson}, {Oesch}, {Pacifici}, {Pan}, {Papovich}, {Setton}, {Shapley}, {Smit}, {Stefanon}, {Taylor}, \& {Williams}}]{Bezanson2024}
{Bezanson}, R., {Labbe}, I., {Whitaker}, K.~E., {et~al.} 2024, \bibinfo{title}{{The JWST UNCOVER Treasury Survey: Ultradeep NIRSpec and NIRCam Observations before the Epoch of Reionization},} \apj, 974, 92, \dodoi{10.3847/1538-4357/ad66cf}

\bibitem[{M. {Brazzini} {et~al.}(2025){Brazzini}, {D'Eugenio}, {Maiolino}, {Juod{\v{z}}balis}, {Ji}, \& {Scholtz}}]{Brazzini2025}
{Brazzini}, M., {D'Eugenio}, F., {Maiolino}, R., {et~al.} 2025, \bibinfo{title}{{Ruling out dominant electron scattering in Little Red Dots' Rosetta Stone using multiple hydrogen lines},} arXiv e-prints, arXiv:2507.08929, \dodoi{10.48550/arXiv.2507.08929}

\bibitem[{M. {Carranza-Escudero} {et~al.}(2025){Carranza-Escudero}, {Conselice}, {Adams}, {Harvey}, {Austin}, {Behroozi}, {Ferreira}, {Ormerod}, {Duan}, {Trussler}, {Li}, {Westcott}, {Windhorst}, {Coe}, {Cohen}, {Cheng}, {Driver}, {Frye}, {Furtak}, {Grogin}, {Hathi}, {Jansen}, {Koekemoer}, {Marshall}, {O'Brien}, {Pirzkal}, {Polletta}, {Robotham}, {Rutkowski}, {Summers}, {Wilkins}, {Willmer}, {Yan}, \& {Zitrin}}]{Carranza-Escudero2025}
{Carranza-Escudero}, M., {Conselice}, C.~J., {Adams}, N., {et~al.} 2025, \bibinfo{title}{{Lonely Little Red Dots: Challenges to the AGN-nature of little red dots through their clustering and spectral energy distributions},} arXiv e-prints, arXiv:2506.04004, \dodoi{10.48550/arXiv.2506.04004}

\bibitem[{C.~M. {Casey} {et~al.}(2025){Casey}, {Akins}, {Finkelstein}, {Franco}, {Fujimoto}, {Liu}, {Long}, {Magdis}, {Manning}, {McKinney}, {Shuntov}, \& {Tanaka}}]{Casey2025}
{Casey}, C.~M., {Akins}, H.~B., {Finkelstein}, S.~L., {et~al.} 2025, \bibinfo{title}{{An upper limit of 10$^6$ M$_\odot$ in dust from ALMA observations in 60 Little Red Dots},} arXiv e-prints, arXiv:2505.18873, \dodoi{10.48550/arXiv.2505.18873}

\bibitem[{R. {Cayrel} {et~al.}(2004){Cayrel}, {Depagne}, {Spite}, {Hill}, {Spite}, {Fran{\c{c}}ois}, {Plez}, {Beers}, {Primas}, {Andersen}, {Barbuy}, {Bonifacio}, {Molaro}, \& {Nordstr{\"o}m}}]{Cayrel2004}
{Cayrel}, R., {Depagne}, E., {Spite}, M., {et~al.} 2004, \bibinfo{title}{{First stars V - Abundance patterns from C to Zn and supernova yields in the early Galaxy},} \aap, 416, 1117, \dodoi{10.1051/0004-6361:20034074}

\bibitem[{S.-J. {Chang} {et~al.}(2025){Chang}, {Gronke}, {Matthee}, \& {Mason}}]{Chang2025}
{Chang}, S.-J., {Gronke}, M., {Matthee}, J., \& {Mason}, C. 2025, \bibinfo{title}{{Impact of Resonance, Raman, and Thomson Scattering on Hydrogen Line Formation in Little Red Dots},} arXiv e-prints, arXiv:2508.08768, \dodoi{10.48550/arXiv.2508.08768}

\bibitem[{C.-H. {Chen} {et~al.}(2025{\natexlab{a}}){Chen}, {Ho}, {Li}, \& {Inayoshi}}]{Chen2025b}
{Chen}, C.-H., {Ho}, L.~C., {Li}, R., \& {Inayoshi}, K. 2025{\natexlab{a}}, \bibinfo{title}{{The Physical Nature of the Off-centered Extended Emission Associated with the Little Red Dots},} arXiv e-prints, arXiv:2505.03183, \dodoi{10.48550/arXiv.2505.03183}

\bibitem[{C.-H. {Chen} {et~al.}(2025{\natexlab{b}}){Chen}, {Ho}, {Li}, \& {Zhuang}}]{Chen2025a}
{Chen}, C.-H., {Ho}, L.~C., {Li}, R., \& {Zhuang}, M.-Y. 2025{\natexlab{b}}, \bibinfo{title}{{The Host Galaxy (If Any) of the Little Red Dots},} \apj, 983, 60, \dodoi{10.3847/1538-4357/ada93a}

\bibitem[{K. {Chen} {et~al.}(2025){Chen}, {Li}, {Inayoshi}, \& {Ho}}]{Chen2025}
{Chen}, K., {Li}, Z., {Inayoshi}, K., \& {Ho}, L.~C. 2025, \bibinfo{title}{{Dust Budget Crisis in Little Red Dots},} arXiv e-prints, arXiv:2505.22600.
\newblock \doarXiv{2505.22600}

\bibitem[{E.~R. {Coughlin} \& M.~C. {Begelman}(2024){Coughlin} \& {Begelman}}]{Coughlin2024}
{Coughlin}, E.~R., \& {Begelman}, M.~C. 2024, \bibinfo{title}{{Quasi-stars as a Means of Rapid Black Hole Growth in the Early Universe},} \apj, 970, 158, \dodoi{10.3847/1538-4357/ad5723}

\bibitem[{A. {de Graaff} {et~al.}(2025{\natexlab{a}}){de Graaff}, {Rix}, {Naidu}, {Labbe}, {Wang}, {Leja}, {Matthee}, {Katz}, {Greene}, {Hviding}, {Baggen}, {Bezanson}, {Boogaard}, {Brammer}, {Dayal}, {van Dokkum}, {Goulding}, {Hirschmann}, {Maseda}, {McConachie}, {Miller}, {Nelson}, {Oesch}, {Setton}, {Shivaei}, {Weibel}, {Whitaker}, \& {Williams}}]{deGraaff2025}
{de Graaff}, A., {Rix}, H.-W., {Naidu}, R.~P., {et~al.} 2025{\natexlab{a}}, \bibinfo{title}{{A remarkable Ruby: Absorption in dense gas, rather than evolved stars, drives the extreme Balmer break of a Little Red Dot at $z=3.5$},} arXiv e-prints, arXiv:2503.16600, \dodoi{10.48550/arXiv.2503.16600}

\bibitem[{A. {de Graaff} {et~al.}(2025{\natexlab{b}}){de Graaff}, {Brammer}, {Weibel}, {Lewis}, {Maseda}, {Oesch}, {Bezanson}, {Boogaard}, {Cleri}, {Cooper}, {Gottumukkala}, {Greene}, {Hirschmann}, {Hviding}, {Katz}, {Labb{\'e}}, {Leja}, {Matthee}, {McConachie}, {Miller}, {Naidu}, {Price}, {Rix}, {Setton}, {Suess}, {Wang}, {Whitaker}, \& {Williams}}]{deGraaff2024}
{de Graaff}, A., {Brammer}, G., {Weibel}, A., {et~al.} 2025{\natexlab{b}}, \bibinfo{title}{{RUBIES: A complete census of the bright and red distant Universe with JWST/NIRSpec},} \aap, 697, A189, \dodoi{10.1051/0004-6361/202452186}

\bibitem[{L.~A. {Dreiling} \& R.~A. {Bell}(1980){Dreiling} \& {Bell}}]{Dreiling1980}
{Dreiling}, L.~A., \& {Bell}, R.~A. 1980, \bibinfo{title}{{The chemical composition, gravity and temperature of Vega.},} \apj, 241, 736, \dodoi{10.1086/158385}

\bibitem[{G.~J. Ferland {et~al.}(2023)Ferland, Chatzikos, Gunasekera, \& van Hoof}]{Ferland2023}
Ferland, G.~J., Chatzikos, M., Gunasekera, C.~M., \& van Hoof, P. A.~M. 2023, \bibinfo{title}{Cloudy C23.01 release,}, C23.01 Zenodo, \dodoi{10.5281/zenodo.14142065}

\bibitem[{M. Fouesneau(2025)Fouesneau}]{zenodopyphot}
Fouesneau, M. 2025, \bibinfo{title}{pyphot,}, pyphot\_v1.6.0 Zenodo, \dodoi{10.5281/zenodo.14712174}

\bibitem[{K. {Gordon}(2024){Gordon}}]{Gordon2024JOSS}
{Gordon}, K. 2024, \bibinfo{title}{{dust\_extinction: Interstellar Dust Extinction Models},} The Journal of Open Source Software, 9, 7023, \dodoi{10.21105/joss.07023}

\bibitem[{K.~D. {Gordon} {et~al.}(2024){Gordon}, {Fitzpatrick}, {Massa}, {Bohlin}, {Chastenet}, {Murray}, {Clayton}, {Lennon}, {Misselt}, \& {Sandstrom}}]{Gordon2024}
{Gordon}, K.~D., {Fitzpatrick}, E.~L., {Massa}, D., {et~al.} 2024, \bibinfo{title}{{Expanded Sample of Small Magellanic Cloud Ultraviolet Dust Extinction Curves: Correlations between the 2175 {\r{A}} Bump, q $_{PAH}$, Ultraviolet Extinction Shape, and N(H I)/A(V)},} \apj, 970, 51, \dodoi{10.3847/1538-4357/ad4be1}

\bibitem[{J.~E. {Greene} {et~al.}(2024){Greene}, {Labbe}, {Goulding}, {Furtak}, {Chemerynska}, {Kokorev}, {Dayal}, {Volonteri}, {Williams}, {Wang}, {Setton}, {Burgasser}, {Bezanson}, {Atek}, {Brammer}, {Cutler}, {Feldmann}, {Fujimoto}, {Glazebrook}, {de Graaff}, {Khullar}, {Leja}, {Marchesini}, {Maseda}, {Matthee}, {Miller}, {Naidu}, {Nanayakkara}, {Oesch}, {Pan}, {Papovich}, {Price}, {van Dokkum}, {Weaver}, {Whitaker}, \& {Zitrin}}]{Greene2024}
{Greene}, J.~E., {Labbe}, I., {Goulding}, A.~D., {et~al.} 2024, \bibinfo{title}{{UNCOVER Spectroscopy Confirms the Surprising Ubiquity of Active Galactic Nuclei in Red Sources at z > 5},} \apj, 964, 39, \dodoi{10.3847/1538-4357/ad1e5f}

\bibitem[{C.~R. Harris {et~al.}(2020)Harris, Millman, van~der Walt, Gommers, Virtanen, Cournapeau, Wieser, Taylor, Berg, Smith, Kern, Picus, Hoyer, van Kerkwijk, Brett, Haldane, del R{\'{i}}o, Wiebe, Peterson, G{\'{e}}rard-Marchant, Sheppard, Reddy, Weckesser, Abbasi, Gohlke, \& Oliphant}]{Harris2020}
Harris, C.~R., Millman, K.~J., van~der Walt, S.~J., {et~al.} 2020, \bibinfo{title}{Array programming with {NumPy},} Nature, 585, 357, \dodoi{10.1038/s41586-020-2649-2}

\bibitem[{C. {Hayashi}(1961){Hayashi}}]{Hayashi1961}
{Hayashi}, C. 1961, \bibinfo{title}{{Stellar evolution in early phases of gravitational contraction.},} \pasj, 13, 450

\bibitem[{S. {Hirose} {et~al.}(2022){Hirose}, {Hauschildt}, {Minoshima}, {Tomida}, \& {Sano}}]{Hirose2022}
{Hirose}, S., {Hauschildt}, P., {Minoshima}, T., {Tomida}, K., \& {Sano}, T. 2022, \bibinfo{title}{{OPTAB: Public code for generating gas opacity tables for radiation hydrodynamics simulations},} \aap, 659, A87, \dodoi{10.1051/0004-6361/202141076}

\bibitem[{S. {Hirose} {et~al.}(2009){Hirose}, {Krolik}, \& {Blaes}}]{Hirose2009}
{Hirose}, S., {Krolik}, J.~H., \& {Blaes}, O. 2009, \bibinfo{title}{{Radiation-Dominated Disks are Thermally Stable},} \apj, 691, 16, \dodoi{10.1088/0004-637X/691/1/16}

\bibitem[{H. {Holweger} \& E.~A. {Mueller}(1974){Holweger} \& {Mueller}}]{Holweger1974}
{Holweger}, H., \& {Mueller}, E.~A. 1974, \bibinfo{title}{{The Photospheric Barium Spectrum: Solar Abundance and Collision Broadening of Ba II Lines by Hydrogen},} \solphys, 39, 19, \dodoi{10.1007/BF00154968}

\bibitem[{T. {Hosokawa} {et~al.}(2013){Hosokawa}, {Yorke}, {Inayoshi}, {Omukai}, \& {Yoshida}}]{Hosokawa2013}
{Hosokawa}, T., {Yorke}, H.~W., {Inayoshi}, K., {Omukai}, K., \& {Yoshida}, N. 2013, \bibinfo{title}{{Formation of Primordial Supermassive Stars by Rapid Mass Accretion},} \apj, 778, 178, \dodoi{10.1088/0004-637X/778/2/178}

\bibitem[{H. {Hu} {et~al.}(2022){Hu}, {Inayoshi}, {Haiman}, {Quataert}, \& {Kuiper}}]{Hu2022}
{Hu}, H., {Inayoshi}, K., {Haiman}, Z., {Quataert}, E., \& {Kuiper}, R. 2022, \bibinfo{title}{{Long-term Evolution of Supercritical Black Hole Accretion with Outflows: A Subgrid Feedback Model for Cosmological Simulations},} \apj, 934, 132, \dodoi{10.3847/1538-4357/ac75d8}

\bibitem[{I. {Hubeny} {et~al.}(2000){Hubeny}, {Agol}, {Blaes}, \& {Krolik}}]{Hubeny2000}
{Hubeny}, I., {Agol}, E., {Blaes}, O., \& {Krolik}, J.~H. 2000, \bibinfo{title}{{Non-LTE Models and Theoretical Spectra of Accretion Disks in Active Galactic Nuclei. III. Integrated Spectra for Hydrogen-Helium Disks},} \apj, 533, 710, \dodoi{10.1086/308708}

\bibitem[{J.~D. Hunter(2007)Hunter}]{Hunter2007}
Hunter, J.~D. 2007, \bibinfo{title}{Matplotlib: A 2D graphics environment,} Computing in Science \& Engineering, 9, 90, \dodoi{10.1109/MCSE.2007.55}

\bibitem[{K. {Inayoshi}(2025){Inayoshi}}]{Inayoshi2025b}
{Inayoshi}, K. 2025, \bibinfo{title}{{Little Red Dots as the Very First Activity of Black Hole Growth},} arXiv e-prints, arXiv:2503.05537, \dodoi{10.48550/arXiv.2503.05537}

\bibitem[{K. {Inayoshi} {et~al.}(2025){Inayoshi}, {Kimura}, \& {Noda}}]{Inayoshi2024}
{Inayoshi}, K., {Kimura}, S.~S., \& {Noda}, H. 2025, \bibinfo{title}{{Weakness of X-rays and variability in high-redshift active galactic nuclei with super-Eddington accretion},} \pasj, 77, 811, \dodoi{10.1093/pasj/psaf050}

\bibitem[{K. {Inayoshi} \& R. {Maiolino}(2025){Inayoshi} \& {Maiolino}}]{Inayoshi2025}
{Inayoshi}, K., \& {Maiolino}, R. 2025, \bibinfo{title}{{Extremely Dense Gas around Little Red Dots and High-redshift Active Galactic Nuclei: A Nonstellar Origin of the Balmer Break and Absorption Features},} \apjl, 980, L27, \dodoi{10.3847/2041-8213/adaebd}

\bibitem[{K. {Inayoshi} {et~al.}(2020){Inayoshi}, {Visbal}, \& {Haiman}}]{Inayoshi2020}
{Inayoshi}, K., {Visbal}, E., \& {Haiman}, Z. 2020, \bibinfo{title}{{The Assembly of the First Massive Black Holes},} \araa, 58, 27, \dodoi{10.1146/annurev-astro-120419-014455}

\bibitem[{X. {Ji} {et~al.}(2025){Ji}, {Maiolino}, {{\"U}bler}, {Scholtz}, {D'Eugenio}, {Sun}, {Perna}, {Turner}, {Arribas}, {Bennett}, {Bunker}, {Carniani}, {Charlot}, {Cresci}, {Curti}, {Egami}, {Fabian}, {Inayoshi}, {Isobe}, {Jones}, {Juod{\v{z}}balis}, {Kumari}, {Lyu}, {Mazzolari}, {Parlanti}, {Robertson}, {Rodr{\'\i}guez Del Pino}, {Schneider}, {Sijacki}, {Tacchella}, {Trinca}, {Valiante}, {Venturi}, {Volonteri}, {Willott}, {Witten}, \& {Witstok}}]{Ji2025}
{Ji}, X., {Maiolino}, R., {{\"U}bler}, H., {et~al.} 2025, \bibinfo{title}{{BlackTHUNDER -- A non-stellar Balmer break in a black hole-dominated little red dot at $z=7.04$},} arXiv e-prints, arXiv:2501.13082, \dodoi{10.48550/arXiv.2501.13082}

\bibitem[{Y.-F. {Jiang}(2021){Jiang}}]{Jiang2021}
{Jiang}, Y.-F. 2021, \bibinfo{title}{{An Implicit Finite Volume Scheme to Solve the Time-dependent Radiation Transport Equation Based on Discrete Ordinates},} \apjs, 253, 49, \dodoi{10.3847/1538-4365/abe303}

\bibitem[{Y.-F. {Jiang}(2022){Jiang}}]{Jiang2022}
{Jiang}, Y.-F. 2022, \bibinfo{title}{{Multigroup Radiation Magnetohydrodynamics Based on Discrete Ordinates including Compton Scattering},} \apjs, 263, 4, \dodoi{10.3847/1538-4365/ac9231}

\bibitem[{Y.-F. {Jiang} {et~al.}(2025){Jiang}, {Blaes}, {Kaul}, \& {Zhang}}]{Jiang2025}
{Jiang}, Y.-F., {Blaes}, O., {Kaul}, I., \& {Zhang}, L. 2025, \bibinfo{title}{{Radiation and Magnetic Pressure Support in Accretion Disks Around Supermassive Black Holes and the Physical Origin of the Extreme-ultraviolet to Soft X-Ray Spectrum},} \apj, 988, 43, \dodoi{10.3847/1538-4357/addecb}

\bibitem[{Y.-F. {Jiang} {et~al.}(2013){Jiang}, {Stone}, \& {Davis}}]{Jiang2013}
{Jiang}, Y.-F., {Stone}, J.~M., \& {Davis}, S.~W. 2013, \bibinfo{title}{{On the Thermal Stability of Radiation-dominated Accretion Disks},} \apj, 778, 65, \dodoi{10.1088/0004-637X/778/1/65}

\bibitem[{Y.-F. {Jiang} {et~al.}(2014){Jiang}, {Stone}, \& {Davis}}]{Jiang2014}
{Jiang}, Y.-F., {Stone}, J.~M., \& {Davis}, S.~W. 2014, \bibinfo{title}{{A Global Three-dimensional Radiation Magneto-hydrodynamic Simulation of Super-Eddington Accretion Disks},} \apj, 796, 106, \dodoi{10.1088/0004-637X/796/2/106}

\bibitem[{Y.-F. {Jiang} {et~al.}(2019){Jiang}, {Stone}, \& {Davis}}]{Jiang2019}
{Jiang}, Y.-F., {Stone}, J.~M., \& {Davis}, S.~W. 2019, \bibinfo{title}{{Super-Eddington Accretion Disks around Supermassive Black Holes},} \apj, 880, 67, \dodoi{10.3847/1538-4357/ab29ff}

\bibitem[{T.~L. {John}(1975){John}}]{John1975}
{John}, T.~L. 1975, \bibinfo{title}{{The continuous absorption coefficient of atomic and molecular negative ions.},} \mnras, 172, 305, \dodoi{10.1093/mnras/172.2.305}

\bibitem[{T.~L. {John}(1988){John}}]{John1988}
{John}, T.~L. 1988, \bibinfo{title}{{Continuous absorption by the negative hydrogen ion reconsidered},} \aap, 193, 189

\bibitem[{I. {Juod{\v{z}}balis} {et~al.}(2024){Juod{\v{z}}balis}, {Ji}, {Maiolino}, {D'Eugenio}, {Scholtz}, {Risaliti}, {Fabian}, {Mazzolari}, {Gilli}, {Prandoni}, {Arribas}, {Bunker}, {Carniani}, {Charlot}, {Curtis-Lake}, {de Graaff}, {Hainline}, {Parlanti}, {Perna}, {P{\'e}rez-Gonz{\'a}lez}, {Robertson}, {Tacchella}, {{\"U}bler}, {Williams}, {Willott}, \& {Witstok}}]{Juodzbalis2024}
{Juod{\v{z}}balis}, I., {Ji}, X., {Maiolino}, R., {et~al.} 2024, \bibinfo{title}{{JADES - the Rosetta stone of JWST-discovered AGN: deciphering the intriguing nature of early AGN},} \mnras, 535, 853, \dodoi{10.1093/mnras/stae2367}

\bibitem[{D. {Kido} {et~al.}(2025){Kido}, {Ioka}, {Hotokezaka}, {Inayoshi}, \& {Irwin}}]{Kido2025}
{Kido}, D., {Ioka}, K., {Hotokezaka}, K., {Inayoshi}, K., \& {Irwin}, C.~M. 2025, \bibinfo{title}{{Black Hole Envelopes in Little Red Dots},} arXiv e-prints, arXiv:2505.06965, \dodoi{10.48550/arXiv.2505.06965}

\bibitem[{A. {King}(2024){King}}]{King2024}
{King}, A. 2024, \bibinfo{title}{{The black hole masses of high-redshift QSOs},} \mnras, 531, 550, \dodoi{10.1093/mnras/stae1171}

\bibitem[{H. {Kobayashi} {et~al.}(2018){Kobayashi}, {Ohsuga}, {Takahashi}, {Kawashima}, {Asahina}, {Takeuchi}, \& {Mineshige}}]{Kobayashi2018}
{Kobayashi}, H., {Ohsuga}, K., {Takahashi}, H.~R., {et~al.} 2018, \bibinfo{title}{{Three-dimensional structure of clumpy outflow from supercritical accretion flow onto black holes},} \pasj, 70, 22, \dodoi{10.1093/pasj/psx157}

\bibitem[{D.~D. {Kocevski} {et~al.}(2023){Kocevski}, {Onoue}, {Inayoshi}, {Trump}, {Arrabal Haro}, {Grazian}, {Dickinson}, {Finkelstein}, {Kartaltepe}, {Hirschmann}, {Aird}, {Holwerda}, {Fujimoto}, {Juneau}, {Amor{\'\i}n}, {Backhaus}, {Bagley}, {Barro}, {Bell}, {Bisigello}, {Calabr{\`o}}, {Cleri}, {Cooper}, {Ding}, {Grogin}, {Ho}, {Hutchison}, {Inoue}, {Jiang}, {Jones}, {Koekemoer}, {Li}, {Li}, {McGrath}, {Molina}, {Papovich}, {P{\'e}rez-Gonz{\'a}lez}, {Pirzkal}, {Wilkins}, {Yang}, \& {Yung}}]{Kocevski2023}
{Kocevski}, D.~D., {Onoue}, M., {Inayoshi}, K., {et~al.} 2023, \bibinfo{title}{{Hidden Little Monsters: Spectroscopic Identification of Low-mass, Broad-line AGNs at z > 5 with CEERS},} \apjl, 954, L4, \dodoi{10.3847/2041-8213/ace5a0}

\bibitem[{D.~D. {Kocevski} {et~al.}(2024){Kocevski}, {Finkelstein}, {Barro}, {Taylor}, {Calabr{\`o}}, {Laloux}, {Buchner}, {Trump}, {Leung}, {Yang}, {Dickinson}, {P{\'e}rez-Gonz{\'a}lez}, {Pacucci}, {Inayoshi}, {Somerville}, {McGrath}, {Akins}, {Bagley}, {Bisigello}, {Bowler}, {Carnall}, {Casey}, {Cheng}, {Cleri}, {Costantin}, {Cullen}, {Davis}, {Donnan}, {Dunlop}, {Ellis}, {Ferguson}, {Fujimoto}, {Fontana}, {Giavalisco}, {Grazian}, {Grogin}, {Hathi}, {Hirschmann}, {Huertas-Company}, {Holwerda}, {Illingworth}, {Juneau}, {Kartaltepe}, {Koekemoer}, {Li}, {Lucas}, {Magee}, {Mason}, {McLeod}, {McLure}, {Napolitano}, {Papovich}, {Pirzkal}, {Rodighiero}, {Santini}, {Wilkins}, \& {Yung}}]{Kocevski2024}
{Kocevski}, D.~D., {Finkelstein}, S.~L., {Barro}, G., {et~al.} 2024, \bibinfo{title}{{The Rise of Faint, Red AGN at $z>4$: A Sample of Little Red Dots in the JWST Extragalactic Legacy Fields},} arXiv e-prints, arXiv:2404.03576, \dodoi{10.48550/arXiv.2404.03576}

\bibitem[{V. {Kokorev} {et~al.}(2024){Kokorev}, {Caputi}, {Greene}, {Dayal}, {Trebitsch}, {Cutler}, {Fujimoto}, {Labb{\'e}}, {Miller}, {Iani}, {Navarro-Carrera}, \& {Rinaldi}}]{Kokorev2024}
{Kokorev}, V., {Caputi}, K.~I., {Greene}, J.~E., {et~al.} 2024, \bibinfo{title}{{A Census of Photometrically Selected Little Red Dots at 4 < z < 9 in JWST Blank Fields},} \apj, 968, 38, \dodoi{10.3847/1538-4357/ad4265}

\bibitem[{M. {Kokubo} \& Y. {Harikane}(2024){Kokubo} \& {Harikane}}]{Kokubo2024}
{Kokubo}, M., \& {Harikane}, Y. 2024, \bibinfo{title}{{Challenging the AGN scenario for JWST/NIRSpec broad H$\alpha$ emitters/Little Red Dots in light of non-detection of NIRCam photometric variability and X-ray},} arXiv e-prints, arXiv:2407.04777, \dodoi{10.48550/arXiv.2407.04777}

\bibitem[{R.~L. Kurucz(2019)Kurucz}]{kurucz_linelist}
Kurucz, R.~L. 2019, \bibinfo{title}{Line Lists,}, \url{http://kurucz.harvard.edu/linelists/}

\bibitem[{I. {Labb{\'e}} {et~al.}(2023){Labb{\'e}}, {van Dokkum}, {Nelson}, {Bezanson}, {Suess}, {Leja}, {Brammer}, {Whitaker}, {Mathews}, {Stefanon}, \& {Wang}}]{Labbe2023}
{Labb{\'e}}, I., {van Dokkum}, P., {Nelson}, E., {et~al.} 2023, \bibinfo{title}{{A population of red candidate massive galaxies 600 Myr after the Big Bang},} \nat, 616, 266, \dodoi{10.1038/s41586-023-05786-2}

\bibitem[{I. {Labbe} {et~al.}(2024){Labbe}, {Greene}, {Matthee}, {Treiber}, {Kokorev}, {Miller}, {Kramarenko}, {Setton}, {Ma}, {Goulding}, {Bezanson}, {Naidu}, {Williams}, {Atek}, {Brammer}, {Cutler}, {Chemerynska}, {Cloonan}, {Dayal}, {de Graaff}, {Fudamoto}, {Fujimoto}, {Furtak}, {Glazebrook}, {Heintz}, {Leja}, {Marchesini}, {Nanayakkara}, {Nelson}, {Oesch}, {Pan}, {Price}, {Shivaei}, {Sobral}, {Suess}, {van Dokkum}, {Wang}, {Weaver}, {Whitaker}, \& {Zitrin}}]{Labbe2024}
{Labbe}, I., {Greene}, J.~E., {Matthee}, J., {et~al.} 2024, \bibinfo{title}{{An unambiguous AGN and a Balmer break in an Ultraluminous Little Red Dot at z=4.47 from Ultradeep UNCOVER and All the Little Things Spectroscopy},} arXiv e-prints, arXiv:2412.04557, \dodoi{10.48550/arXiv.2412.04557}

\bibitem[{I. {Labbe} {et~al.}(2025){Labbe}, {Greene}, {Bezanson}, {Fujimoto}, {Furtak}, {Goulding}, {Matthee}, {Naidu}, {Oesch}, {Atek}, {Brammer}, {Chemerynska}, {Coe}, {Cutler}, {Dayal}, {Feldmann}, {Franx}, {Glazebrook}, {Leja}, {Maseda}, {Marchesini}, {Nanayakkara}, {Nelson}, {Pan}, {Papovich}, {Price}, {Suess}, {Wang}, {Weaver}, {Whitaker}, {Williams}, \& {Zitrin}}]{Labbe2025}
{Labbe}, I., {Greene}, J.~E., {Bezanson}, R., {et~al.} 2025, \bibinfo{title}{{UNCOVER: Candidate Red Active Galactic Nuclei at 3 < z < 7 with JWST and ALMA},} \apj, 978, 92, \dodoi{10.3847/1538-4357/ad3551}

\bibitem[{E. {Lambrides} {et~al.}(2024){Lambrides}, {Garofali}, {Larson}, {Ptak}, {Chiaberge}, {Long}, {Hutchison}, {Norman}, {McKinney}, {Akins}, {Berg}, {Chisholm}, {Civano}, {Cloonan}, {Endsley}, {Faisst}, {Gilli}, {Gillman}, {Hirschmann}, {Kartaltepe}, {Kocevski}, {Kokorev}, {Pacucci}, {Richardson}, {Stiavelli}, \& {Whalen}}]{Lambrides2024}
{Lambrides}, E., {Garofali}, K., {Larson}, R., {et~al.} 2024, \bibinfo{title}{{The Case for Super-Eddington Accretion: Connecting Weak X-ray and UV Line Emission in JWST Broad-Line AGN During the First Gyr of Cosmic Time},} arXiv e-prints, arXiv:2409.13047, \dodoi{10.48550/arXiv.2409.13047}

\bibitem[{H.-W. {Lee}(2005){Lee}}]{Lee2005}
{Lee}, H.-W. 2005, \bibinfo{title}{{Exact low-energy expansion of the Rayleigh scattering cross-section by atomic hydrogen},} \mnras, 358, 1472, \dodoi{10.1111/j.1365-2966.2005.08859.x}

\bibitem[{P. {Lenzuni} {et~al.}(1991){Lenzuni}, {Chernoff}, \& {Salpeter}}]{Lenzuni1991}
{Lenzuni}, P., {Chernoff}, D.~F., \& {Salpeter}, E.~E. 1991, \bibinfo{title}{{Rosseland and Planck Mean Opacities of a Zero-Metallicity Gas},} \apjs, 76, 759, \dodoi{10.1086/191580}

\bibitem[{R. {Li} {et~al.}(2025){Li}, {Ho}, \& {Chen}}]{Li2025}
{Li}, R., {Ho}, L.~C., \& {Chen}, C.-H. 2025, \bibinfo{title}{{The Dichotomy in the Nuclear and Host Galaxy Properties of High-redshift Quasars},} arXiv e-prints, arXiv:2505.12867, \dodoi{10.48550/arXiv.2505.12867}

\bibitem[{A.~P. {Lightman} \& D.~M. {Eardley}(1974){Lightman} \& {Eardley}}]{Lightman1974}
{Lightman}, A.~P., \& {Eardley}, D.~M. 1974, \bibinfo{title}{{Black Holes in Binary Systems: Instability of Disk Accretion},} \apjl, 187, L1, \dodoi{10.1086/181377}

\bibitem[{X. {Lin} {et~al.}(2025{\natexlab{a}}){Lin}, {Fan}, {Cai}, {Bian}, {Liu}, {Sun}, {Ma}, {Greene}, {Strauss}, {Green}, {Lyu}, {Champagne}, {Goulding}, {Inayoshi}, {Jin}, {Leung}, {Li}, {Liu}, {Mao}, {Pudoka}, {Tee}, {Wang}, {Wang}, {Wu}, {Yang}, {Zhang}, \& {Zhu}}]{Lin2025b}
{Lin}, X., {Fan}, X., {Cai}, Z., {et~al.} 2025{\natexlab{a}}, \bibinfo{title}{{The Discovery of Little Red Dots in the Local Universe: Signatures of Cool Gas Envelopes},} arXiv e-prints, arXiv:2507.10659, \dodoi{10.48550/arXiv.2507.10659}

\bibitem[{X. {Lin} {et~al.}(2025{\natexlab{b}}){Lin}, {Fan}, {Sun}, {Zhang}, {Egami}, {Helton}, {Wang}, {Zhang}, {Bunker}, {Cai}, {Ji}, {Jin}, {Maiolino}, {Pudoka}, {Rinaldi}, {Robertson}, {Tacchella}, {Tee}, {Sun}, {Willmer}, {Willott}, \& {Zhu}}]{Lin2025}
{Lin}, X., {Fan}, X., {Sun}, F., {et~al.} 2025{\natexlab{b}}, \bibinfo{title}{{The Large-scale Environments of Low-luminosity AGNs at $3.9 < z < 6$ and Implications for Their Host Dark Matter Halos from a Complete NIRCam Grism Redshift Survey},} arXiv e-prints, arXiv:2505.02896, \dodoi{10.48550/arXiv.2505.02896}

\bibitem[{A. {Lupi} {et~al.}(2024){Lupi}, {Trinca}, {Volonteri}, {Dotti}, \& {Mazzucchelli}}]{Lupi2024}
{Lupi}, A., {Trinca}, A., {Volonteri}, M., {Dotti}, M., \& {Mazzucchelli}, C. 2024, \bibinfo{title}{{Size matters: are we witnessing super-Eddington accretion in high-redshift black holes from JWST?},} \aap, 689, A128, \dodoi{10.1051/0004-6361/202451249}

\bibitem[{Y. {Ma} {et~al.}(2025){Ma}, {Greene}, {Setton}, {Volonteri}, {Leja}, {Wang}, {Bezanson}, {Brammer}, {Cutler}, {Dayal}, {van Dokkum}, {Furtak}, {Glazebrook}, {Goulding}, {de Graaff}, {Kokorev}, {Labbe}, {Pan}, {Price}, {Weaver}, {Williams}, {Whitaker}, \& {Zitrin}}]{Ma2025}
{Ma}, Y., {Greene}, J.~E., {Setton}, D.~J., {et~al.} 2025, \bibinfo{title}{{UNCOVER: 404 Error{\textemdash}Models Not Found for the Triply Imaged Little Red Dot A2744-QSO1},} \apj, 981, 191, \dodoi{10.3847/1538-4357/ada613}

\bibitem[{P. {Madau} \& F. {Haardt}(2024){Madau} \& {Haardt}}]{Madau2024}
{Madau}, P., \& {Haardt}, F. 2024, \bibinfo{title}{{X-Ray Weak Active Galactic Nuclei from Super-Eddington Accretion onto Infant Black Holes},} \apjl, 976, L24, \dodoi{10.3847/2041-8213/ad90e1}

\bibitem[{R. {Maiolino} {et~al.}(2024){Maiolino}, {Scholtz}, {Curtis-Lake}, {Carniani}, {Baker}, {de Graaff}, {Tacchella}, {{\"U}bler}, {D'Eugenio}, {Witstok}, {Curti}, {Arribas}, {Bunker}, {Charlot}, {Chevallard}, {Eisenstein}, {Egami}, {Ji}, {Jones}, {Lyu}, {Rawle}, {Robertson}, {Rujopakarn}, {Perna}, {Sun}, {Venturi}, {Williams}, \& {Willott}}]{Maiolino2024}
{Maiolino}, R., {Scholtz}, J., {Curtis-Lake}, E., {et~al.} 2024, \bibinfo{title}{{JADES: The diverse population of infant black holes at 4 < z < 11: Merging, tiny, poor, but mighty},} \aap, 691, A145, \dodoi{10.1051/0004-6361/202347640}

\bibitem[{R. {Maiolino} {et~al.}(2025){Maiolino}, {Risaliti}, {Signorini}, {Trefoloni}, {Juod{\v{z}}balis}, {Scholtz}, {{\"U}bler}, {D'Eugenio}, {Carniani}, {Fabian}, {Ji}, {Mazzolari}, {Bertola}, {Brusa}, {Bunker}, {Charlot}, {Comastri}, {Cresci}, {DeCoursey}, {Egami}, {Fiore}, {Gilli}, {Perna}, {Tacchella}, \& {Venturi}}]{Maiolino2025}
{Maiolino}, R., {Risaliti}, G., {Signorini}, M., {et~al.} 2025, \bibinfo{title}{{JWST meets Chandra: a large population of Compton thick, feedback-free, and intrinsically X-ray weak AGN, with a sprinkle of SNe},} \mnras, 538, 1921, \dodoi{10.1093/mnras/staf359}

\bibitem[{J. {Matthee} {et~al.}(2024){Matthee}, {Naidu}, {Brammer}, {Chisholm}, {Eilers}, {Goulding}, {Greene}, {Kashino}, {Labbe}, {Lilly}, {Mackenzie}, {Oesch}, {Weibel}, {Wuyts}, {Xiao}, {Bordoloi}, {Bouwens}, {van Dokkum}, {Illingworth}, {Kramarenko}, {Maseda}, {Mason}, {Meyer}, {Nelson}, {Reddy}, {Shivaei}, {Simcoe}, \& {Yue}}]{Matthee2024}
{Matthee}, J., {Naidu}, R.~P., {Brammer}, G., {et~al.} 2024, \bibinfo{title}{{Little Red Dots: An Abundant Population of Faint Active Galactic Nuclei at z {\ensuremath{\sim}} 5 Revealed by the EIGER and FRESCO JWST Surveys},} \apj, 963, 129, \dodoi{10.3847/1538-4357/ad2345}

\bibitem[{J. {Matthee} {et~al.}(2025){Matthee}, {Naidu}, {Kotiwale}, {Furtak}, {Kramarenko}, {Mackenzie}, {Greene}, {Adamo}, {Bouwens}, {Di Cesare}, {Eilers}, {de Graaff}, {Heintz}, {Kashino}, {Maseda}, {Tacchella}, \& {Torralba}}]{Matthee2024b}
{Matthee}, J., {Naidu}, R.~P., {Kotiwale}, G., {et~al.} 2025, \bibinfo{title}{{Environmental Evidence for Overly Massive Black Holes in Low-mass Galaxies and a Black Hole{\textendash}Halo Mass Relation at z {\ensuremath{\sim}} 5},} \apj, 988, 246, \dodoi{10.3847/1538-4357/ade886}

\bibitem[{J.~C. {McKinney} {et~al.}(2014){McKinney}, {Tchekhovskoy}, {Sadowski}, \& {Narayan}}]{McKinney2014}
{McKinney}, J.~C., {Tchekhovskoy}, A., {Sadowski}, A., \& {Narayan}, R. 2014, \bibinfo{title}{{Three-dimensional general relativistic radiation magnetohydrodynamical simulation of super-Eddington accretion, using a new code HARMRAD with M1 closure},} \mnras, 441, 3177, \dodoi{10.1093/mnras/stu762}

\bibitem[{D. {Mihalas}(1965){Mihalas}}]{Mihalas1965}
{Mihalas}, D. 1965, \bibinfo{title}{{Model Atmospheres and Line Profiles for Early-Type Stars.},} \apjs, 9, 321, \dodoi{10.1086/190104}

\bibitem[{R.~P. {Naidu} {et~al.}(2025){Naidu}, {Matthee}, {Katz}, {de Graaff}, {Oesch}, {Smith}, {Greene}, {Brammer}, {Weibel}, {Hviding}, {Chisholm}, {Labb\textbackslash'e}, {Simcoe}, {Witten}, {Atek}, {Baggen}, {Belli}, {Bezanson}, {Boogaard}, {Bose}, {Covelo-Paz}, {Dayal}, {Fudamoto}, {Furtak}, {Giovinazzo}, {Goulding}, {Gronke}, {Heintz}, {Hirschmann}, {Illingworth}, {Inoue}, {Johnson}, {Leja}, {Leonova}, {McConachie}, {Maseda}, {Natarajan}, {Nelson}, {Setton}, {Shivaei}, {Sobral}, {Stefanon}, {Tacchella}, {Toft}, {Torralba}, {van Dokkum}, {van der Wel}, {Volonteri}, {Walter}, {Wang}, \& {Watson}}]{Naidu2025}
{Naidu}, R.~P., {Matthee}, J., {Katz}, H., {et~al.} 2025, \bibinfo{title}{{A ``Black Hole Star'' Reveals the Remarkable Gas-Enshrouded Hearts of the Little Red Dots},} arXiv e-prints, arXiv:2503.16596, \dodoi{10.48550/arXiv.2503.16596}

\bibitem[{T. {Ohmura} \& H. {Ohmura}(1960){Ohmura} \& {Ohmura}}]{Ohmura1960}
{Ohmura}, T., \& {Ohmura}, H. 1960, \bibinfo{title}{{Electron-Hydrogen Scattering at Low Energies},} Physical Review, 118, 154, \dodoi{10.1103/PhysRev.118.154}

\bibitem[{K. {Ohsuga} {et~al.}(2005){Ohsuga}, {Mori}, {Nakamoto}, \& {Mineshige}}]{Ohsuga2005}
{Ohsuga}, K., {Mori}, M., {Nakamoto}, T., \& {Mineshige}, S. 2005, \bibinfo{title}{{Supercritical Accretion Flows around Black Holes: Two-dimensional, Radiation Pressure-dominated Disks with Photon Trapping},} \apj, 628, 368, \dodoi{10.1086/430728}

\bibitem[{S.~P. {Owocki} {et~al.}(2004){Owocki}, {Gayley}, \& {Shaviv}}]{Owocki2004}
{Owocki}, S.~P., {Gayley}, K.~G., \& {Shaviv}, N.~J. 2004, \bibinfo{title}{{A Porosity-Length Formalism for Photon-Tiring-limited Mass Loss from Stars above the Eddington Limit},} \apj, 616, 525, \dodoi{10.1086/424910}

\bibitem[{F. {Pacucci} \& R. {Narayan}(2024){Pacucci} \& {Narayan}}]{Pacucci2024}
{Pacucci}, F., \& {Narayan}, R. 2024, \bibinfo{title}{{Mildly Super-Eddington Accretion onto Slowly Spinning Black Holes Explains the X-Ray Weakness of the Little Red Dots},} \apj, 976, 96, \dodoi{10.3847/1538-4357/ad84f7}

\bibitem[{ {Planck Collaboration} {et~al.}(2016){Planck Collaboration}, {Ade}, {Aghanim}, {Arnaud}, {Ashdown}, {Aumont}, {Baccigalupi}, {Banday}, {Barreiro}, {Bartlett}, {Bartolo}, {Battaner}, {Battye}, {Benabed}, {Beno{\^\i}t}, {Benoit-L{\'e}vy}, {Bernard}, {Bersanelli}, {Bielewicz}, {Bock}, {Bonaldi}, {Bonavera}, {Bond}, {Borrill}, {Bouchet}, {Boulanger}, {Bucher}, {Burigana}, {Butler}, {Calabrese}, {Cardoso}, {Catalano}, {Challinor}, {Chamballu}, {Chary}, {Chiang}, {Chluba}, {Christensen}, {Church}, {Clements}, {Colombi}, {Colombo}, {Combet}, {Coulais}, {Crill}, {Curto}, {Cuttaia}, {Danese}, {Davies}, {Davis}, {de Bernardis}, {de Rosa}, {de Zotti}, {Delabrouille}, {D{\'e}sert}, {Di Valentino}, {Dickinson}, {Diego}, {Dolag}, {Dole}, {Donzelli}, {Dor{\'e}}, {Douspis}, {Ducout}, {Dunkley}, {Dupac}, {Efstathiou}, {Elsner}, {En{\ss}lin}, {Eriksen}, {Farhang}, {Fergusson}, {Finelli}, {Forni}, {Frailis}, {Fraisse}, {Franceschi}, {Frejsel}, {Galeotta}, {Galli}, {Ganga}, {Gauthier}, {Gerbino}, {Ghosh}, {Giard}, {Giraud-H{\'e}raud}, {Giusarma}, {Gjerl{\o}w}, {Gonz{\'a}lez-Nuevo}, {G{\'o}rski}, {Gratton}, {Gregorio}, {Gruppuso}, {Gudmundsson}, {Hamann}, {Hansen}, {Hanson}, {Harrison}, {Helou}, {Henrot-Versill{\'e}}, {Hern{\'a}ndez-Monteagudo}, {Herranz}, {Hildebrandt}, {Hivon}, {Hobson}, {Holmes}, {Hornstrup}, {Hovest}, {Huang}, {Huffenberger}, {Hurier}, {Jaffe}, {Jaffe}, {Jones}, {Juvela}, {Keih{\"a}nen}, {Keskitalo}, {Kisner}, {Kneissl}, {Knoche}, {Knox}, {Kunz}, {Kurki-Suonio}, {Lagache}, {L{\"a}hteenm{\"a}ki}, {Lamarre}, {Lasenby}, {Lattanzi}, {Lawrence}, {Leahy}, {Leonardi}, {Lesgourgues}, {Levrier}, {Lewis}, {Liguori}, {Lilje}, {Linden-V{\o}rnle}, {L{\'o}pez-Caniego}, {Lubin}, {Mac{\'\i}as-P{\'e}rez}, {Maggio}, {Maino}, {Mandolesi}, {Mangilli}, {Marchini}, {Maris}, {Martin}, {Martinelli}, {Mart{\'\i}nez-Gonz{\'a}lez}, {Masi}, {Matarrese}, {McGehee}, {Meinhold}, {Melchiorri}, {Melin}, {Mendes}, {Mennella}, {Migliaccio}, {Millea}, {Mitra}, {Miville-Desch{\^e}nes}, {Moneti}, {Montier}, {Morgante}, {Mortlock}, {Moss}, {Munshi}, {Murphy}, {Naselsky}, {Nati}, {Natoli}, {Netterfield}, {N{\o}rgaard-Nielsen}, {Noviello}, {Novikov}, {Novikov}, {Oxborrow}, {Paci}, {Pagano}, {Pajot}, {Paladini}, {Paoletti}, {Partridge}, {Pasian}, {Patanchon}, {Pearson}, {Perdereau}, {Perotto}, {Perrotta}, {Pettorino}, {Piacentini}, {Piat}, {Pierpaoli}, {Pietrobon}, {Plaszczynski}, {Pointecouteau}, {Polenta}, {Popa}, {Pratt}, \& {Pr{\'e}zeau}}]{PlanckColl2016}
{Planck Collaboration}, {Ade}, P.~A.~R., {Aghanim}, N., {et~al.} 2016, \bibinfo{title}{{Planck 2015 results. XIII. Cosmological parameters},} \aap, 594, A13, \dodoi{10.1051/0004-6361/201525830}

\bibitem[{P. {Rinaldi} {et~al.}(2024){Rinaldi}, {Bonaventura}, {Rieke}, {Alberts}, {Caputi}, {Baker}, {Baum}, {Bhatawdekar}, {Bunker}, {Carniani}, {Curtis-Lake}, {D'Eugenio}, {Egami}, {Ji}, {Hainline}, {Helton}, {Lin}, {Lyu}, {Johnson}, {Ma}, {Maiolino}, {P{\'e}rez-Gonz{\'a}lez}, {Rieke}, {Robertson}, {Shivaei}, {Stone}, {Sun}, {Tacchella}, {{\"U}bler}, {Williams}, {Willmer}, {Willott}, {Zhang}, \& {Zhu}}]{Rinaldi2024}
{Rinaldi}, P., {Bonaventura}, N., {Rieke}, G.~H., {et~al.} 2024, \bibinfo{title}{{Not Just a Dot: the complex UV morphology and underlying properties of Little Red Dots},} arXiv e-prints, arXiv:2411.14383, \dodoi{10.48550/arXiv.2411.14383}

\bibitem[{R.~D. {Rohrmann}(2018){Rohrmann}}]{Rohrmann2018}
{Rohrmann}, R.~D. 2018, \bibinfo{title}{{Rayleigh scattering in dense fluid helium},} \mnras, 473, 457, \dodoi{10.1093/mnras/stx2440}

\bibitem[{V. {Rusakov} {et~al.}(2025){Rusakov}, {Watson}, {Nikopoulos}, {Brammer}, {Gottumukkala}, {Harvey}, {Heintz}, {Nielsen}, {Sim}, {Sneppen}, {Vijayan}, {Adams}, {Austin}, {Conselice}, {Goolsby}, {Toft}, \& {Witstok}}]{Rusakov2025}
{Rusakov}, V., {Watson}, D., {Nikopoulos}, G.~P., {et~al.} 2025, \bibinfo{title}{{JWST's little red dots: an emerging population of young, low-mass AGN cocooned in dense ionized gas},} arXiv e-prints, arXiv:2503.16595, \dodoi{10.48550/arXiv.2503.16595}

\bibitem[{G.~B. {Rybicki} \& A.~P. {Lightman}(1979){Rybicki} \& {Lightman}}]{Rybicki1979}
{Rybicki}, G.~B., \& {Lightman}, A.~P. 1979, {Radiative processes in astrophysics}

\bibitem[{A. {Sacchi} \& A. {Bogdan}(2025){Sacchi} \& {Bogdan}}]{Sacchi2025}
{Sacchi}, A., \& {Bogdan}, A. 2025, \bibinfo{title}{{Chandra Rules Out Super-Eddington Accretion For Little Red Dots},} arXiv e-prints, arXiv:2505.09669, \dodoi{10.48550/arXiv.2505.09669}

\bibitem[{A. Sadowski {et~al.}(2014)Sadowski, Narayan, McKinney, \& Tchekhovskoy}]{Sadowski2014}
Sadowski, A., Narayan, R., McKinney, J.~C., \& Tchekhovskoy, A. 2014, \bibinfo{title}{Numerical simulations of super-critical black hole accretion flows in general relativity,} Monthly Notices of the Royal Astronomical Society, 439, 503, \dodoi{10.1093/mnras/stt2479}

\bibitem[{A. {Sadowski} {et~al.}(2015){Sadowski}, {Narayan}, {Tchekhovskoy}, {Abarca}, {Zhu}, \& {McKinney}}]{Sadowski2015}
{Sadowski}, A., {Narayan}, R., {Tchekhovskoy}, A., {et~al.} 2015, \bibinfo{title}{{Global simulations of axisymmetric radiative black hole accretion discs in general relativity with a mean-field magnetic dynamo},} \mnras, 447, 49, \dodoi{10.1093/mnras/stu2387}

\bibitem[{A. {Schuster}(1905){Schuster}}]{Schuster1905}
{Schuster}, A. 1905, \bibinfo{title}{{Radiation Through a Foggy Atmosphere},} \apj, 21, 1, \dodoi{10.1086/141186}

\bibitem[{D.~J. {Setton} {et~al.}(2024){Setton}, {Greene}, {de Graaff}, {Ma}, {Leja}, {Matthee}, {Bezanson}, {Boogaard}, {Cleri}, {Katz}, {Labbe}, {Maseda}, {McConachie}, {Miller}, {Price}, {Suess}, {van Dokkum}, {Wang}, {Weibel}, {Whitaker}, \& {Williams}}]{Setton2024}
{Setton}, D.~J., {Greene}, J.~E., {de Graaff}, A., {et~al.} 2024, \bibinfo{title}{{Little Red Dots at an Inflection Point: Ubiquitous ``V-Shaped'' Turnover Consistently Occurs at the Balmer Limit},} arXiv e-prints, arXiv:2411.03424, \dodoi{10.48550/arXiv.2411.03424}

\bibitem[{D.~J. {Setton} {et~al.}(2025){Setton}, {Greene}, {Spilker}, {Williams}, {Labb{\'e}}, {Ma}, {Wang}, {Whitaker}, {Leja}, {de Graaff}, {Alberts}, {Bezanson}, {Boogaard}, {Brammer}, {Cutler}, {Cleri}, {Cooper}, {Dayal}, {Fujimoto}, {Furtak}, {Goulding}, {Hirschmann}, {Kokorev}, {Maseda}, {McConachie}, {Matthee}, {Miller}, {Naidu}, {Oesch}, {Pan}, {Price}, {Suess}, {Weaver}, {Xiao}, {Zhang}, \& {Zitrin}}]{Setton2025}
{Setton}, D.~J., {Greene}, J.~E., {Spilker}, J.~S., {et~al.} 2025, \bibinfo{title}{{A Confirmed Deficit of Hot and Cold Dust Emission in the Most Luminous Little Red Dots},} \apjl, 991, L10, \dodoi{10.3847/2041-8213/ade78b}

\bibitem[{N.~I. {Shakura} \& R.~A. {Sunyaev}(1973){Shakura} \& {Sunyaev}}]{Shakura1973}
{Shakura}, N.~I., \& {Sunyaev}, R.~A. 1973, \bibinfo{title}{{Black holes in binary systems. Observational appearance.},} \aap, 24, 337

\bibitem[{J.~W. {Stock} {et~al.}(2022){Stock}, {Kitzmann}, \& {Patzer}}]{Stock2022}
{Stock}, J.~W., {Kitzmann}, D., \& {Patzer}, A. B.~C. 2022, \bibinfo{title}{{FASTCHEM 2 : an improved computer program to determine the gas-phase chemical equilibrium composition for arbitrary element distributions},} \mnras, 517, 4070, \dodoi{10.1093/mnras/stac2623}

\bibitem[{J.~M. Stone {et~al.}(2020)Stone, Tomida, White, \& Felker}]{Stone2020}
Stone, J.~M., Tomida, K., White, C.~J., \& Felker, K.~G. 2020, \bibinfo{title}{The Athena$\mathplus$$\mathplus$ Adaptive Mesh Refinement Framework: Design and Magnetohydrodynamic Solvers,} The Astrophysical Journal Supplement Series, 249, 4, \dodoi{10.3847/1538-4365/ab929b}

\bibitem[{S.~P. {Tarafdar} \& M.~S. {Vardya}(1973){Tarafdar} \& {Vardya}}]{Tarafdar1973}
{Tarafdar}, S.~P., \& {Vardya}, M.~S. 1973, \bibinfo{title}{{The importance of molecular Rayleigh scattering in the atmospheres of very late-type stars.},} \mnras, 163, 261, \dodoi{10.1093/mnras/163.3.261}

\bibitem[{A.~J. {Taylor} {et~al.}(2025){Taylor}, {Kokorev}, {Kocevski}, {Akins}, {Cullen}, {Dickinson}, {Finkelstein}, {Arrabal Haro}, {Bromm}, {Giavalisco}, {Inayoshi}, {Juneau}, {Leung}, {Perez-Gonzalez}, {Somerville}, {Trump}, {Amorin}, {Barro}, {Burgarella}, {Brooks}, {Carnall}, {Casey}, {Cheng}, {Chisholm}, {Chworowsky}, {Davis}, {Donnan}, {Dunlop}, {Ellis}, {Fernandez}, {Fujimoto}, {Grogin}, {Gupta}, {Hathi}, {Jung}, {Hirschmann}, {Kartaltepe}, {Koekemoer}, {Larson}, {Leung}, {Llerena}, {Lucas}, {McLeod}, {McLure}, {Napolitano}, {Papovich}, {Stanton}, {Tripodi}, {Wang}, {Wilkins}, {Yung}, \& {Zavala}}]{Taylor2025}
{Taylor}, A.~J., {Kokorev}, V., {Kocevski}, D.~D., {et~al.} 2025, \bibinfo{title}{{CAPERS-LRD-z9: A Gas Enshrouded Little Red Dot Hosting a Broad-line AGN at z=9.288},} arXiv e-prints, arXiv:2505.04609, \dodoi{10.48550/arXiv.2505.04609}

\bibitem[{A. {Trinca} {et~al.}(2024){Trinca}, {Valiante}, {Schneider}, {Juod{\v{z}}balis}, {Maiolino}, {Graziani}, {Lupi}, {Natarajan}, {Volonteri}, \& {Zana}}]{Trinca2024}
{Trinca}, A., {Valiante}, R., {Schneider}, R., {et~al.} 2024, \bibinfo{title}{{Episodic super-Eddington accretion as a clue to Overmassive Black Holes in the early Universe},} arXiv e-prints, arXiv:2412.14248, \dodoi{10.48550/arXiv.2412.14248}

\bibitem[{H. {{\"U}bler} {et~al.}(2023){{\"U}bler}, {Maiolino}, {Curtis-Lake}, {P{\'e}rez-Gonz{\'a}lez}, {Curti}, {Perna}, {Arribas}, {Charlot}, {Marshall}, {D'Eugenio}, {Scholtz}, {Bunker}, {Carniani}, {Ferruit}, {Jakobsen}, {Rix}, {Rodr{\'\i}guez Del Pino}, {Willott}, {Boeker}, {Cresci}, {Jones}, {Kumari}, \& {Rawle}}]{Ubler2023}
{{\"U}bler}, H., {Maiolino}, R., {Curtis-Lake}, E., {et~al.} 2023, \bibinfo{title}{{GA-NIFS: A massive black hole in a low-metallicity AGN at z {\ensuremath{\sim}} 5.55 revealed by JWST/NIRSpec IFS},} \aap, 677, A145, \dodoi{10.1051/0004-6361/202346137}

\bibitem[{A.~B. {Underhill}(1949){Underhill}}]{Underhill1949}
{Underhill}, A.~B. 1949, \bibinfo{title}{{Transfer Problems in an Atmosphere with Continuous Scattering and Continuous Absorption.},} \apj, 110, 340, \dodoi{10.1086/145211}

\bibitem[{P.~A.~M. {van Hoof} {et~al.}(2014){van Hoof}, {Williams}, {Volk}, {Chatzikos}, {Ferland}, {Lykins}, {Porter}, \& {Wang}}]{vanHoof2014}
{van Hoof}, P.~A.~M., {Williams}, R.~J.~R., {Volk}, K., {et~al.} 2014, \bibinfo{title}{{Accurate determination of the free-free Gaunt factor - I. Non-relativistic Gaunt factors},} \mnras, 444, 420, \dodoi{10.1093/mnras/stu1438}

\bibitem[{D.~A. {Verner} {et~al.}(1996){Verner}, {Ferland}, {Korista}, \& {Yakovlev}}]{Verner1996}
{Verner}, D.~A., {Ferland}, G.~J., {Korista}, K.~T., \& {Yakovlev}, D.~G. 1996, \bibinfo{title}{{Atomic Data for Astrophysics. II. New Analytic Fits for Photoionization Cross Sections of Atoms and Ions},} \apj, 465, 487, \dodoi{10.1086/177435}

\bibitem[{D.~A. {Verner} \& D.~G. {Yakovlev}(1995){Verner} \& {Yakovlev}}]{Verner1995}
{Verner}, D.~A., \& {Yakovlev}, D.~G. 1995, \bibinfo{title}{{Analytic FITS for partial photoionization cross sections.},} \aaps, 109, 125

\bibitem[{P. Virtanen {et~al.}(2020)Virtanen, Gommers, Oliphant, Haberland, Reddy, Cournapeau, Burovski, Peterson, Weckesser, Bright, {van der Walt}, Brett, Wilson, Millman, Mayorov, Nelson, Jones, Kern, Larson, Carey, Polat, Feng, Moore, {VanderPlas}, Laxalde, Perktold, Cimrman, Henriksen, Quintero, Harris, Archibald, Ribeiro, Pedregosa, {van Mulbregt}, \& {SciPy 1.0 Contributors}}]{Scipy2020}
Virtanen, P., Gommers, R., Oliphant, T.~E., {et~al.} 2020, \bibinfo{title}{{{SciPy} 1.0: Fundamental Algorithms for Scientific Computing in Python},} Nature Methods, 17, 261, \dodoi{10.1038/s41592-019-0686-2}

\bibitem[{M. {Volonteri} {et~al.}(2021){Volonteri}, {Habouzit}, \& {Colpi}}]{Volonteri2021}
{Volonteri}, M., {Habouzit}, M., \& {Colpi}, M. 2021, \bibinfo{title}{{The origins of massive black holes},} Nature Reviews Physics, 3, 732, \dodoi{10.1038/s42254-021-00364-9}

\bibitem[{B. {Wang} {et~al.}(2024){Wang}, {Leja}, {de Graaff}, {Brammer}, {Weibel}, {van Dokkum}, {Baggen}, {Suess}, {Greene}, {Bezanson}, {Cleri}, {Hirschmann}, {Labb{\'e}}, {Matthee}, {McConachie}, {Naidu}, {Nelson}, {Oesch}, {Setton}, \& {Williams}}]{Wang2024}
{Wang}, B., {Leja}, J., {de Graaff}, A., {et~al.} 2024, \bibinfo{title}{{RUBIES: Evolved Stellar Populations with Extended Formation Histories at z {\ensuremath{\sim}} 7{\textendash}8 in Candidate Massive Galaxies Identified with JWST/NIRSpec},} \apjl, 969, L13, \dodoi{10.3847/2041-8213/ad55f7}

\bibitem[{B. {Wang} {et~al.}(2025){Wang}, {de Graaff}, {Davies}, {Greene}, {Leja}, {Brammer}, {Goulding}, {Miller}, {Suess}, {Weibel}, {Williams}, {Bezanson}, {Boogaard}, {Cleri}, {Hirschmann}, {Katz}, {Labb{\'e}}, {Maseda}, {Matthee}, {McConachie}, {Naidu}, {Oesch}, {Rix}, {Setton}, \& {Whitaker}}]{Wang2025}
{Wang}, B., {de Graaff}, A., {Davies}, R.~L., {et~al.} 2025, \bibinfo{title}{{RUBIES: JWST/NIRSpec Confirmation of an Infrared-luminous, Broad-line Little Red Dot with an Ionized Outflow},} \apj, 984, 121, \dodoi{10.3847/1538-4357/adc1ca}

\bibitem[{M.~L. Waskom(2021)Waskom}]{Waskom2021}
Waskom, M.~L. 2021, \bibinfo{title}{seaborn: statistical data visualization,} Journal of Open Source Software, 6, 3021, \dodoi{10.21105/joss.03021}

\bibitem[{C.~C. {Williams} {et~al.}(2024){Williams}, {Alberts}, {Ji}, {Hainline}, {Lyu}, {Rieke}, {Endsley}, {Suess}, {Sun}, {Johnson}, {Florian}, {Shivaei}, {Rujopakarn}, {Baker}, {Bhatawdekar}, {Boyett}, {Bunker}, {Cameron}, {Carniani}, {Charlot}, {Curtis-Lake}, {DeCoursey}, {de Graaff}, {Egami}, {Eisenstein}, {Gibson}, {Hausen}, {Helton}, {Maiolino}, {Maseda}, {Nelson}, {P{\'e}rez-Gonz{\'a}lez}, {Rieke}, {Robertson}, {Saxena}, {Tacchella}, {Willmer}, \& {Willott}}]{Williams2024}
{Williams}, C.~C., {Alberts}, S., {Ji}, Z., {et~al.} 2024, \bibinfo{title}{{The Galaxies Missed by Hubble and ALMA: The Contribution of Extremely Red Galaxies to the Cosmic Census at 3 < z < 8},} \apj, 968, 34, \dodoi{10.3847/1538-4357/ad3f17}

\bibitem[{M. {Xiao} {et~al.}(2025){Xiao}, {Oesch}, {Bing}, {Elbaz}, {Matthee}, {Fudamoto}, {Fujimoto}, {Marques-Chaves}, {Williams}, {Dessauges-Zavadsky}, {Valentino}, {Brammer}, {Covelo-Paz}, {Daddi}, {Fynbo}, {Gillman}, {Ginolfi}, {Giovinazzo}, {Greene}, {Gu}, {Illingworth}, {Inayoshi}, {Kokorev}, {Meyer}, {Naidu}, {Reddy}, {Schaerer}, {Shapley}, {Stefanon}, {Steinhardt}, {Setton}, {Vestergaard}, \& {Wang}}]{Xiao2025}
{Xiao}, M., {Oesch}, P.~A., {Bing}, L., {et~al.} 2025, \bibinfo{title}{{No [CII] or dust detection in two Little Red Dots at z$_{\rm spec}$ > 7},} arXiv e-prints, arXiv:2503.01945, \dodoi{10.48550/arXiv.2503.01945}

\bibitem[{M. {Yan} {et~al.}(2001){Yan}, {Sadeghpour}, \& {Dalgarno}}]{Yan2001}
{Yan}, M., {Sadeghpour}, H.~R., \& {Dalgarno}, A. 2001, \bibinfo{title}{{Erratum: Photoionization Cross Sections of He and H$_{2}$},} \apj, 559, 1194, \dodoi{10.1086/322775}

\bibitem[{M. {Yue} {et~al.}(2024){Yue}, {Eilers}, {Ananna}, {Panagiotou}, {Kara}, \& {Miyaji}}]{Yue2024}
{Yue}, M., {Eilers}, A.-C., {Ananna}, T.~T., {et~al.} 2024, \bibinfo{title}{{Stacking X-Ray Observations of ``Little Red Dots'': Implications for Their Active Galactic Nucleus Properties},} \apjl, 974, L26, \dodoi{10.3847/2041-8213/ad7eba}

\bibitem[{Z. {Zhang} {et~al.}(2025){Zhang}, {Jiang}, {Liu}, \& {Ho}}]{ZhangZ2025}
{Zhang}, Z., {Jiang}, L., {Liu}, W., \& {Ho}, L.~C. 2025, \bibinfo{title}{{Analysis of Multi-epoch JWST Images of {\ensuremath{\sim}}300 Little Red Dots: Tentative Detection of Variability in a Minority of Sources},} \apj, 985, 119, \dodoi{10.3847/1538-4357/adcb3e}

\bibitem[{M.-Y. {Zhuang} {et~al.}(2025){Zhuang}, {Li}, {Shen}, {Lin}, {Shapley}, {Wang}, {Wu}, \& {Yang}}]{Zhuang2025}
{Zhuang}, M.-Y., {Li}, J., {Shen}, Y., {et~al.} 2025, \bibinfo{title}{{NEXUS: A Spectroscopic Census of Broad-line AGNs and Little Red Dots at $3\lesssim z\lesssim 6$},} arXiv e-prints, arXiv:2505.20393, \dodoi{10.48550/arXiv.2505.20393}

\end{thebibliography}
\bibliographystyle{aasjournalv7}
\end{CJK*}

\end{document}